\documentclass[fleqn,usenatbib]{mnras}

\usepackage{newtxtext,newtxmath}

\usepackage[T1]{fontenc}

\DeclareRobustCommand{\VAN}[3]{#2}
\let\VANthebibliography\thebibliography
\def\thebibliography{\DeclareRobustCommand{\VAN}[3]{##3}\VANthebibliography}

\usepackage{graphicx}	
\usepackage{amsmath}	
\usepackage{tabularx}

\newcommand{\change}{\color{black}}
\newcommand{\MSUN}{\rm{M}_{\odot}}
\newcommand{\logmstar}{\log(M_{\star}/\mathrm{M}_{\odot})}

\usepackage[svgnames]{xcolor}
\usepackage{xstring}
\usepackage{soul}

\title[Comparing IllustrisTNG and HSC galaxies]{ERGO-ML: Comparing IllustrisTNG and HSC galaxy images via contrastive learning}

\author[L. Eisert et al.]{Lukas Eisert,$^{1}$\thanks{E-mail: eisert@mpia.de}
Connor Bottrell,$^{2,3,4}$ 
Annalisa Pillepich,$^{1}$  Rhythm Shimakawa,$^{5,6}$ Vicente Rodriguez-Gomez,$^{7}$
\newauthor
Dylan Nelson,$^{8}$ Eirini Angeloudi,$^{9}$ and Marc Huertas-Company$^{9}$
\\\\
$^{1}$ Max-Planck-Institut f{\"u}r Astronomie, K{\"o}nigstuhl 17, 69117 Heidelberg, Germany\\
$^{2}$ International Centre for Radio Astronomy Research, University of Western Australia, Stirling Hwy, Crawley, WA 6009, Australia \\ 
$^{3}$ Kavli Institute for the Physics and Mathematics of the Universe (WPI), UTIAS, University of Tokyo, Kashiwa, Chiba 277-8583, Japan\\
$^{4}$ Center for Data-Driven Discovery, Kavli IPMU (WPI), UTIAS, The University of Tokyo, Kashiwa, Chiba 277-8583, Japan\\
$^{5}$ Waseda Institute for Advanced Study (WIAS), Waseda University, Nishi Waseda, Shinjuku, Tokyo 169-0051, Japan\\
$^{6}$ Center for Data Science, Waseda University, 1-6-1, Nishi-Waseda, Shinjuku, Tokyo 169-0051, Japan\\
$^{7}$ Instituto de Radioastronom\'ia y Astrof\'isica, Universidad Nacional Aut\'onoma de M\'exico, Apdo. Postal 72-3, 58089 Morelia, Mexico\\
$^{8}$ Universit\"{a}t Heidelberg, Zentrum f\"{u}r Astronomie, Institut f\"{u}r theoretische Astrophysik, Albert-Ueberle-Str. 2, 69120 Heidelberg, Germany\\
$^{9}$ Departamento de Astrof\'{i}sica, Instituto de Astrof\'{i}sica de Canarias, Universidad de La Laguna, E-38200 La Laguna, Spain \\
}

\date{Accepted XXX. Received YYY; in original form ZZZ}

\pubyear{2023}

\begin{document}
\label{firstpage}
\pagerange{\pageref{firstpage}--\pageref{lastpage}}
\maketitle

\begin{abstract}
Modern cosmological hydrodynamical galaxy simulations provide tens of thousands of reasonably realistic synthetic galaxies across cosmic time. However, quantitatively assessing the level of realism of simulated universes in comparison to the real one is difficult. In this paper of the ERGO-ML series (Extracting Reality from Galaxy Observables with Machine Learning), we utilize contrastive learning to directly compare a large sample of simulated and observed galaxies based on their stellar-light images. This eliminates the need to specify summary statistics and allows to exploit the whole information content of the observations. We produce survey-realistic galaxy mock datasets resembling real Hyper Suprime-Cam (HSC) observations using the cosmological simulations TNG50 and TNG100. Our focus is on galaxies with stellar masses between $10^9$ and $10^{12} M_\odot$ at $z=0.1-0.4$. This allows us to evaluate the realism of the simulated TNG galaxies in comparison to actual HSC observations. We apply the self-supervised contrastive learning method NNCLR to the images from both simulated and observed datasets (g, r, i - bands). This results in a 256-dimensional representation space, encoding all relevant observable galaxy properties. Firstly, this allows us to identify simulated galaxies that closely resemble real ones by seeking similar images in this multi-dimensional space. Even more powerful, we quantify the alignment between the representations of these two image sets, finding that the majority ($\gtrsim 70$ per cent) of the TNG galaxies align well with observed HSC images. However, a subset of simulated galaxies with larger sizes, steeper Sersic profiles, smaller Sersic ellipticities, and larger asymmetries appears unrealistic. We also demonstrate the utility of our derived image representations by inferring properties of real HSC galaxies using simulated TNG galaxies as the ground truth.
\end{abstract}



\section{Introduction}
\label{sec:intro}
The morphological structures of observed galaxies are rich in both complexity and diversity. In the nearby Universe, the stellar structures of some galaxies can be well-characterized by simple analytic models, e.g. smooth ellipticals and flattened disks. Meanwhile, the morphological menagerie also includes galaxies with tightly/loosely wound spiral structures, stellar streams and shells, lopsidedness, and other generally complicated structures. The categorization and characterization of these stellar morphologies, in conjunction with theoretical models, has yielded significant insight into their evolutionary pathways \citep{Buta2013,2014ARA&A..52..291C,Masters_2020}.

Modern cosmological hydrodynamical simulations of galaxies, like e.g. IllustrisTNG (see below), EAGLE \citep{Eagle_1, Eagle_2}, HorizonAGN \citep{HorizonAGN}, and SIMBA \citep{SIMBA}, are nowadays able to produce diverse morphological galaxy populations that resemble those seen in observations. However, directly and quantitatively comparing the morphologies of simulated galaxies to observed ones is not a simple task. 

Traditionally the comparison between simulated and observed galaxies has been done by deriving scalar properties or profiles that summarize the high dimensional observed data, e.g. images and spectra. These scalar properties can be measured for the simulated galaxies and then validated against the set of observed ones \citep[e.g.][]{Vogelsberger_2014, Genel_2014, Eagle_1, Eagle_2}.
However, whereas the structures and summary statistics of observed galaxies are derived from images that are based on light and that necessarily include observational limitations (such as depth), additional steps are required to extract equivalent and self-consistently derived structural quantities from the simulation data. 
This has been achieved in recent years by constructing mocks of simulated galaxies as if they were observed in actual surveys \citep[e.g.][]{Torrey_2015, Bottrell2017, Gomez_2019, Bignone_2020, Sarmiento2023, Snyder_2023}. In turn, these mocks have been used to extract scalar properties and summary statistics using the same methods as observers  \citep[e.g.][]{Snyder_2015, Lange_2019, Merritt_2020, Graaff_2022, Ghosh_2023, Ortega_2023, Alonso_2023, Ardila_2021}. 

Notwithstanding the progress, the `traditional way' of comparing simulated and observed galaxies still suffers from the limitation of restricting the comparisons to a limited set of scalar properties or summary statistics: important information content or cross correlations in the data may be missed. 
%
This caveat can be avoided at once by comparing simulations and observations directly at the map level: i.e. by contrasting full images of real galaxies to survey-realistic mocks of simulated galaxies. This ensures not only that all the available information contained in the images is used, but also bypasses the non-trivial choice and construction of summary statistics on which to frame the comparison: e.g., galaxy sizes \citep{van_der_Wel_2014}, light profiles \citep[e.g. Sersic profiles,][]{Sersic_1963, Graham_2005}, morphological estimators \citep[e.g. Gini-M20 parameters,][]{Lotz_2008}, stellar colors \citep{Kawinwanichakij_2021}, etc.

Recently, machine learning methods have been proposed to cope with the complex task of evaluating and comparing simulated and observed images \citep[e.g.][]{Huertas_2019, Pearson_2019, Wang_2020, Margalef_2020, Zanisi_2021}. Especially of interest are the class of so called `contrastive learning' methods \citep[see][for an overview]{Company_2023}, which are self-supervised algorithms able to identify features of interest from e.g. images without prior labeling of the galaxies \citep{Vega-Ferrero_2023}. In doing so, it is therefore possible to circumvent the difficult task of manually identifying and calculating labels overall. 

In this paper, we build upon the prior successes of machine learning applied to galaxy images by implementing a novel contrastive learning method on a vastly larger set of observed and simulated galaxies. Our method accomplishes this by clustering images that have common features and repelling images which do not. As such, the model naturally identifies points of tension (i.e. contrasts) and commonalities within the imaging data. 
We simultaneously apply the method to images of galaxies from the Public Data Release 3 of the Hyper Suprime-Cam Subaru Strategic Program (HSC-SSP PDR3, \citealt{HSC_DR3}) and from HSC-SSP mock images of galaxies selected from the IllustrisTNG cosmological magneto-hydrodynamical simulations \citep[TNG hereafter, ][]{Springel_2017, naiman2017results, Marinacci_2018, nelson2017results, Nelson_2019, Pillepich_2018, Pillepich_2019}.
By exposing the model to both real galaxy images and mock images of simulated galaxies simultaneously, instead of training the model on simulations and observations separately, we give the model every opportunity to focus on the potential differences between observations and simulations, as the model is designed to seek points of contrast between images. In practice, in our methodology, each galaxy image is placed into a high-dimension representation space that can then be explored for comparing the simulated and observed data sets. We do so by introducing an Out-of-Domain score for each image and by quantifying and comparing the score distributions across the galaxy populations. We also further reduce the representation spaces of the galaxy datasets into 2-dimensional UMAPs and visually inspect and compare them and connect them to known galaxy properties.

The method developed and discussed in this paper is especially important for the broader scope of the ERGO-ML (Extracting Reality from Galaxy Observables with Machine Learning) project, whereby we aim at inferring unobservable physical features of galaxies from observations. In two previous papers we have in fact shown that it is possible to infer posteriors for unobservable features, such as the mass fraction of accreted vs. in-situ stars \citep{Shi_2022, Angeloudi_2023}, by using a limited set of observable galaxy scalars \citep{Eisert_2023}. We have shown that this is possible in the context of simulated galaxies, from e.g. the TNG100 simulation of the aforementioned TNG project. In order to extend this harvesting opportunity to real galaxies and to their image data, we first need to investigate and evaluate how similar images from simulations and observations are, i.e. if their domains are comparable.

\begin{figure*}
	\centering
	\includegraphics[width=15cm]{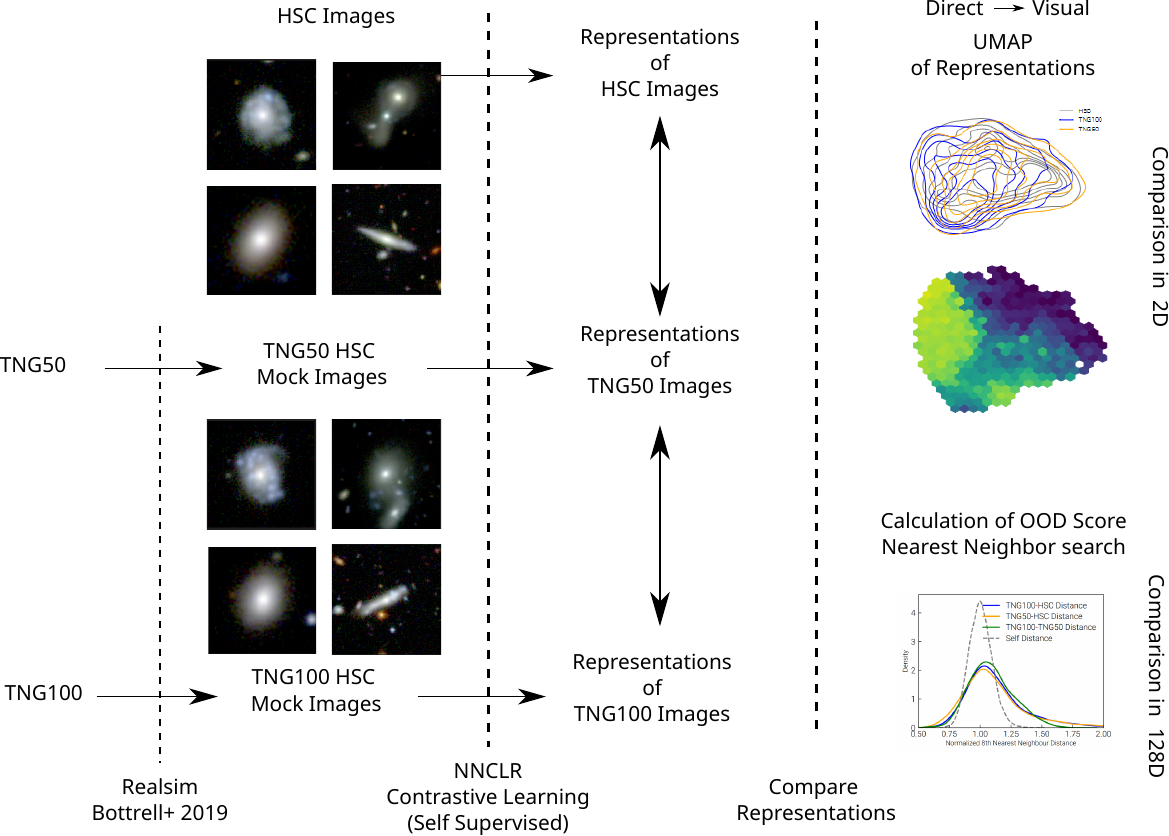}
	\caption{{\bf Overview over the method proposed in this paper to compare observed and simulated galaxies at the image level.} We use a set of Hyper Suprime Cam (HSC) galaxy images and survey-realistic mocks of TNG50 and TNG100 simulated galaxies. We then use them together to train the contrastive learning method NNCLR, in order to map all images into the same representation space. This allows us in the following to compare the samples either in the representation space itself, via an Out-Of-Domain score and its distributions, or visually in a further dimensionally-reduced 2D UMAP representation.}
	\label{fig:concept}
\end{figure*}

The overall scope of this work is shown in Figure \ref{fig:concept}. 
First we introduce the data sets and the preparation steps in Section \ref{sec:data}. We then introduce in Section \ref{sec:methods} the Machine Learning Methods used throughout. We hence present the results of our study in the following Sections: an overview of the galaxy features that have been learnt by our self-supervised model in Section \ref{sec:umap} and the quantitative comparison between TNG (simulated) and HSC (observed) galaxies in Sections~\ref{sec:similarity} and \ref{sec:discussion}. We summarize ideas and findings in Section~\ref{sec:summary}.

In this work we use a cosmology consistent with the results from \citep[][]{Planck_2015}, $\Omega_{\Lambda,0} = 0.6911$, $\Omega_{m,0} = 0.3089$,
$\Omega_{b,0}= 0.0486$, $\sigma_8 = 0.8159$, $n_s = 0.9667$ and $h = 0.6774$.

\section{Galaxy Image Data}
\label{sec:data}
Three distinct sets of galaxy images are used in this work:
\begin{enumerate}
    \item Observed Hyper-Supreme-Cam images of real galaxies: Section~\ref{sec:hsc}.
    \item Survey-realistic mock images of galaxies simulated within the TNG100 simulation: Section~\ref{sec:tng}.
    \item The same as above but using galaxies from the higher-resolution TNG50 simulation, which adds galaxies especially at the low-mass end of our sample: these are also explained in Section~\ref{sec:tng}.
\end{enumerate}
By using two flagship runs of the IllustrisTNG simulation model with different resolutions, we can a) build intuition about the meaning of the results via relative comparisons and b) investigate how simulation resolution affects the visual appearances of galaxies. In the following subsections, we detail the datasets and steps necessary to prepare the raw data for our machine learning model. It is worth noting that we apply exactly the same steps to all three datasets: this ensures that we do not introduce any artificial biases into the data.

\subsection{Subaru Hyper Suprime-Cam images}
\label{sec:hsc}
We obtain a large sample of galaxy images from the Public Data Release 3 of the Hyper Suprime-Cam Subaru Strategic Program (HSC-SSP; \citealt{HSC_DR3}). 

The HSC-SSP is a 330 night observing program with the Subaru 8.2m telescope on Mauna Kea in Hawaii \citep{Miyazaki_2012, Miyazaki_2018, Komiyama_2018, Furusawa_2018}. The survey comprises several layers ranging in survey footprint size and depth. We focus on the Wide Layer, which has median $i$-band seeing of 0.61 arcseconds, $5\sigma$ point-source depth of $26.2$ AB mag, and final survey area of 1200 square degrees -- of these, 670 square degrees are provided in PDR3. Individual exposures are obtained in $5$ optical broad-band filters $grizy$ \citep{Kawanomoto_2018} and co-added to produce final depth images. 

For this paper, we select a galaxy sample from HSC-SSP using a similar process to that described in \cite{Shimakawa_2022} -- which cross-matches sources in the HSC-SSP Wide Layer to their counterparts in the Sloan Digital Sky Survey (DR16, \citealt{SDSS_DR_16}). The selection first takes all SDSS DR16 objects that (1) have Petrosian magnitudes $r_{\rm petro} < 20$ mag and (2) are within the HSC-SSP footprint. This list is then cross-matched to HSC-SSP sources with detections in each of the $grizy$ broad bands using a 1 arcsec matching tolerance. Exactly as in \cite{Shimakawa_2022}, additional quality constraints are then placed on the cross-matched sample to e.g. remove objects whose photometry is contaminated by bright stars, bad pixels and cosmic rays \citep{2018PASJ...70S...7C,Bosch_2018}, and those that do not satisfy a conservative star/galaxy separation cut \citep{Strauss_2002, Baldry_2010}. This selection method is distinguished from the original one of \cite{Shimakawa_2022} by their use of the SDSS galaxy spec-$z$ sample \citep{Strauss_2002} as a reference catalogue. We thereby lose the more reliable spec-$z$ information from SDSS but gain a much larger dataset for our machine learning approach.
In this work we use redshifts mainly for matching the observed galaxy populations to the ones from the simulations at the same cosmic epoch -- we focus throughout on the interval $0.1 \leq z \leq 0.4$. According to Figure 4 of \cite{2018PASJ...70S...9T}, the estimated mean and dispersion of the photo-$z$ bias, $(z_{\mathrm{phot}} - z_{\mathrm{spec}}) / (1+z_{\mathrm{spec}})$, over our chosen redshift interval are $<0.01$ and approximately $0.06$, respectively. We therefore adopt the less accurate, but to this aim \emph{sufficiently} accurate, photo-$z$s from \cite{2018PASJ...70S...9T} for epoch-matching between simulated and observed galaxies.

We start from cutouts centered at each galaxy center with a fixed edge-size of 50 arcseconds including galaxies with photometric redshifts between $z=[0.1, 0.4]$. In total our sample contains r,g and i-band images of 768484 galaxies. 

\subsection{TNG simulated galaxies and their HSC mock images}
\label{sec:tng}
We use galaxies from the IllustrisTNG simulations \citep[][]{Springel_2017, naiman2017results, Marinacci_2018, nelson2017results, Nelson_2019, Pillepich_2018, Pillepich_2019} for our contrastive comparison to real galaxies from the HSC-SSP described above. 

TNG is a suite of full cosmological magneto-hydro simulations of galaxies based on the moving-mesh code \textsc{AREPO} \citep{Springel_2010}, which traces the formation and evolution of galaxies based on cold dark-matter particles and baryonic matter (gas cells, stars and black holes particles) by including forces between them, star formation, and feedback processes. The periodic-boundary cubic simulation boxes are evolved from redshift $z=127$ to $z=0$ with initial conditions given by the Zeldovich approximation \citep[via the N-GenIC code][]{2005Natur.435..629S}. The underlying cosmology is given by Planck \citep{Planck_2015}.

In this work we use the outcome of the two full-physics runs called TNG100 and TNG50, which consist of simulation boxes with co-moving edge-lengths of roughly 100 Mpc and 50 Mpc \citep{nelson2019illustristng}. While the smaller volume of TNG50 allows a higher resolution with respect to TNG100, the number of galaxies therein simulated is, for the same reason, smaller, especially at the high-mass end. Therefore we also incorporate TNG100 to boost the sample size especially for galaxies with stellar mass $\logmstar>10$.

TNG does not explicitly model light from stars nor the interaction of such light with gas and dust. It is nonetheless highly desirable to generate images based on light from the stellar/gas particle data that are compatible with images of observed galaxies taken with real instruments. Previous works showed the utmost importance of pushing images into the same domain as the observations \citep[e.g.][]{Gomez_2019, Huertas_2019, Bottrell_2019, Su_2020, Ciprijanovic_2021}. This is especially important for a contrastive learning application like the one pursued in this work, as the model is designed to find differences in images. For example, a non-realistic background could be identified as the most significant difference between images even if the imaged galaxies were fully consistent. We therefore use synthetic HSC-SSP images of TNG galaxies produced using dust radiative transfer and injected into real HSC-SSP cutouts: these have been developed by \cite{Bottrell_2023} and are succinctly described in the following.

\subsubsection{Forward modeling of simulated galaxies into HSC observations}
Galaxies from the TNG50 and TNG100 runs and found at the snapshots spanning $0\leq z \leq0.7$ were forward-modeled into synthetic HSC-SSP images by \cite{Bottrell_2023}. First, dust radiative transfer post-processing with SKIRT\footnote{\url{https://skirt.ugent.be}} \citep{Baes_2011,camps2015,Baes_2020} was used to model the emission, absorption, and scattering of stellar light in the HSC $grizy$ bands\footnote{\url{https://www.tng-project.org/explore/gallery/bottrell23i}} \citep{kawanomoto2018} and several supplementary filters spanning $0.3-5$ microns. Light from particles representing stellar populations older than 10 Myr was modeled using the \cite{bruzual2003} stellar population synthesis (SPS) library assuming a \cite{chabrier2003} initial mass function. Continuum and nebular line emission from young stellar populations embedded in birth clouds (star particles ages $<10$ Myr) was modeled using the MAPPINGS III library \citep{groves2008}. To model attenuation/scattering by dust, which isn't explicitly tracked in TNG, gas cells were ascribed dust densities using the post-processing method from \cite{popping2022}. In this dust model, the dust-to-gas mass ratio scales with gas metallicity \citep{2014A&A...563A..31R} and further assumes a \cite{weingartner2001} Milky-Way dust grain composition and size distribution. 

The images generated by SKIRT are \emph{idealized}, i.e. they contain no noise/background and have been produced with high spatial resolution: 100 pc/pixel. The idealized SKIRT images were then inserted into real HSC-SSP fields using an adapted version of $\mathtt{RealSim}$\footnote{\url{https://github.com/cbottrell/RealSim}}. This step handles (1) cosmological surface-brightness dimming and physical-to-angular size scaling according to the snapshot redshift of each galaxy; and (2) injection into real HSC-SSP cutouts to match statistics of depth, seeing, and crowding by foreground/background objects and artefacts from the actual observations.\footnote{\url{https://www.tng-project.org/explore/gallery/bottrell23}}.

\subsubsection{Simulated galaxy sample used in this work}
Throughout this paper, we use the HSC-realistic images of \cite{Bottrell_2023} generated for TNG50 and TNG100 galaxies in all 20 simulation snapshots between $z=0.1$ and $z=0.4$ (snapshots $72-91$). In particular, we focus on TNG50 galaxies with total stellar mass $\logmstar>9$ and TNG100 galaxies with $\logmstar>10$ over this redshift interval. These limits are chosen to ensure that all galaxies (1) comprise at least $10^4$ stellar particles and hence have reasonably-resolved stellar structures and (2) are detectable with respect to the surface-brightness limits of the HSC-SSP. Each TNG50 and TNG100 galaxy was imaged along $4$ lines of sight, oriented along the arms of a tetrahedron with respect to the simulation volume. In this way, the image sample is $4$ times larger than the galaxy sample, giving us a synthetic image dataset of 234948 images for TNG50 and 498748 for TNG100.

\subsection{Data preparation and homogenization}
\label{sec:preparation}
While we ensure that the HSC mocks of TNG galaxies are as realistic as possible in comparison to real HSC images, a few further preparatory steps are needed to guarantee that the real and synthetic galaxy datasets are compared on an even ground. In particular, we have to tackle the following tasks:
\begin{enumerate}
    \item Choosing a common approach to set the field of view for the images and resize them to the same dimensions.
    \item Matching the three galaxy populations in terms of basic observable properties so that we can implement an approximate common selection function for all datasets.
    \item Deciding which bands to use and how to stretch them, i.e. to find a mapping between the raw image intensities and the normalized values the model works with.
    \item Splitting the datasets into training, validation, and test sets. 
\end{enumerate}
It is of utmost importance to apply exactly the same preparation steps to all three sets of galaxy images, to ensure that we do not introduce any systematic differences that could then be picked up, spuriously, by our contrastive learning approach.

\begin{table*}
\centering
\begin{tabularx}{\linewidth}{{> {\hsize=0.18\hsize}X>
	                          {\hsize=0.35\hsize}X>                           
                                {\hsize=0.35\hsize}X>
                                {\hsize=0.4\hsize}X}}
Parameter & Description  & HSC & TNG50 \& TNG100\\
\hline
Redshift & Redshift of the galaxies relative to the observer & Photometric measurement \citep{2018PASJ...70S...9T} & Determined from the discrete snapshot number (i.e. simulation time) the galaxy is taken from\\
$i$-band Magnitude & Apparent I-Band magnitude of stars & HSC-SSP Photometric Pipeline \citep{Bosch_2018} & Integrated stellar light incl. dust attenuation \citep{Nelson_2018} \\
Petrosian Radius & Radius enclosing $90$ per cent of the overall light that is originating from the respective galaxy. Determined from an isotropic Petrosian fit. & Derived via Petrofit in the FOV calculation & Derived via Petrofit in the FOV calculation              
\end{tabularx}
\caption{Overview over the fundamental galaxy properties used to match the observed and simulated galaxy samples and for validation tests.}
\label{tab:matching}
\end{table*}

\subsubsection{FOV selection}
The FOV is an important factor when preparing images for any machine learning algorithm. In our case, if the fiducial FOV is too small, the model might miss important features around the galaxy, e.g. loosely-wound spiral arms, or extended stellar halo features. On the other hand, if the FOV is too large the model may pay more attention to background than to the target galaxy itself. 

The field of view (FOV) for the HSC-realistic TNG mocks is large ($50-500$ kpc) and scales with the physical size of each simulated galaxy. Meanwhile the HSC cutouts have a fixed angular size.
To ensure that we get good results across the full range of galaxy masses and sizes, we choose an adaptive approach to crop the large FOV of the simulated images. Namely, we perform a Petrosian fit with Petrofit \citep{Geda_2022} and measure the Petrosian radius $r_{p90}$ containing 90 per cent of the light for the central object in each image. We then choose $4 \times r_{p90}$ as FOV and crop the images accordingly. The cropped images are then resized to a common grid of $256^2$ pixels. Because the resolution of the original image data is fairly high (0.168 arcsec/pixel), the angular resolution of the cropped/resized images is almost always lower than the resolution of the original data. While we did not perform any exhaustive tests regarding the Petrosian fitting, crosschecks with existing size measurements available for TNG and HSC suggests that our fitting procedure is consistent. Additionally, by applying the fitting to all datasets with the same parameters, we ensure that possible biases and errors are contained in similar fashions in all sets.

\subsubsection{Galaxy-population matching}
\label{sec:matching}
When comparing distinct sets of data, especially between observations and simulations, we have to ensure that the distribution of their intrinsic properties are the same. For example, observations are limited by exposure i.e. surface brightness cuts, in addition to survey area, whereas the galaxy populations from uniform-volume cosmological simulations are volume limited and limited e.g. by a mass cut. Therefore, the prior distributions of basic quantities of observed and simulated galaxy populations may be different even if the simulated galaxies were perfectly realistic.

To match the three galaxy data sets we first need to identify properties that are derivable from both the simulations and the observations. We choose photometric redshift, $i$-band Magnitude and galaxy stellar size i.e. Petrosian radius, as summarized in Table \ref{tab:matching}.

Then, to enforce a common selection function for both simulated and observed galaxies, we draw galaxies from the HSC-SSP set to match the redshifts, apparent magnitudes, and angular sizes of  galaxies from the TNG50 and TNG100 samples. The specific algorithm is as follows. For each TNG(50/100) galaxy, we identify a match-tolerance window in the space of redshift ($\pm 0.035$), apparent $i$-band magnitude ($\pm0.05$ mag) and Petrosian radius ($\pm 0.4$ arcsec). These tolerances are chosen based on typical observational measurement uncertainties in each of these quantities. We then draw a random HSC galaxy from within the tolerance window and assign it uniquely to the TNG(50/100) galaxy being matched (without replacement). TNG(50/100) galaxies for which no HSC-SSP galaxy offers a match are discarded from our analysis. Because of the different volume and resolution of the parent simulations, the intrinsic galaxy distributions from TNG50 and TNG100 are different, and each is hence matched individually to HSC galaxies.

The three tolerances are simply conservative estimates and are practically free parameters. We chose them such that they are well within the stated errors for the HSC-SSP photometry pipeline \citep{Bosch_2018} and MIZUKI photo-$z$ measurements \citep{2018PASJ...70S...9T} while also ensuring that we don't restrict ourselves to too small of a window, which would lead to larger numbers of unmatched TNG galaxies and wasted data. This is especially problematic for the redshift, as the TNG galaxies are only available for a few discrete snapshots in time, rather than a smooth and continuous range.
\begin{figure*}
	\centering
        \includegraphics[width=0.32\linewidth]{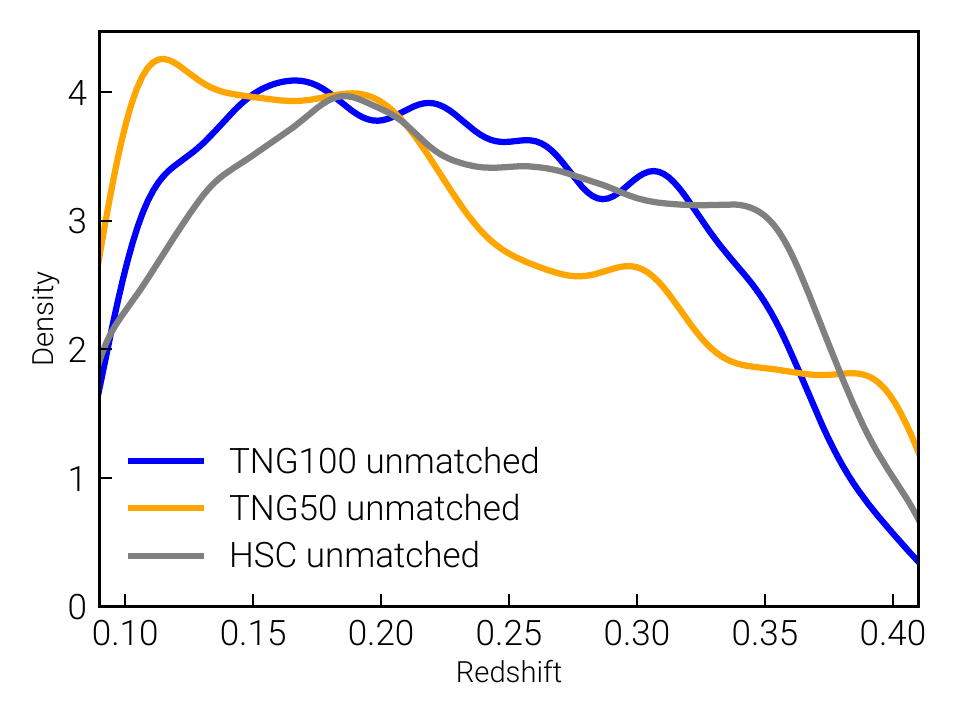}
        \includegraphics[width=0.32\linewidth]{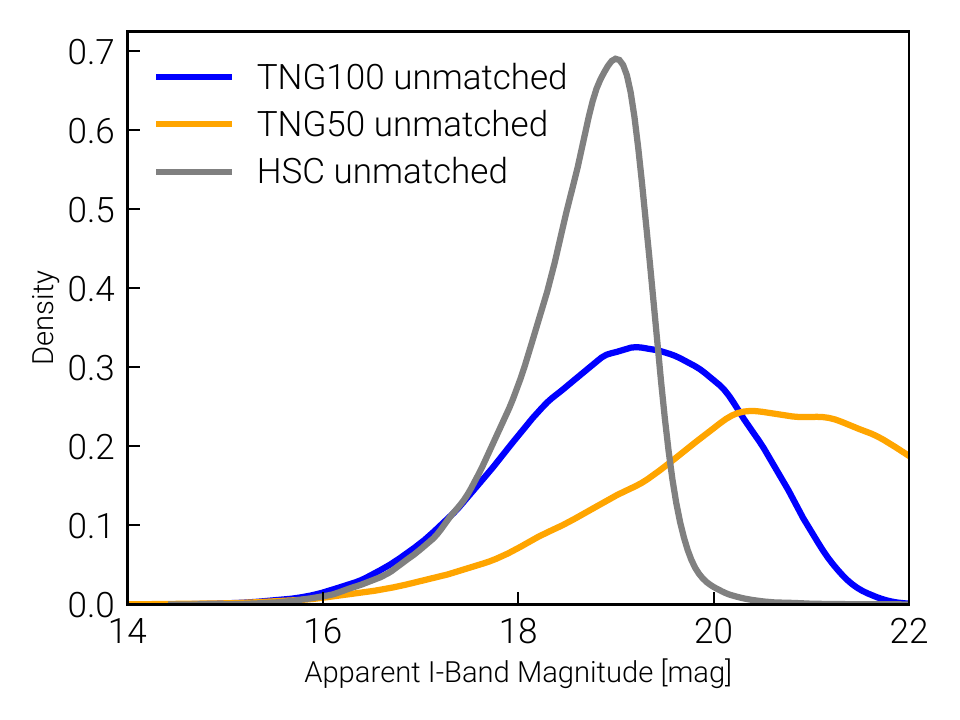}
        \includegraphics[width=0.32\linewidth]{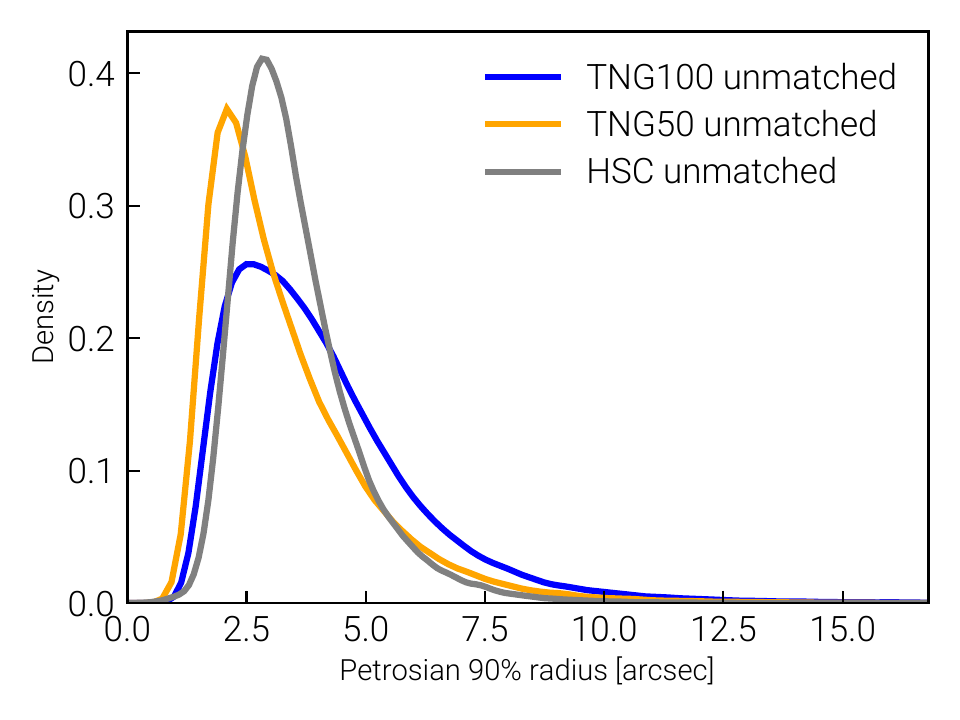}
        \includegraphics[width=0.32\linewidth]{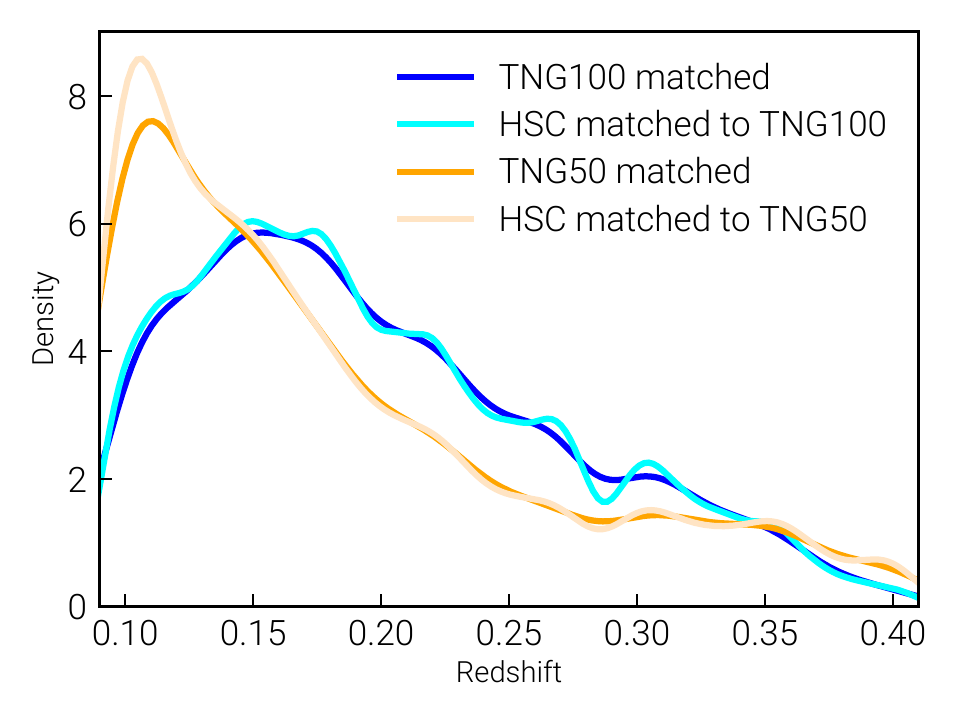}
        \includegraphics[width=0.32\linewidth]{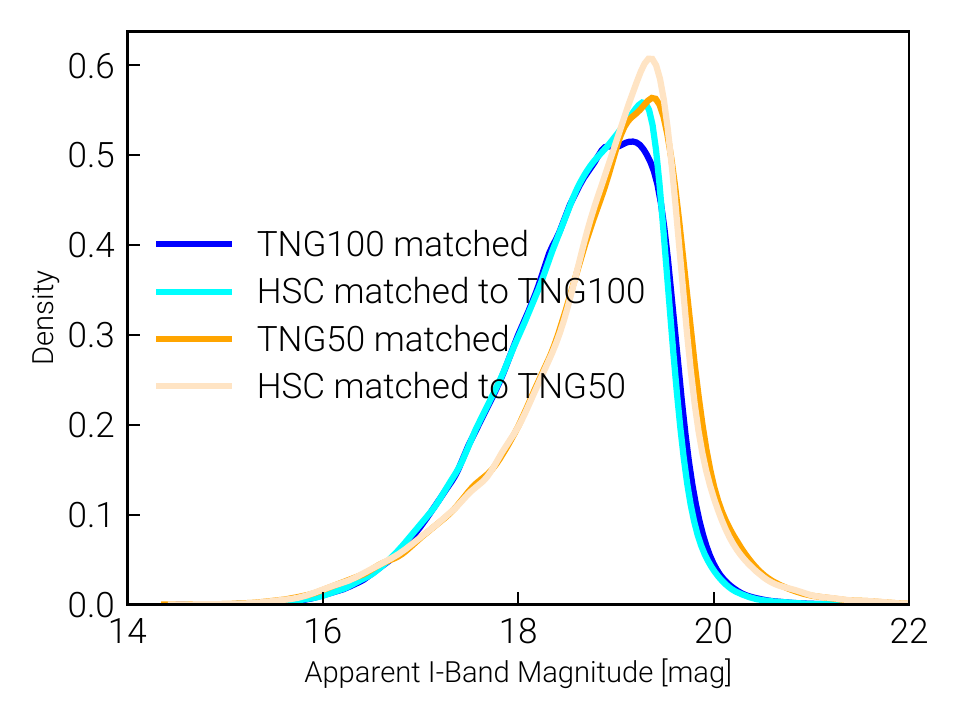}
        \includegraphics[width=0.32\linewidth]{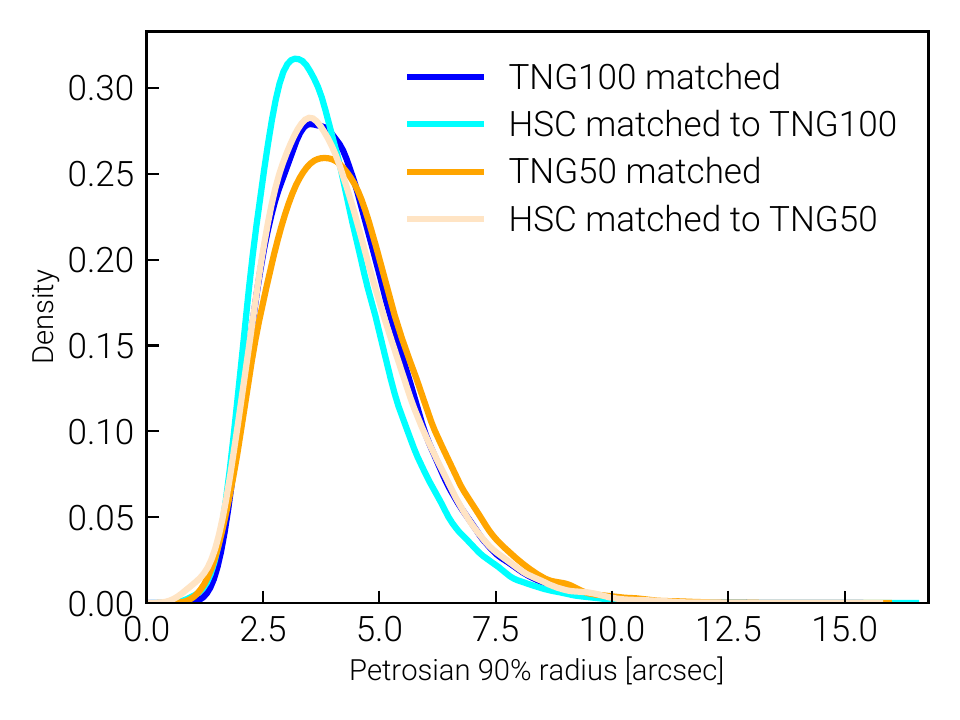}
	\caption{{\bf Matching the observed HSC dataset to the two simulated TNG50 and TNG100 samples of galaxy images, respectively.} We show the distribution of all three sets in photometric redshift, apparent I-band magnitude and petrosian radius before (top) and after (bottom) the matching (refer to Table~\ref{tab:matching} for definitions). We split the set of matched HSC galaxies into two subsets: one that is matched to TNG100 and one that is matched to TNG50. This is necessary as the prior distributions between the two simulated image samples are  different, because of the different volume and resolution of the simulations. Note that we apply a larger smoothing kernel for the redshift: this is necessary as the simulated TNG galaxies are given only for a set of discrete snapshots.}
    \label{fig:matching}
\end{figure*}

We show the resulting distributions of matched galaxy samples in Figure~\ref{fig:matching}. The unmatched i.e. intrinsic galaxy distributions are shown in the upper panels whereas the bottom panels show the galaxy distributions after the matching. As mentioned above, as the intrinsic distributions differ between TNG50 and TNG100, we show two sets of galaxy samples: one that has been matched to TNG50 and one that has been matched to TNG100.

We see that, naturally, the main bias in the unmatched sets is the $i$-band magnitude, as there is a surface brightness cut for HSC but no such a limitation for the simulated galaxies. However the apparent brightness is heavily linked to the redshift:
(1) At low redshift there are more bright TNG galaxies than HSC galaxies. This is a volume effect as the simulated volume is for low redshift larger than the observed. (2) At 'higher' redshifts (i.e. up to $z=0.4$) there are fewer faint HSC than TNG galaxies owing to our HSC Petrosian magnitude and SDSS counterpart detection criteria for the HSC sample. Due to this offset, when matching the samples we remove many faint high-redshift TNG galaxies. The matched samples therefore lean towards the low-redshift regime. One could counteract this by decreasing the matching window for the redshift; however this would lead to an even more depleted TNG sample.

Moreover, there is a correlation between the luminosity and the Petrosian radius: TNG (50 and 100) galaxies tend to be increasingly larger than HSC galaxies for increasing luminosity. It is for this reason that we need to match simulated and observed samples also with respect to galaxy sizes. This offset was already found in previous studies that compared TNG100 and Pan-STARRS galaxies \citep{Gomez_2019}. In fact, this is not visible in Figure~\ref{fig:matching} as the unmatched TNG population is biased towards smaller galaxies because of the large fraction of low-luminosity galaxies. As a consequence of the matching in stellar sizes, we improve the domain agreement also in terms of image resolution and FOV distribution.
After the matching process, we have 628710 images (incl. HSC, TNG50 and TNG100) for training, validation and testing the machine learning model.
From now on, we omit the specification ``matched'' when referring to the samples of TNG100/TNG50 images matched to HSC.

\subsubsection{Band selection and stretching}
While in principle the model can work with any number of luminosity bands or channels, we decide to start with an RGB representation that is most intuitive for a visual inspection of the results. For the RGB images we use the respective HSC i, r, and g bands, respectively. For those bands we have, on the one hand, large amounts of HSC data with sufficient depth and data quality. Furthermore they roughly correspond to the RGB color scheme at the rest frame that covers most of the visual spectrum. 

We then apply a stretching of the images based on \cite{Bottrell_2019} to ensure that galaxy features across the full dynamic range of surface brightness are well captured. Without this stretching a machine learning model might just focus on the much brighter galaxy core, neglecting features in the outskirts.
To each channel we apply the following steps separately:
\begin{enumerate}
    \item Perform a logarithm on the intensities.
    \item Set the minimum brightness to the median of all pixels. We ignore all pixels with less than $10^{-7}$ for the calculation of the median. 
    \item Set the maximum brightness to the $99$ per-cent quantile of the central $20 \times 20$ pixels.
    \item Normalize the logarithmic images with the given minimal and maximal brightness.
\end{enumerate}
The pixel values of the stretched images are therefore between $0$ (background level) and $1$ (center of the galaxies). We want to note that we experimented with multiple stretching mechanisms. For example, a arcsinh stretch better emphasizes the noise and thereby leads to a model putting more importance into the noise and, subsequently, the redshift of a given galaxy. The band selection and stretching therefore also depends sensitively on the scientific question at hand.

\subsubsection{Splitting of the sample(s)}
ML models may become so well-tuned to the dataset on which they are trained that they may generalize poorly to new data, even from the same domain (i.e. over-fitting). It is therefore common practice to split the full dataset in three fixed and distinct subsets. These are not to be mixed during the training and testing procedure.
We split the combined HSC and TNG sets in (1) a training set ($80$ per cent of the overall sample), which is used for the training of the model only, (2) a validation set ($10$ per cent), which is used to assess the model performance, i.e. the generalization capability during the training, and (3) a test set ($10$ per cent) to which the model is never exposed in training and that is used to perform all analysis and tests (Section \ref{sec:umap} and following).

As discovered in previous work \citep{Eisert_2023}, it is important to ensure that there are no progenitor/descendants of a (simulated) galaxy in multiple subsets. This is to avoid an artificial improvement of the test results later on.
Although this was shown for scalar properties only, we keep this conservative approach and ensure that we split all TNG galaxies along with their root descendant; i.e. the galaxy they will finally merge into. In this way, all progenitors and descendants of a given galaxy are contained within only one of the three sets, exclusively.

\section{The Machine Learning Methods}
\label{sec:methods}
To capture the potentially complex features that distinguish galaxies from each other (including real and simulated galaxies), we use the self-supervised contrastive learning method. This model is designed to extract features from the images and encode them as high-dimensional, vectorized \emph{representations}. Through training of the model, these representations should come to encode detailed information about each galaxy image. We use a Residual Network described in detail in Section \ref{sec:resnet} to map the images into this representation space. This network is then trained by the contrastive learning method NNCLR described in Section \ref{sec:contrastive_learning}. The details of the training procedure is explained in Section \ref{sec:training}.

\subsection{E(2) equivariant Residual Networks}
 \label{sec:resnet}
The input images have $256^2$ pixels in 3 bands. This corresponds to $196608$ degrees of freedom for each image. To efficiently learn features from this high dimensional image data, we use a (Wide-)ResNet Architecture \cite{zagoruyko2017wide} to extract information from the images and map this information into a lower, i.e. $256$, dimensional representation space. Instead of standard convolutional layers, so called "E(2)-Equivariant Steerable Convolutional" layers \cite{weiler2021general} are used in our ResNet model. In doing so we explicitly pay attention to the fact that the orientation of the galaxies on the image plane should be irrelevant -- all convolution layers, and therefore the mapping, are equivariant regarding transformations of the dihedral group D(8). Simply speaking, these are discrete rotations $360^\circ / 8 = 45^\circ$ and reflections on the images that should not alter the properties of the galaxies shown in the images. Tests performed with a group of higher symmetry like $D(16)$ doesn't improve results but does increase the computation time. On the other hand, the model converges faster and towards a lower validation loss with D(8) than without any equivariance i.e. using conventional convolutional layers. For the fiducial model we use a ResNet depth of 16 and a width of $2$. The ResNet layers are equivariant regarding D(8) and are regularized using a Dropout of 30 per cent of the nodes.
The output from the ResNet is a flattened feature vector. To give the model additional flexibility regarding the representations, we add a fully-connected network, comprising 2 layers with 256 nodes each, at the end of the ResNet to map the feature vector into the final representation space.

\subsection{Contrastive Learning: NNCLR}
\label{sec:contrastive_learning}
The ResNet architecture itself maps the images into a representation space -- however, the task at hand is to ensure that these representations are meaningful for our purposes, i.e., they must encode the visible features of the galaxies in the images. To achieve this, the ResNet is trained using the self-supervised method 'Nearest neighbour Contrastive Learning' (NNCLR) \citep{https://doi.org/10.48550/arxiv.2104.14548} which is based on SimCLR \citep{https://doi.org/10.48550/arxiv.2002.05709}. SimCLR aims to 'teach' the ResNet to attract transformed images of one and the same galaxy and discriminate against transformed images of other galaxies. 
For this work we use the following (random) transformations: 
\begin{itemize}
    \item Reflections along the horizontal and vertical axis
    \item Rotations from $-15^\circ -  + 15^\circ$
    \item Central zooms of max 10 per cent  i.e. the center of the image stays in the center of the crop while zooming
    \item Affine translations by up to 32 pixels
    \item Gaussian blur with a max std of 0.1 pixel
    \item Gaussian noise with a max std of 0.08 (note that the images are normalized to 1).
\end{itemize}
We perform these transformations on the original $256^2$ pixel images as described in Section \ref{sec:data} and then rebin them to $128^2$ pixel for the ResNet.
As a result, the ResNet is automatically focusing at commonalities and differences between images and placing them in distinct spaces of the representation space: e.g. grouping elliptical or disk galaxies together.

While this describes the functioning of SimCLR on a first order, the NNCLR method we apply in this work takes another step. In addition to the transformations described above, we search for the nearest neighbour (hence the name) of that transformed image in the representation space and use that neighbour image as the transformed 'same' galaxy. In doing so, one can reach a higher generalization ability, as the network interpolates between the training galaxies through this additional augmentation step. We tested and compared the SimCLR method with NNLCR and choose NNCLR as the fiducial choice for this reason.

For the nearest neighbouring search we are \emph{not} using all images of the training set as this would be too computational expensive. Instead we use a queue: at each training step we push the newest batch of input images and their representations into the queue and remove the oldest batch. In total we keep $512$ batches of images and their representations in that queue from which to match nearest neighbours. 

Finally, we do not run the contrastive learning directly on the representation space but we adding a so called 'Projection Network', which consists of 2 fully connected layers with 512 nodes each. This projects the representation space into a space with the same number of dimensions (i.e. 256) on which the contrastive loss is calculated and the network trained \citep[refer for details to][]{https://doi.org/10.48550/arxiv.2104.14548}. When we test/apply the network to data in the next sections, we remove this projection network and only use the representation space. This means that the projection network is just a part of the training. 

\subsection{Training of the Residual Network via NNCLR}
\label{sec:training}
For the training, we use the training sets of all galaxy samples at the same time: the simulated TNG100 and TNG50 galaxies and the observed HSC galaxies. In doing so we allow the model to find differences and commonalities among all sets of images. We further elaborate on this choice in Section \ref{sec:discussion}.

We use an Adam optimizer \citep{Kingma_2014} to back-propagate the NNCLR loss of the common training set through the network. We do this in random batches of $32$ images with 2 transformations for each image. After $350$ random batches (which we call an epoch) we use the images in the validation set to validate the model. We set an initial learning rate of 0.001, which is reduced by a factor of 0.8 if the validation loss is not decreasing for 3 epochs. Training is stopped if the validation loss has not decreased for 15 epochs. All hyperparameters mentioned in this section (including the network architecture) have been optimized with respect of the validation loss using the Optuna framework \citep{Optuna}. We present the results of our model exploration and a discussion on parameter choices in Appendix \ref{sec:optuna}.

\begin{figure*}
	\centering
        \includegraphics[width=0.45\linewidth]{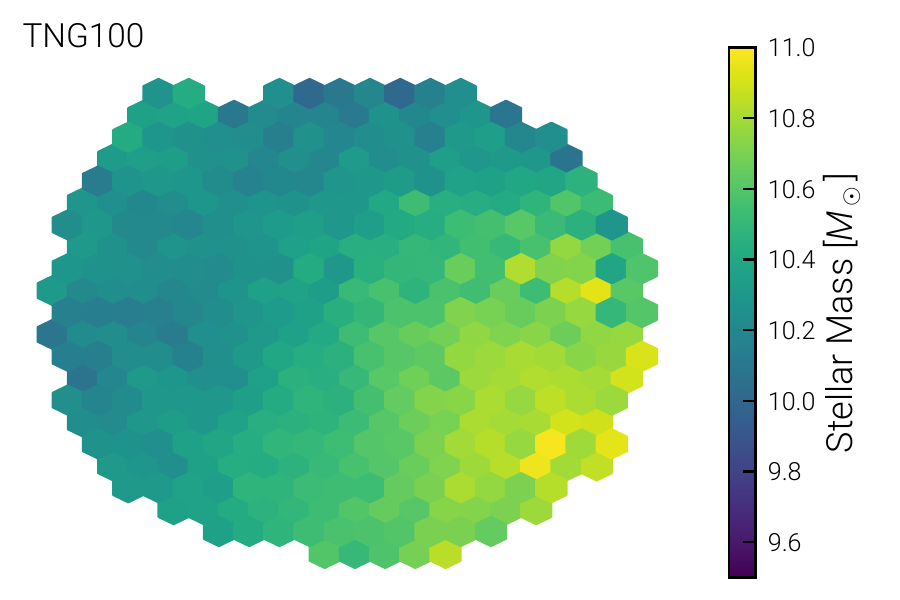}
        \includegraphics[width=0.45\linewidth]{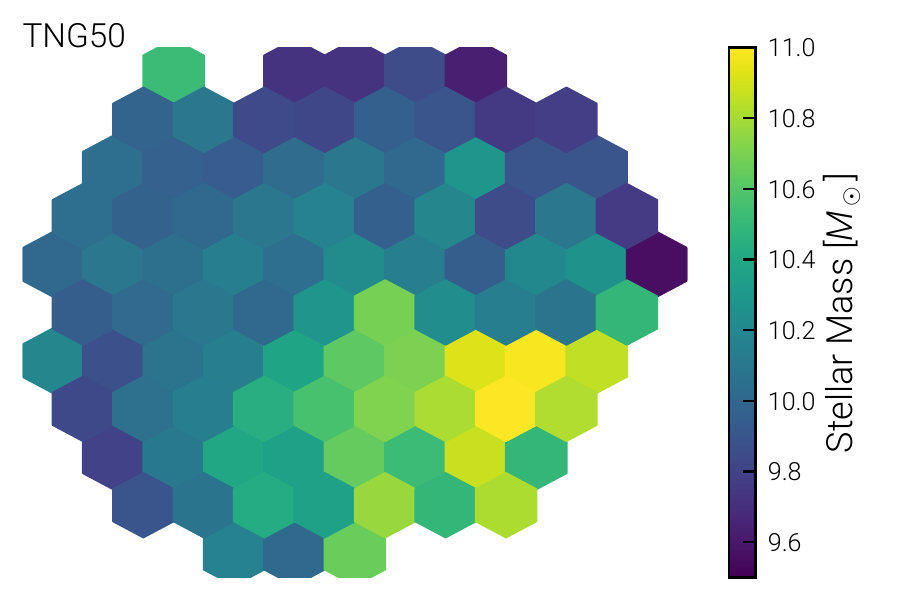}
        \includegraphics[width=0.45\linewidth]{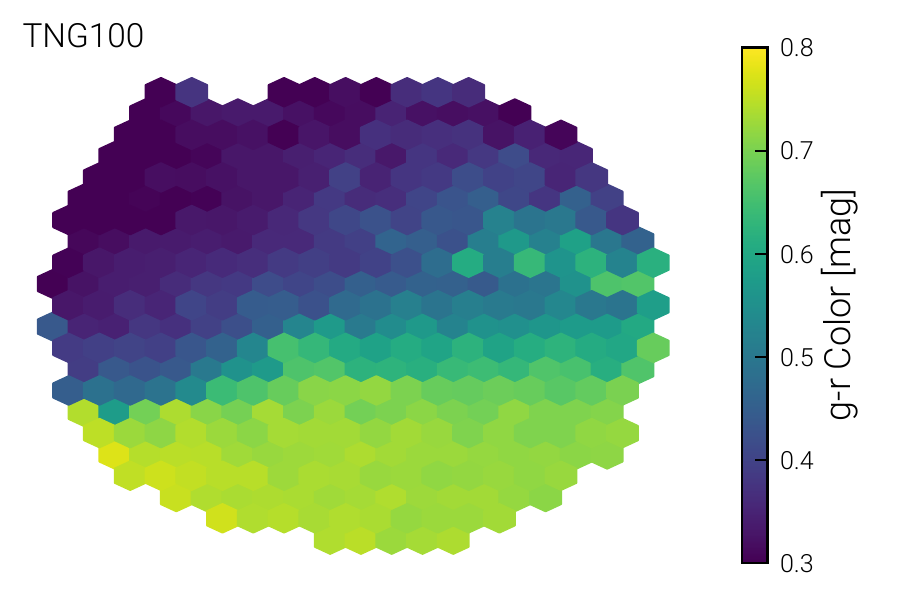}
        \includegraphics[width=0.45\linewidth]{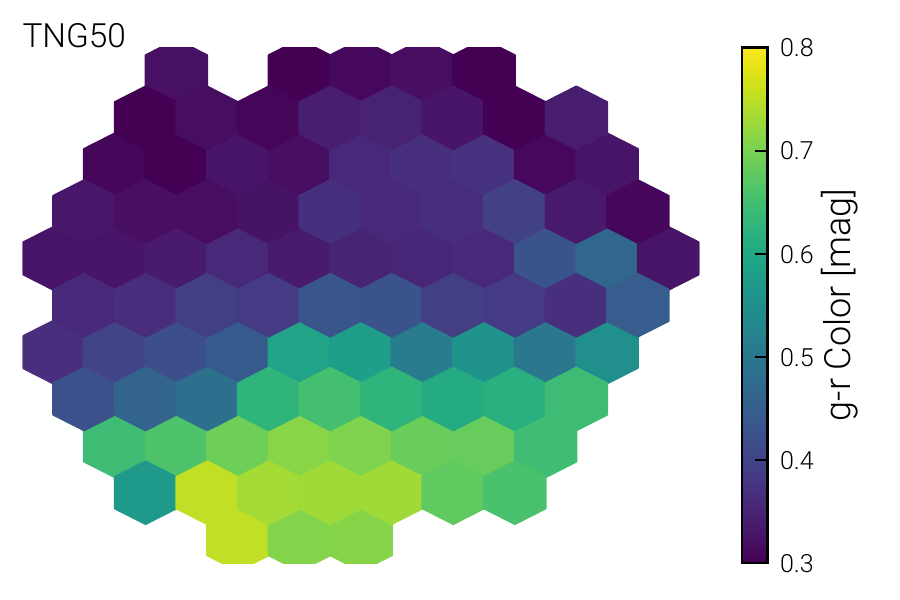}
        \includegraphics[width=0.45\linewidth]{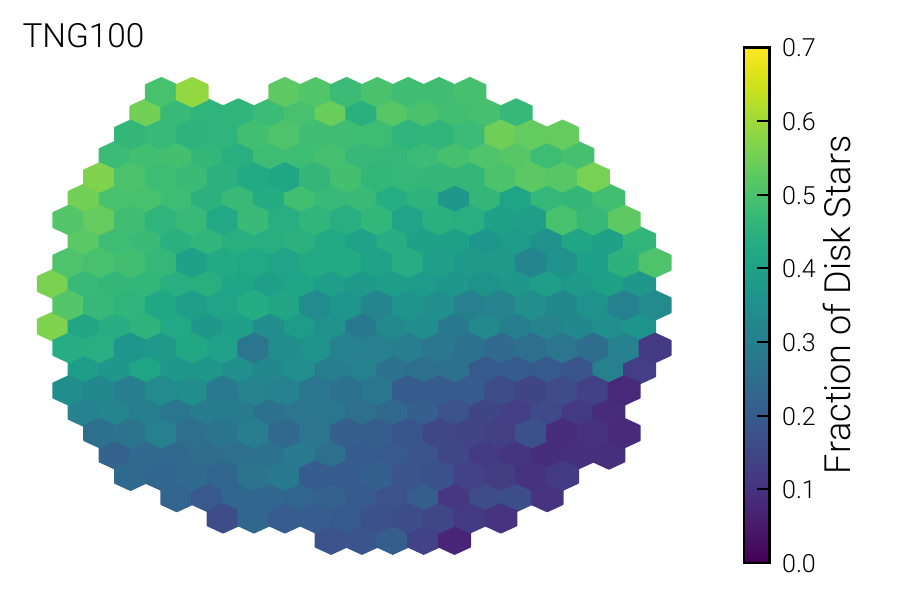}
        \includegraphics[width=0.45\linewidth]{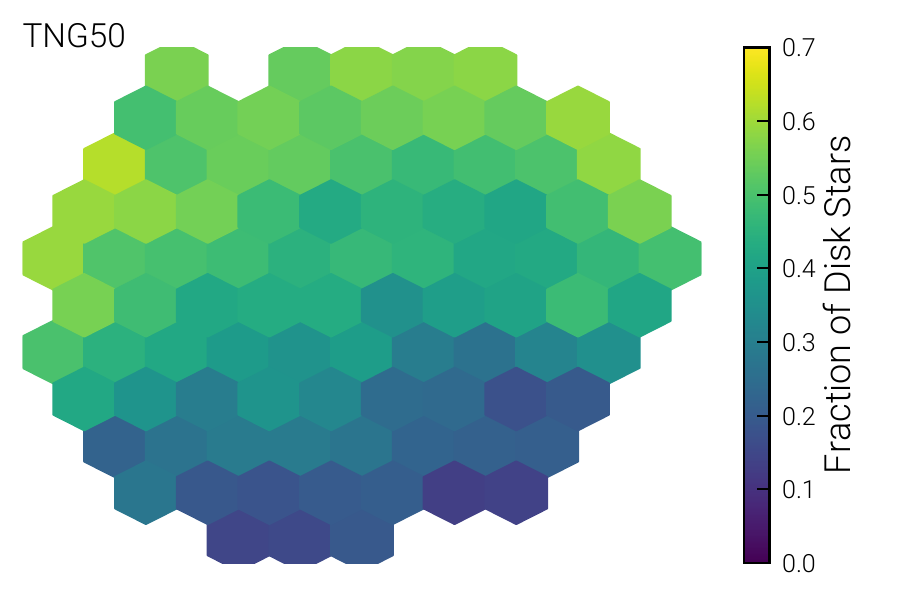}
        \includegraphics[width=0.45\linewidth]{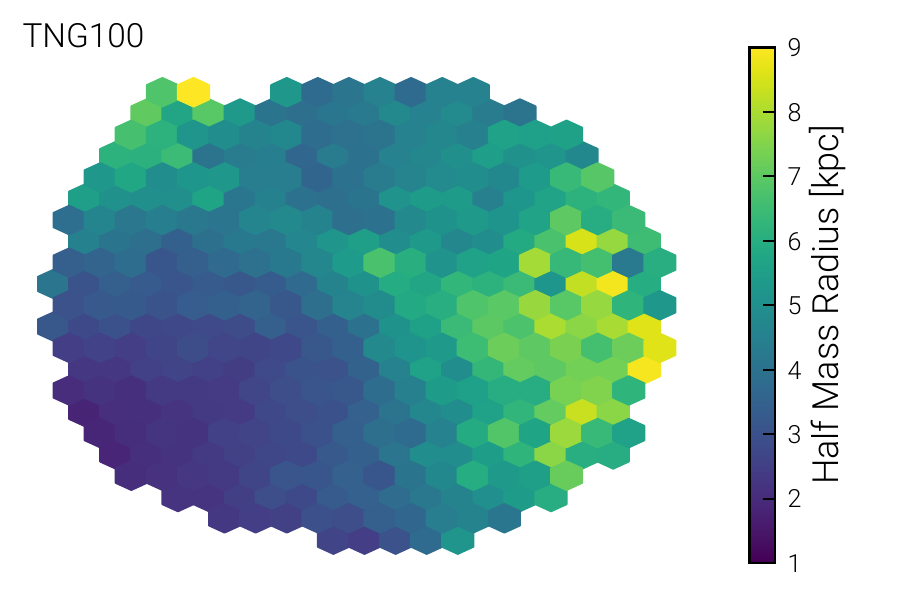}
        \includegraphics[width=0.45\linewidth]{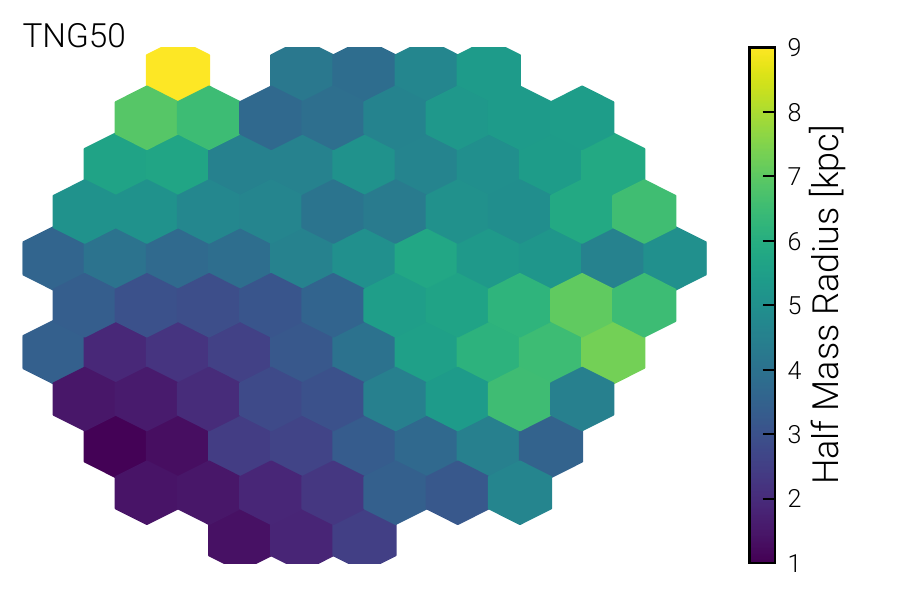}
	\caption{{\bf Are the representations related to observable features?} Each panel shows a 2D hexbin histogram of TNG100 galaxy images (left panels) and TNG50 galaxy images (right panels) in the 2D UMAP parameter space. The UMAP mapping itself was trained using the 256-dimensional representations of TNG50/100 and HSC test galaxies. Note that because of the smaller sample size of TNG50, we choose a larger bin size in the right panels. The bins are coloured according to the median image/galaxy property in each bin. We show from top to bottom, and in the continuation figures: total stellar mass, fraction of disk stars, integrated galaxy colour index (g-r), stellar half-mass radius, Sérsic half-light radius, Sérsic index, Sérsic ellipticity, {\change point asymmetry} of the light distribution, concentration of the light, smoothness of the light distribution, Gini-M20 bulge parameter and Gini-M20 merger parameter. The UMAP, and therefore the representations upon which it is based, are clearly related to observable properties of the TNG50/100 galaxies.}
	\label{fig:tng_umaps}
\end{figure*}
\begin{figure*}
	\centering
        \includegraphics[width=0.45\linewidth]{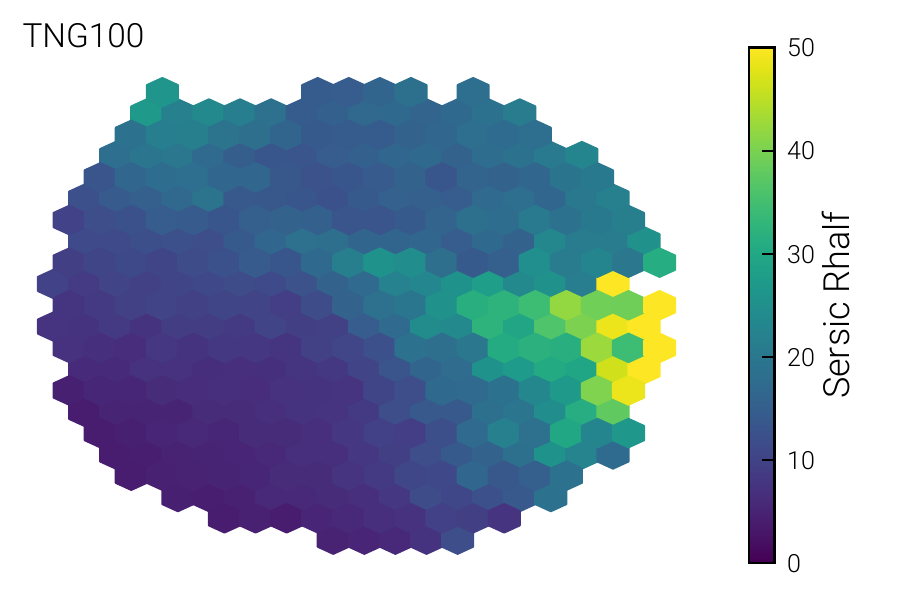}
        \includegraphics[width=0.45\linewidth]{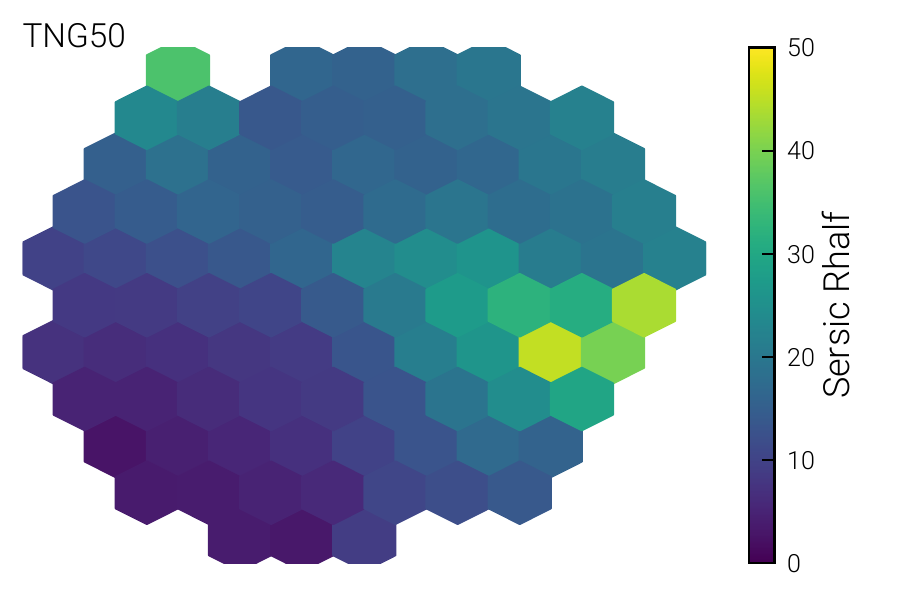}
        \includegraphics[width=0.45\linewidth]{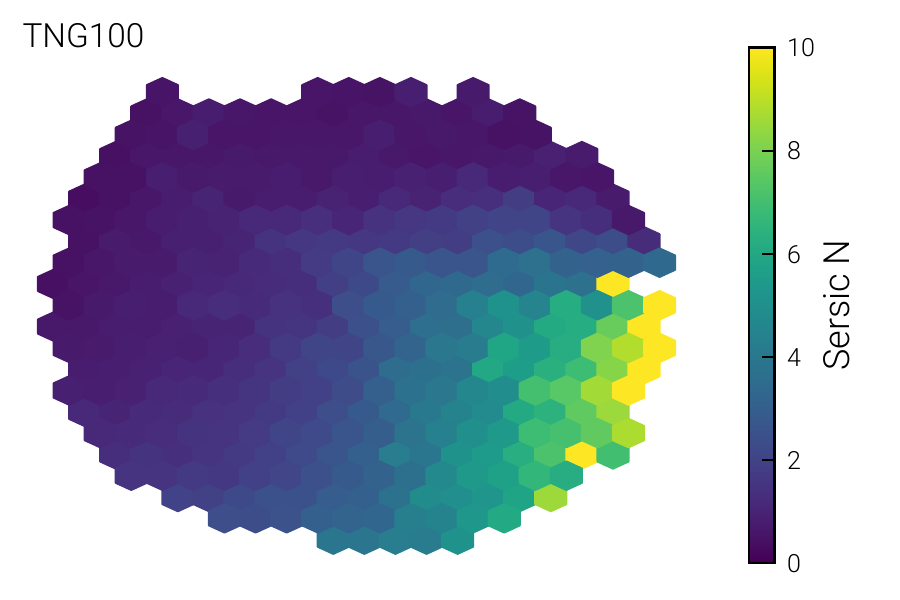}
        \includegraphics[width=0.45\linewidth]{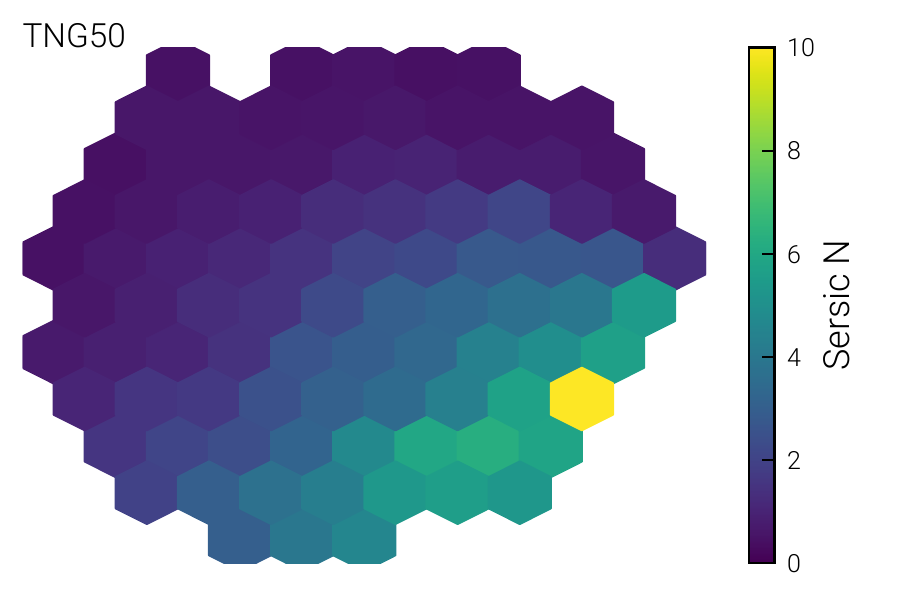}
        \includegraphics[width=0.45\linewidth]{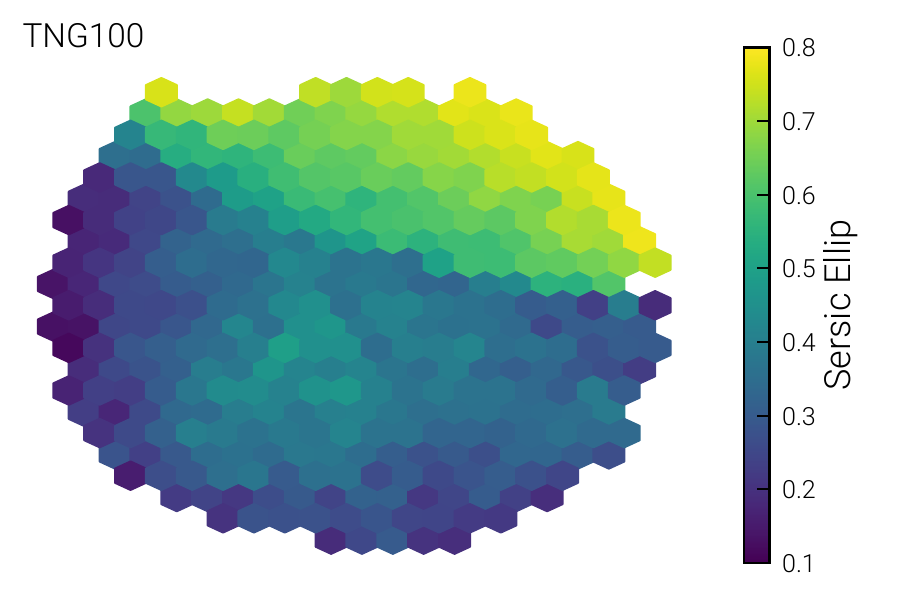}
        \includegraphics[width=0.45\linewidth]{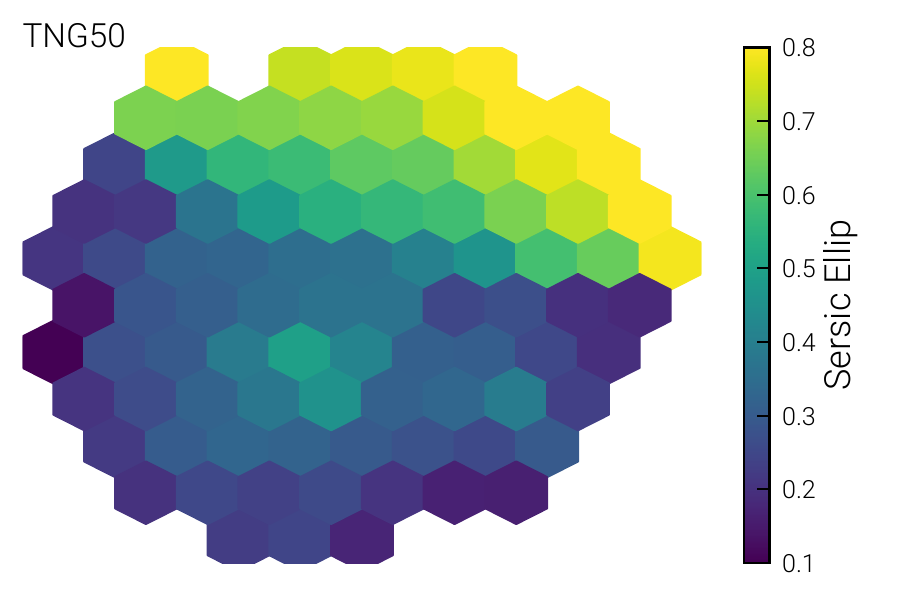}
        \includegraphics[width=0.45\linewidth]{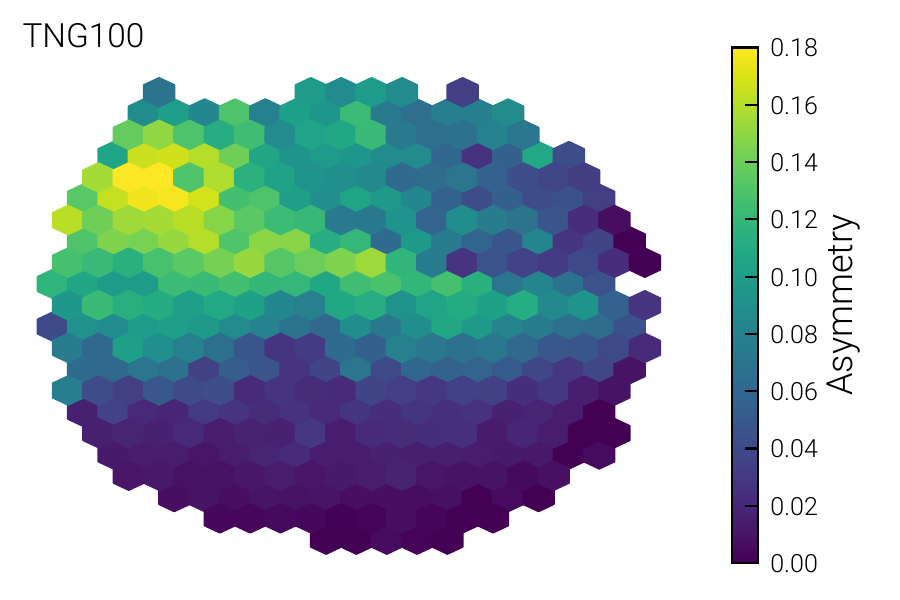}
        \includegraphics[width=0.45\linewidth]{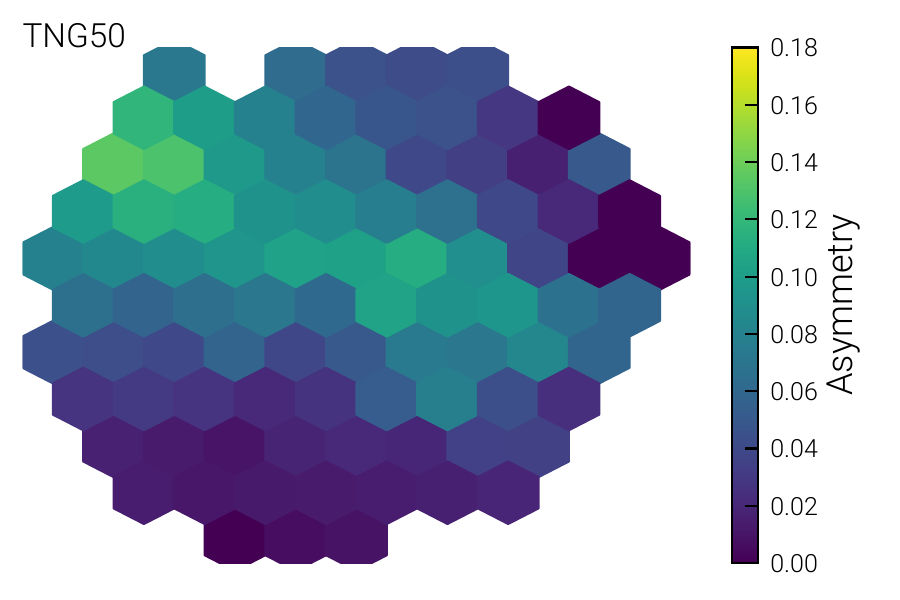}
	\caption{Continuation of Figure \ref{fig:tng_umaps}.}
	\label{fig:tng_umaps_2}
\end{figure*}

\begin{figure*}
	\centering
        \includegraphics[width=0.45\linewidth]{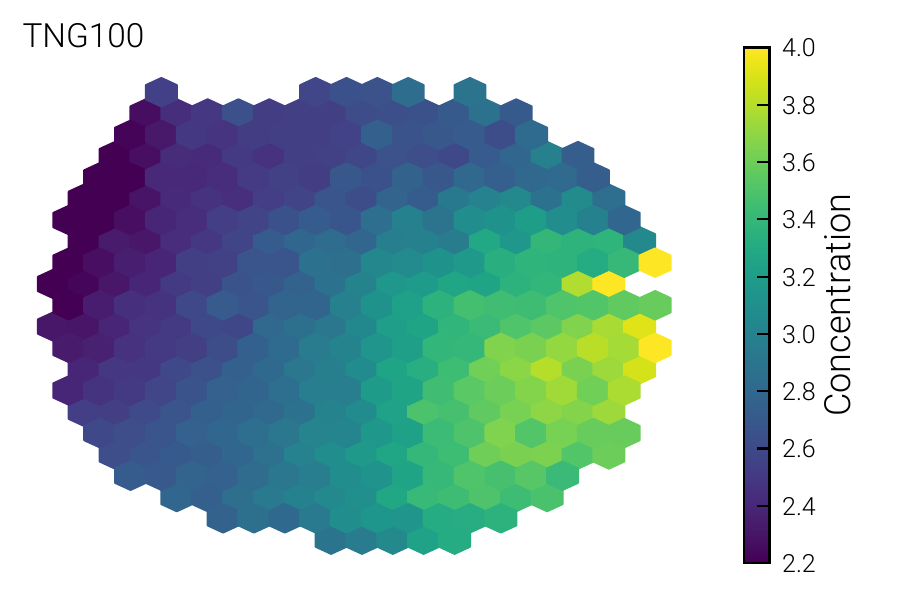}
        \includegraphics[width=0.45\linewidth]{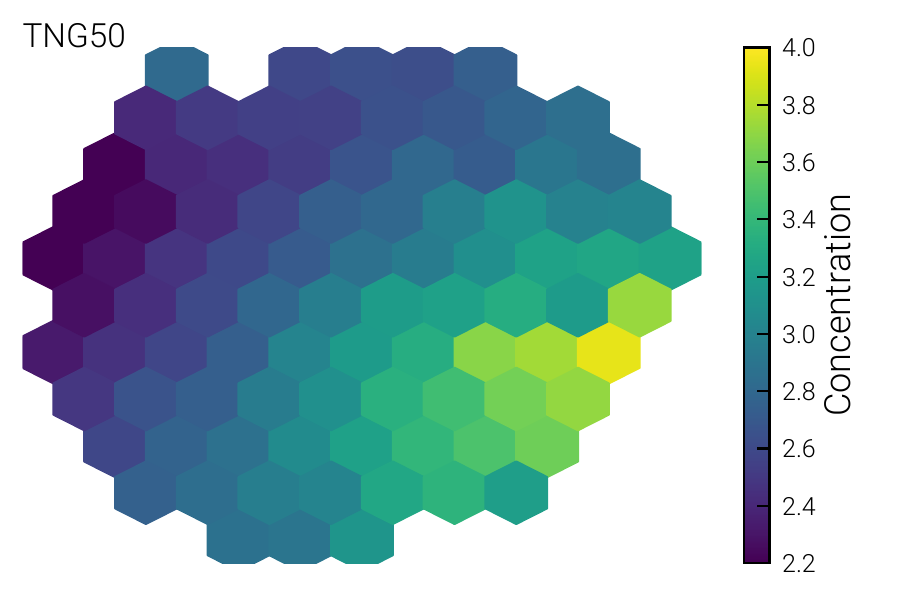}
        \includegraphics[width=0.45\linewidth]{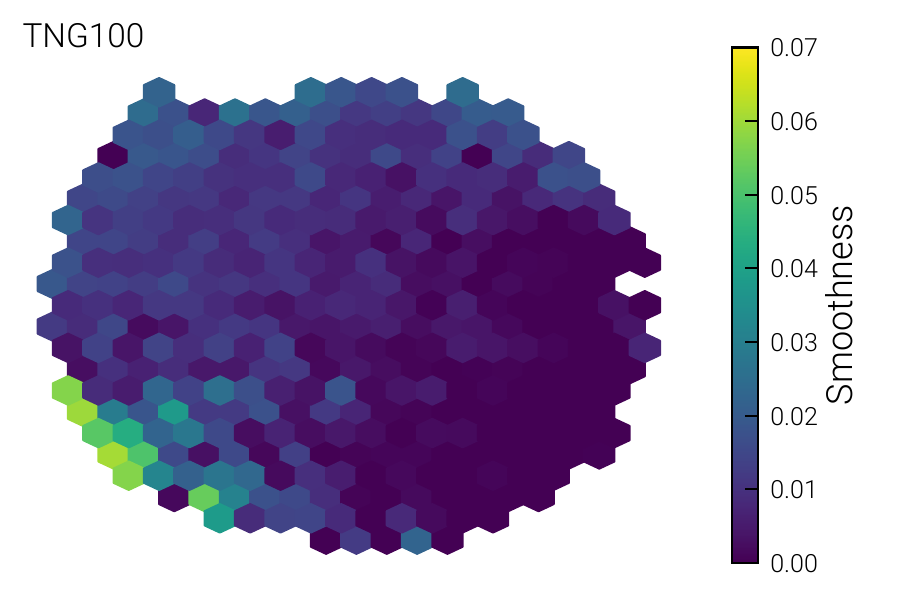}
        \includegraphics[width=0.45\linewidth]{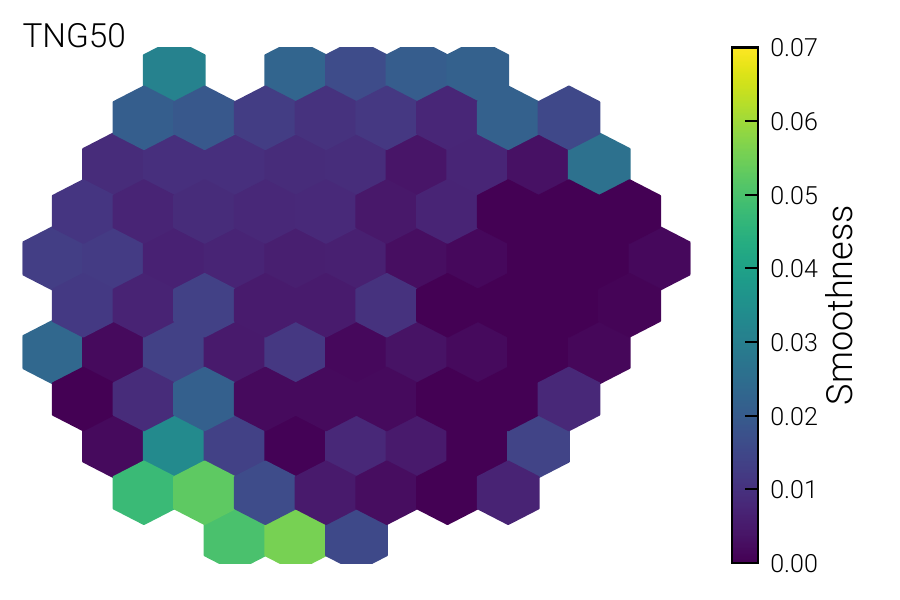}
        \includegraphics[width=0.45\linewidth]{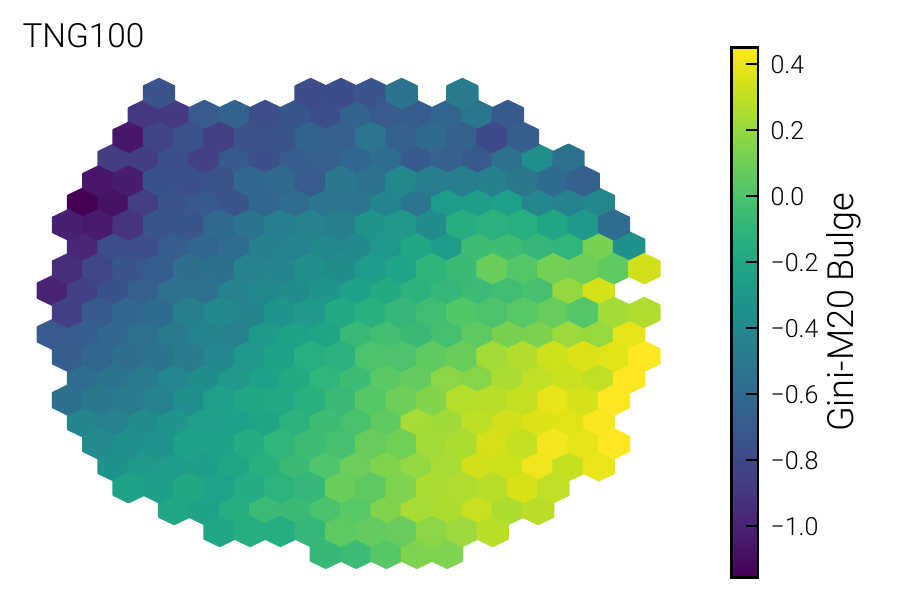}
        \includegraphics[width=0.45\linewidth]{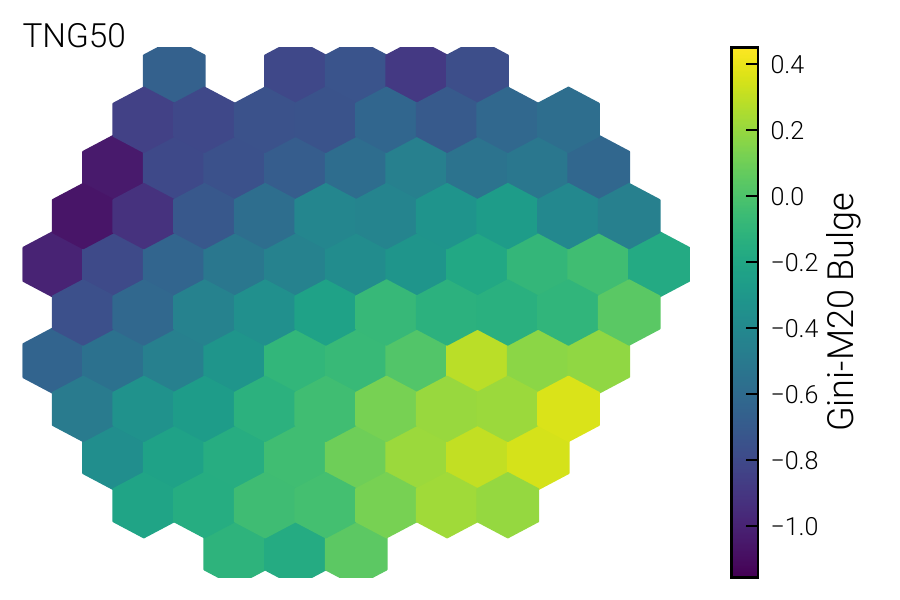}
        \includegraphics[width=0.45\linewidth]{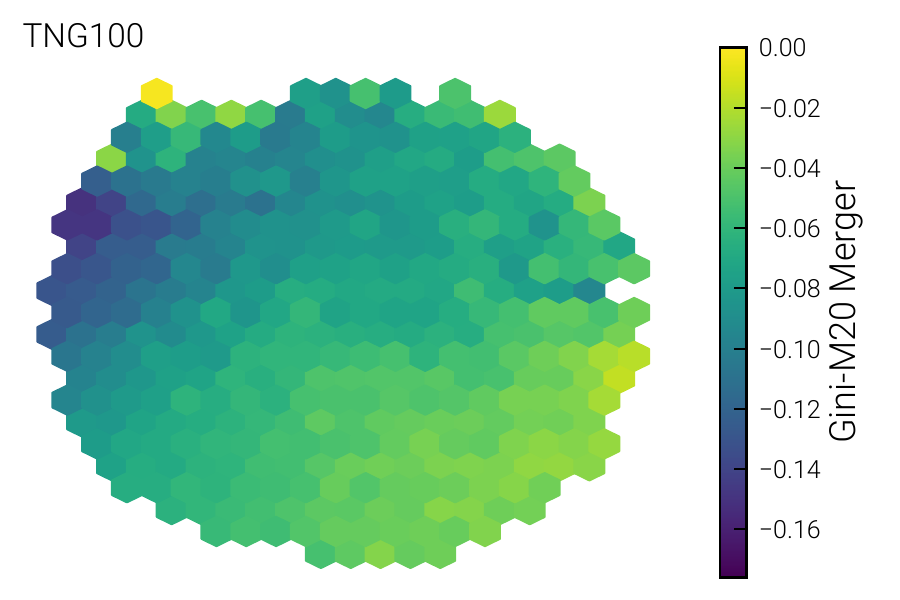}
        \includegraphics[width=0.45\linewidth]{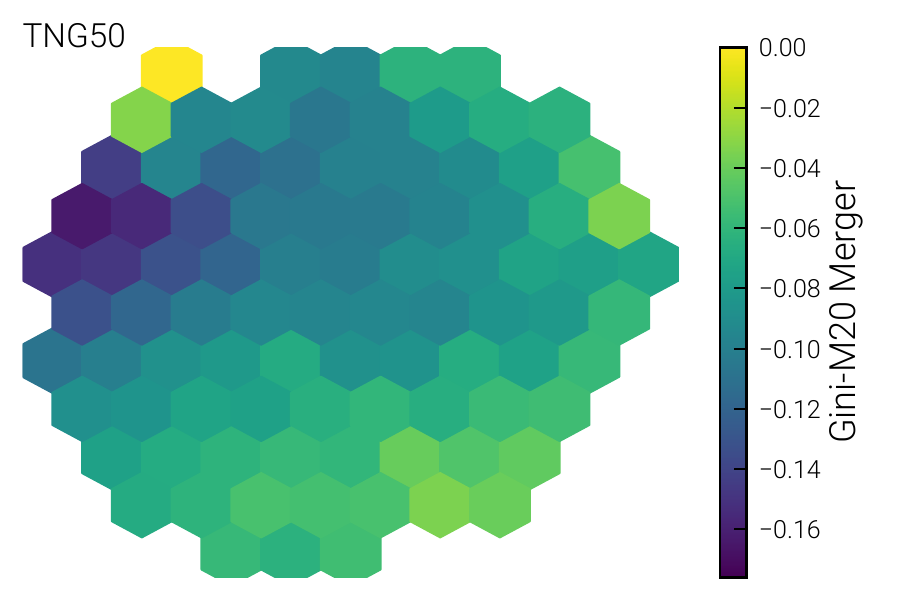}
	\caption{Continuation of Figure \ref{fig:tng_umaps}.}
	\label{fig:tng_umaps_3}
\end{figure*}

\section{Results}
\subsection{Interpreting the representation space with UMAPs}
\label{sec:umap}
After training via the contrastive learning method NNCLR introduced above, the galaxy images from HSC and TNG50/100 should be arranged in a `meaningful' way in representation space. Based on the choices described above, this means that each galaxy is assigned a $256$ dimensional vector that is somehow related to its visual and observational properties in the image. However, a priori the definition of 'meaningful' might differ between a ML network and a human. For example, a model could pay more attention to the image background than to the galaxy in question. In this section, we check whether the representations indeed encode meaningful galaxy properties.

To visualize the 256 dimensional representation space, we utilize the dimensional reduction algorithm UMAP \citep{mcinnes2020umap}, which maps the 256 dimensions non-linearly into two dimensions while conserving the topology of the space as much as possible. To ensure that our network generalizes well to a previously unseen set of images, we train and apply the UMAP method to the representations of the images in the test set of both observations and simulations: HSC and TNG50/100 galaxies. During the training of UMAP, we gauge two parameters of the method: (1) the number of nearest neighbours and (2) the minimal distance between nodes. {\change We do not find any large-scale difference when varying these parameters: namely,  for no combination we see that the observed and simulated domains separate. However, the local structures in the UMAPs do differ strongly for different choices of those parameters. This has to be kept in mind when judging the representation space based on a single 2D UMAP representation. For this work we use a rather smooth UMAP representation, focusing more on the global than the local structure of the representation space, which leads to a fiducial choice of 100 neighbours and a minimal distance of $0.5$.}

In Figures~\ref{fig:tng_umaps}, \ref{fig:tng_umaps_2} and \ref{fig:tng_umaps_3},  we show the UMAPs of the TNG100 (left panels) and TNG50 (right panels) test galaxies, coloured according to a selected (but not exhaustive) set of observable galaxy properties. First, we show UMAPs coloured according to exactly known galaxy properties, i.e. measured from the stellar particle data of the TNG50 and TNG100 simulations considering all stars gravitationally bound to each galaxy according to \textsc{Subfind} \citep{Springel_2001}: total stellar mass, fraction of disk stars \citep{2015ApJ...804L..40G}, integrated galaxy colour \citep{nelson2017results}, half mass radius \citep[refer for definition to][]{Eisert_2023}. We then examine morphological measurements that were derived from the survey-realistic mocks using \textsc{Statmorph} \citep{Gomez_2019} on the r-band of the uncropped TNG mock images: Sérsic half light radius, Sérsic index, Sérsic ellipticity \citep{Sersic_1963, Graham_2005}, {\change point asymmetry} of the light distribution, concentration of the light, smoothness of the light distribution, Gini-M20 bulge parameter and Gini-M20 merger parameter \citep[refer for definition to][]{Gomez_2019}. In doing so we also incorporate statistics as they are measured and used by observers. Note that we remove $4352$ (of overall $25896$) images with unreliable \textsc{Statmorph}-fits when plotting the UMAPs for the \textsc{Statmorph} parameters.

We see that all these observable quantities are very well captured in the representation space: namely, galaxies with similar properties occupy the same regions within the representation space.
Importantly, known and expected relations among properties are conserved, e.g. disk galaxies have on average a lower stellar mass while presenting themselves in a bluer colour due to ongoing star-formation (and therefore a lower g-r colour index). We also can see the expected relationship between the theoretical half-mass radius and the observational Sérsic half-light radius, one being derived just by summing up the known stellar particle masses and the other being based on a profile fit to galaxy light, respectively: despite the different definitions, both quantities align well in the representation. This is especially of interest as the FOV was chosen to be proportional to the Petrosian galaxy sizes, and hence the size is not directly observable from the images.
It is also interesting to note that the Gini-M20 Bulge and Merger parameters are captured, which promises deeper insights in the structural composition of each galaxy and its history.

Comparing the results between TNG50 and TNG100, both sets show qualitatively similar trends: the model is therefore able to map two IllustrisTNG runs with different resolution into the same space. However, this statement is just based on the median of the quantities in a 2D projection. {\change Although wedo not show any information about the scatter and point densities in each of the hexagonal bins, we have checked and indeed the standard deviation of some galaxy properties (e.g. galaxy colors) can be large and varying across the UMAPs, even though the medians exhibit clear trends.} In the sections that follow, we therefore employ additional methods to compare the images based on the full representation space.

\subsection{How similar are the mocks from the TNG simulations to the observed HSC images?}
\label{sec:similarity}
The previous section showed that the ML-constructed representations of galaxy images naturally encode morphological properties and stellar structures. But how can we gauge whether simulated galaxies are realistic in comparison to observed ones? As argued in Section~\ref{sec:intro}, this assessment is especially important for inferring unobservable properties from HSC data. This is only possible when the TNG galaxies that are used to learn the connection between light maps and galaxy properties are sufficiently realistic in comparison to the HSC observations in question. The task ahead is therefore to assess if the network has learned a sensible definition of similarity and to quantify this similarity.

In the following, we proceed with the underlying idea that, if two galaxy images are close in the $256$-dimensional representation space, then they are also ``similar'' in appearance. If the representations of two galaxy samples have disjoint or matching subsets, we can use this to measure for which type of galaxies the mocks from TNG and the HSC images are different or similar.

\subsubsection{Comparing the representation space with UMAPs}
\label{sec:umap_compare}
As a first step we reuse the UMAP method from Section \ref{sec:umap} to investigate the overlap in 2D among the three image sets (TNG50, TNG100 and HSC). 
\begin{figure*}
	\centering
	\includegraphics[width=0.33\linewidth]{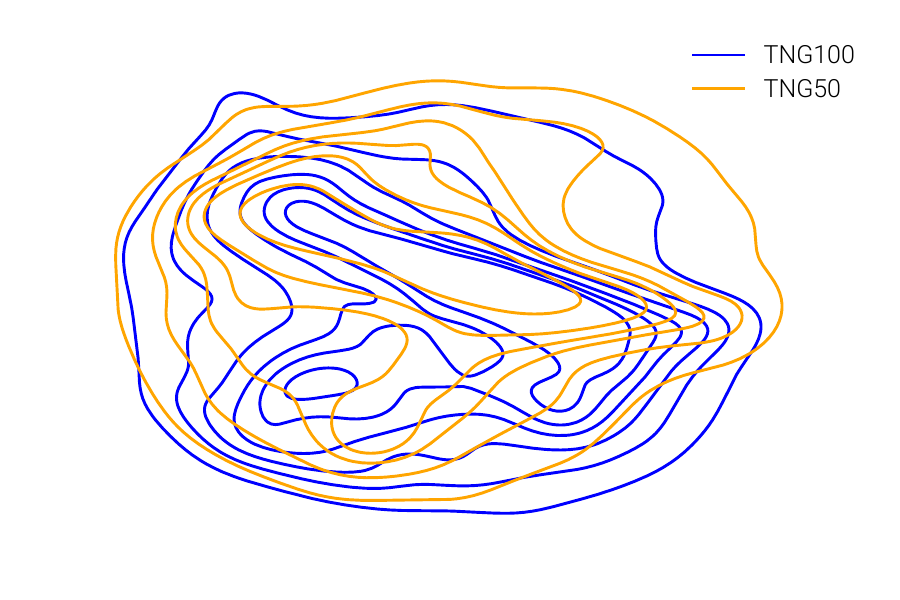}
        \includegraphics[width=0.33\linewidth]{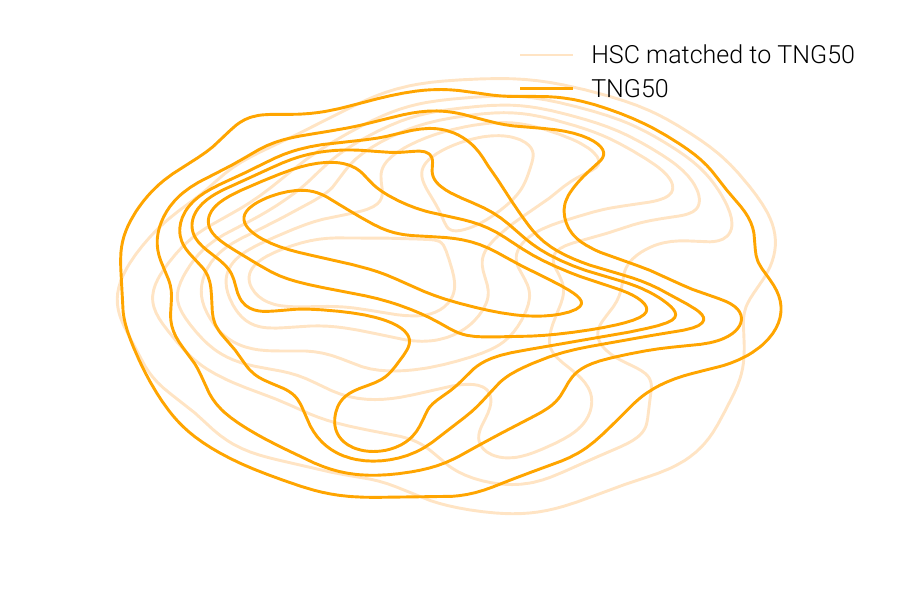}
        \includegraphics[width=0.33\linewidth]{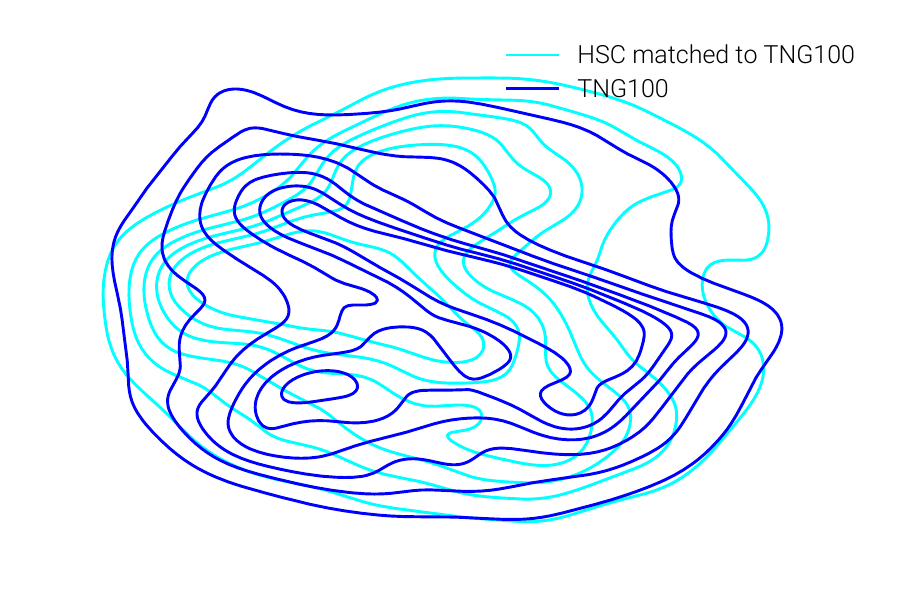}
        \includegraphics[width=0.33\linewidth]{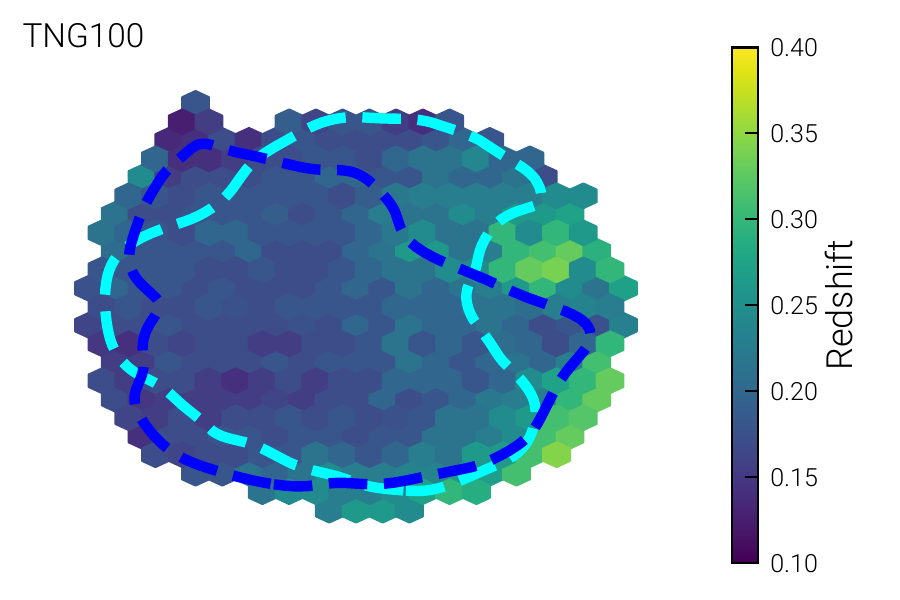}
        \includegraphics[width=0.33\linewidth]{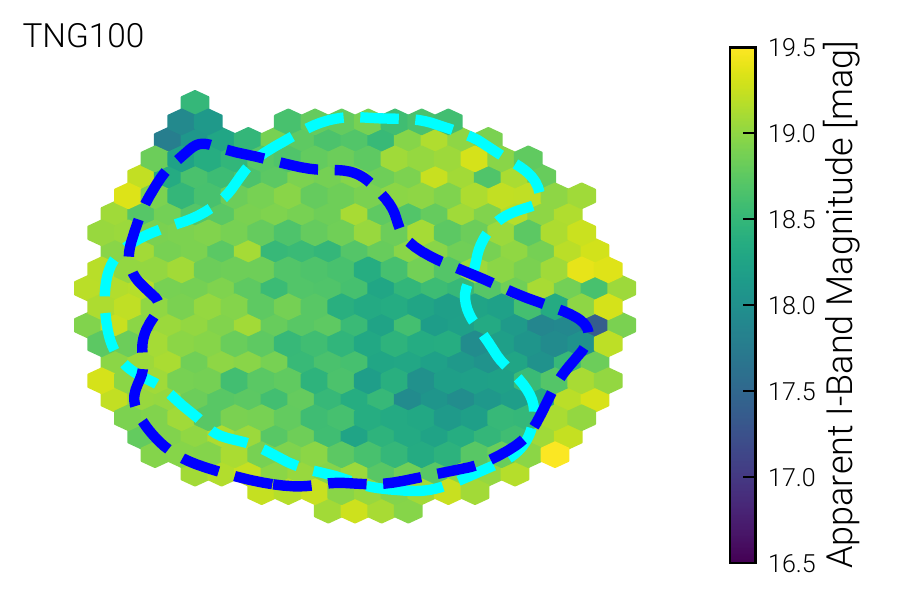}
        \includegraphics[width=0.33\linewidth]{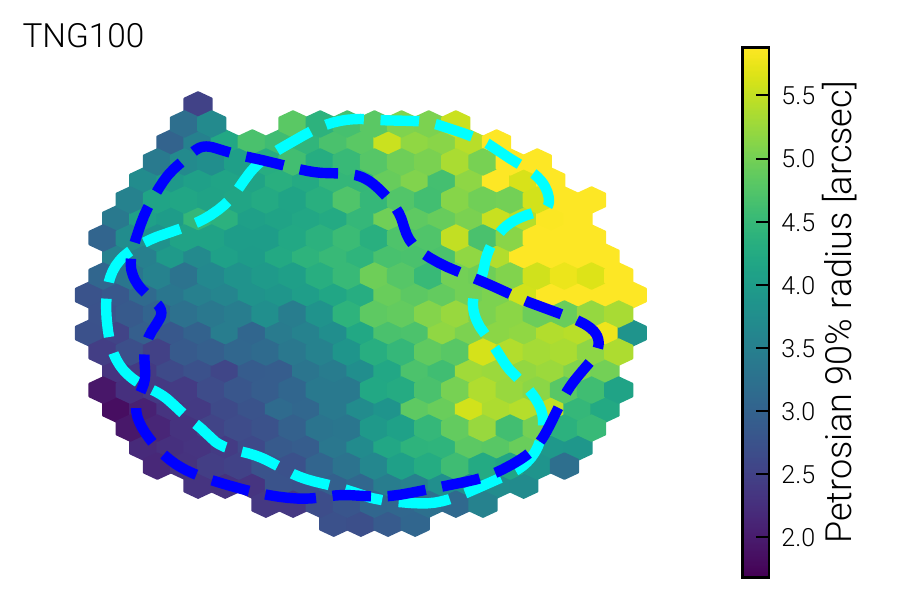}
        \includegraphics[width=0.33\linewidth]{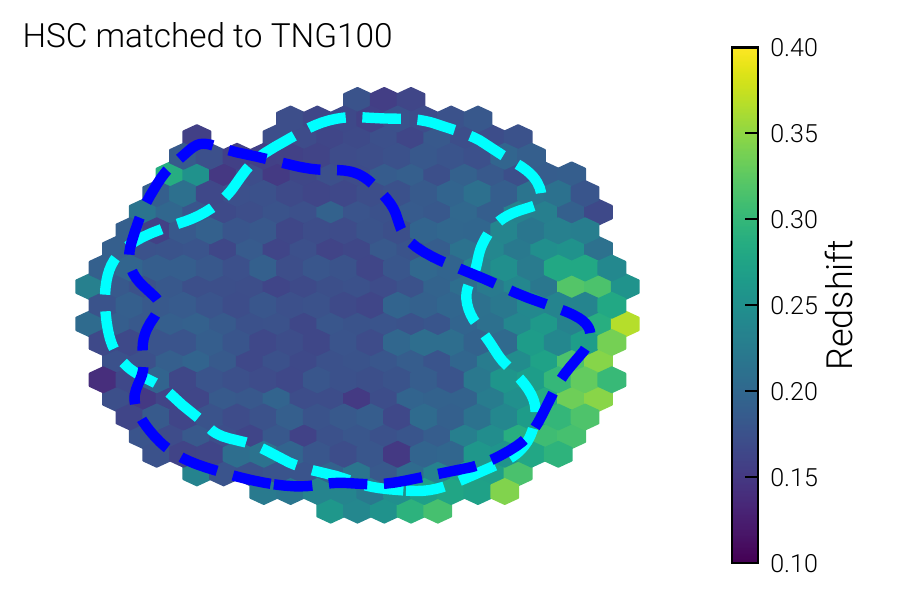}
        \includegraphics[width=0.33\linewidth]{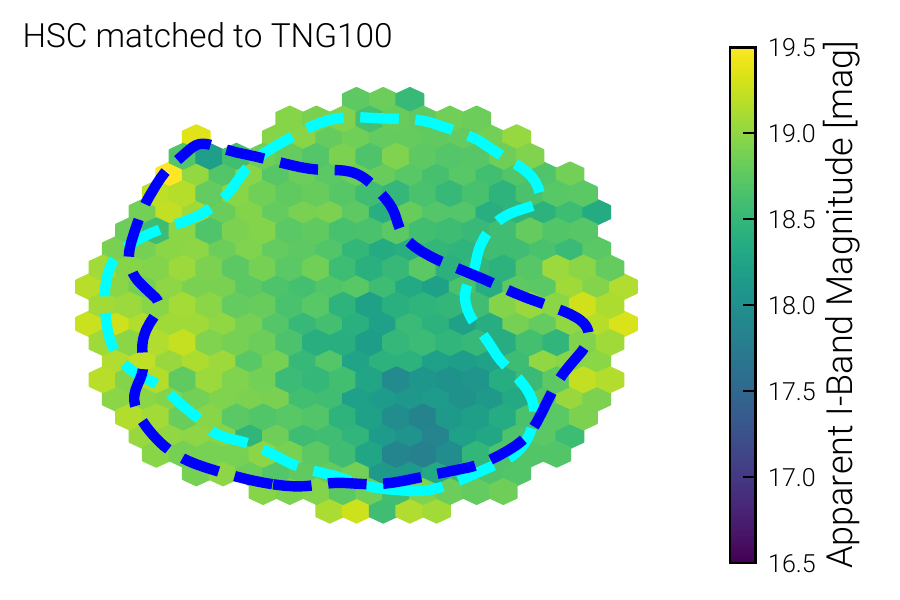}
        \includegraphics[width=0.33\linewidth]{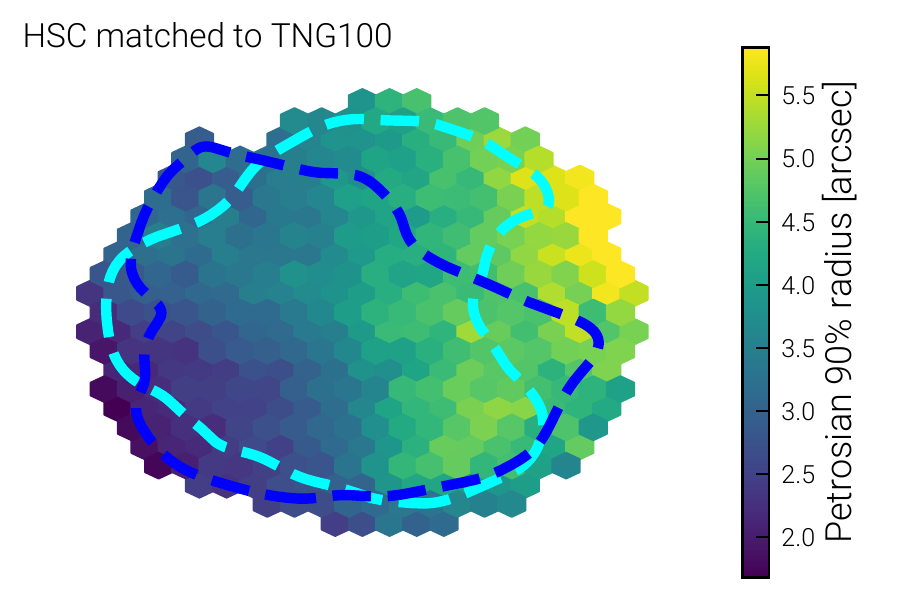}
        \includegraphics[width=0.33\linewidth]{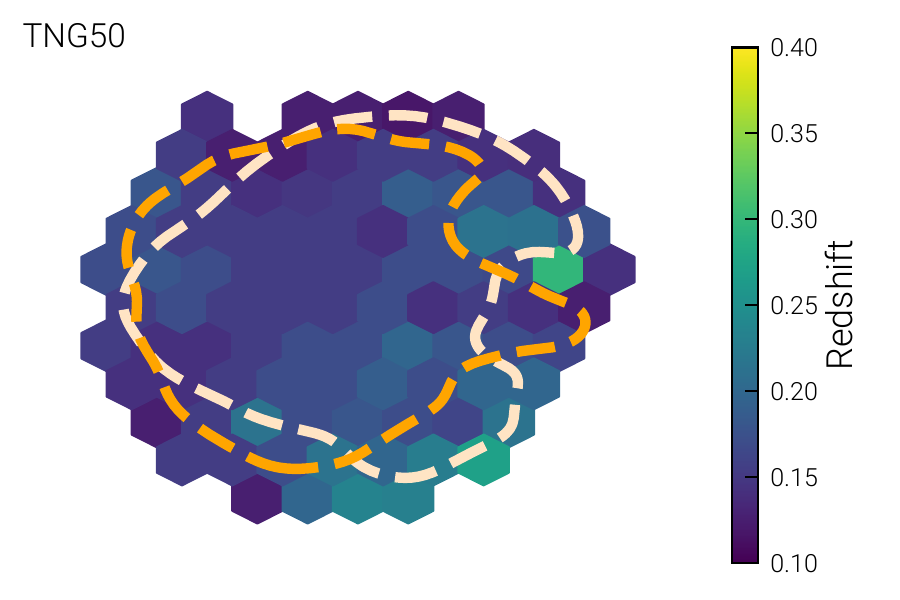}
        \includegraphics[width=0.33\linewidth]{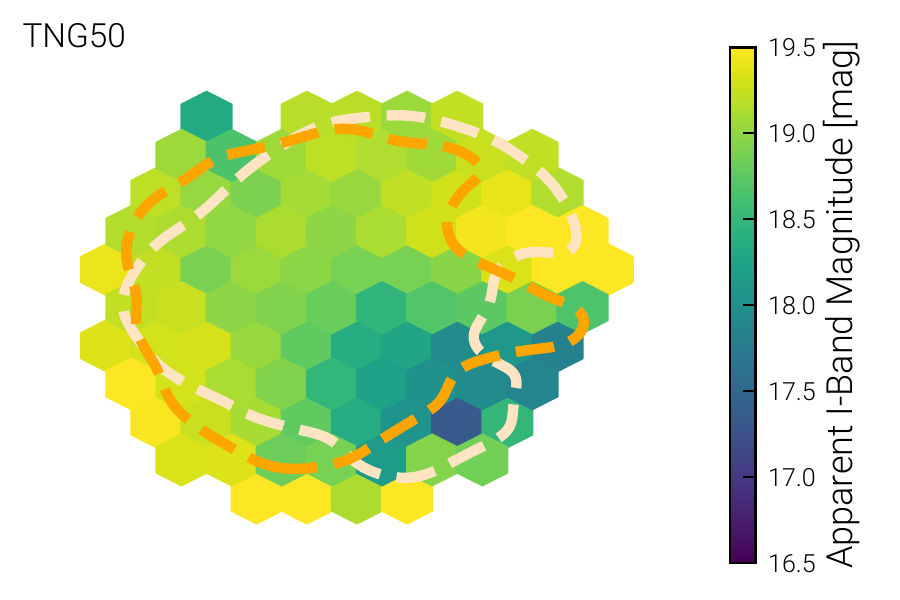}
        \includegraphics[width=0.33\linewidth]{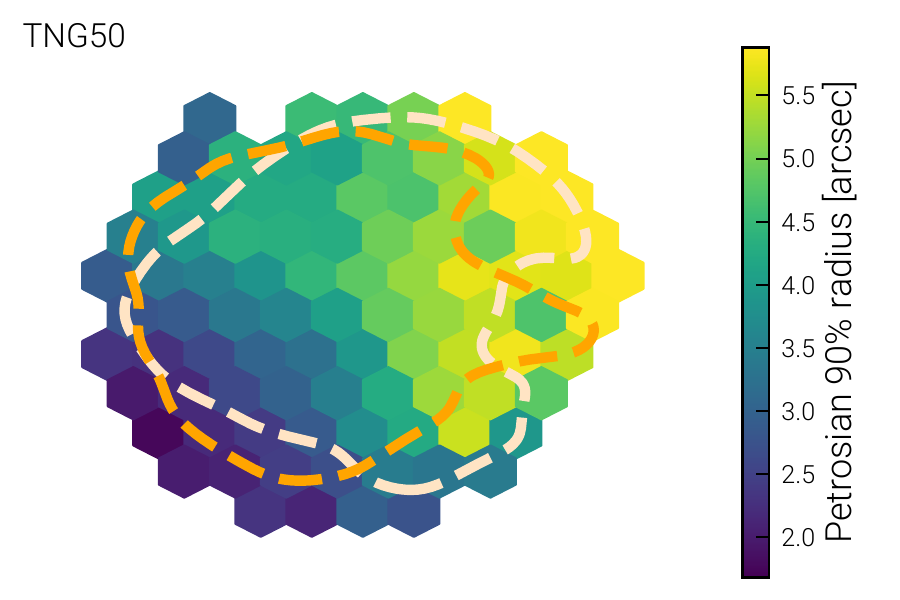}
        \includegraphics[width=0.33\linewidth]{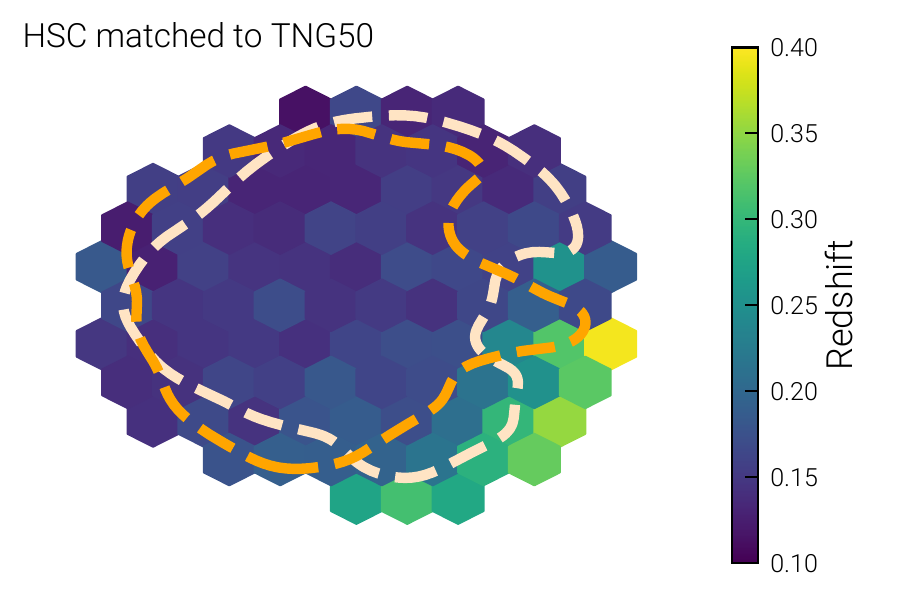}
        \includegraphics[width=0.33\linewidth]{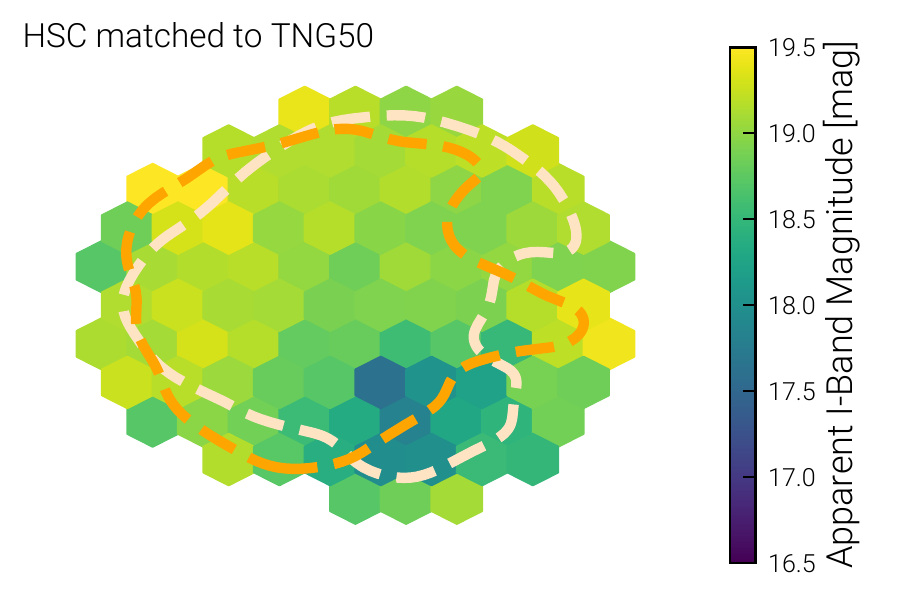}
        \includegraphics[width=0.33\linewidth]{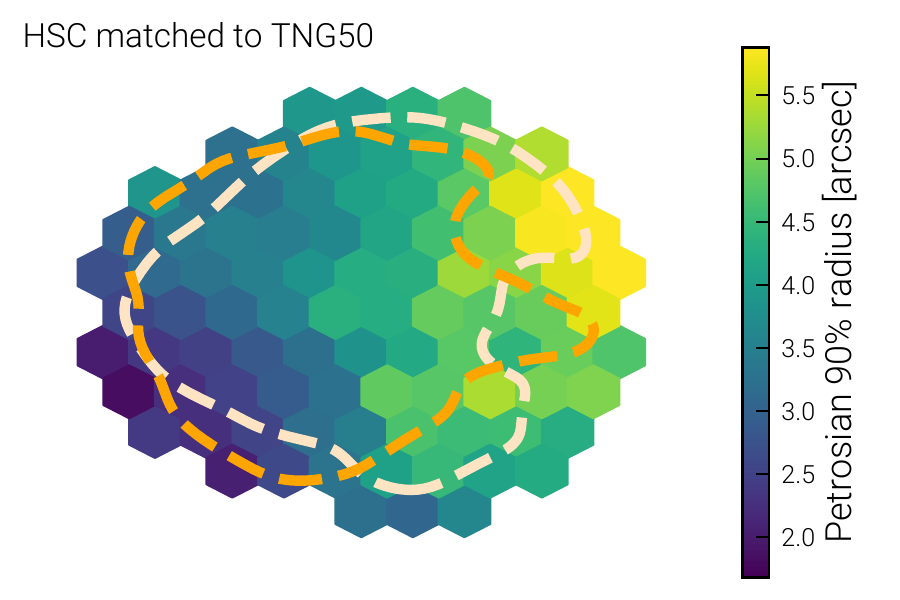}
	\caption{{\bf How well do the representations of the observed and simulated galaxy images align to each other?} We compare the distributions of TNG50/100 and HSC images in the corresponding 2D-UMAP mapping of the 256-dimensional representations. In the uppermost panel we show KDE density plots of TNG100 (TNG50) images in blue (orange). Additionally the HSC sets that are matched against TNG100/TNG50, respectively, are shown in light blue/orange in two additional plots. We see that there is a large overlap among the three sets but also a slight offset and differences in point density. 
    In the lower two panels we further investigate the offset in a visual way: we show the 2D hexbins of the UMAPs coloured bin-wise according to the median value of the three common properties used for the galaxy-sample matching: from left to right, redshift, i-band magnitude, Petrosian radius. In each row we show one of the datasets: from top to bottom, TNG100, HSC matched to TNG100, TNG50, HSC matched to TNG50. We also include the contours covering 80 per cent of the datasets from the uppermost panel: TNG100 in blue, TNG50 in orange and in light orange/light blue the subsets of HSC matched to TNG50/TNG100.}
	\label{fig:kde_umap_compare}
\end{figure*}

The upper panel of Figure \ref{fig:kde_umap_compare} shows the distribution of TNG50/100 and HSC representations in a 2D UMAP. A noteworthy success is that the sets are clearly not disjoint. In other words, there is significant overlap in the image-based representations of TNG50/TNG100/HSC galaxies. Meanwhile, we see that a residual fraction of TNG50/100 and HSC galaxies do not share a common domain in their representations. Also, the point densities do not align between the three sets. For example, the TNG sets both exhibit higher densities (i.e. more galaxy images) along a diagonal segment through the center of the UMAP, while being suppressed in the top right region.

The next four rows of Figure \ref{fig:kde_umap_compare} compare the UMAP embeddings of TNG50/100 images and their matched HSC ones, respectively, with histograms coloured according to the binned median of the properties used to match the galaxy samples (see Section~\ref{sec:matching}). The first two rows show the UMAPs of the TNG100 galaxies and the UMAPs of the HSC galaxies matched to TNG100, respectively. Similarly, the latter two rows show the UMAPs of TNG50 galaxies and the UMAPs of the HSC galaxies matched to TNG50, respectively. From left to right, the color encodes (photometric) redshift, apparent $i$-band magnitude, and Petrosian radius.

Our main finding is that the trends in the UMAP embeddings of the observed quantities are visually well-reproduced across the three datasets -- especially for the Petrosian radii. However for the redshift there is some discrepancy between TNG100/50 and their respective matched HSC sets. This is very pronounced for TNG50, where almost no clear trend between lower- and higher-redshift galaxies is visible. {\change For TNG100 and the matched HSC sets there is however a strong correlation towards the bottom/right of the UMAPs.} During training, we found that an alternative colour stretching, which puts more emphasis on the noise while removing the noise level of the augmentation, enhances the model's ability to distinguish between redshifts. However, this greater sensitivity to redshift compromises the model's sensitivity to other galaxy features. Furthermore, the distribution of redshifts in the sample is very askew. The model therefore might be blinded by the vast amount of galaxies with lower redshifts -- especially for TNG50 for which the sample truncates at higher-redshifts due to a lack of appropriate matches in the magnitude-limited HSC sample (see Figure \ref{fig:matching}).

To summarize in relation to the areas where the distributions are not well aligned (see top most panel), we find the following: (1) for TNG100, the top right corner of the UMAPs suggests issues with low-redshift galaxies; (2) in contrast, in the bottom right corner of the UMAPs, high-redshift galaxies for the TNG50 set are missing.

\begin{figure*}
	\centering
	\includegraphics[width=16cm]{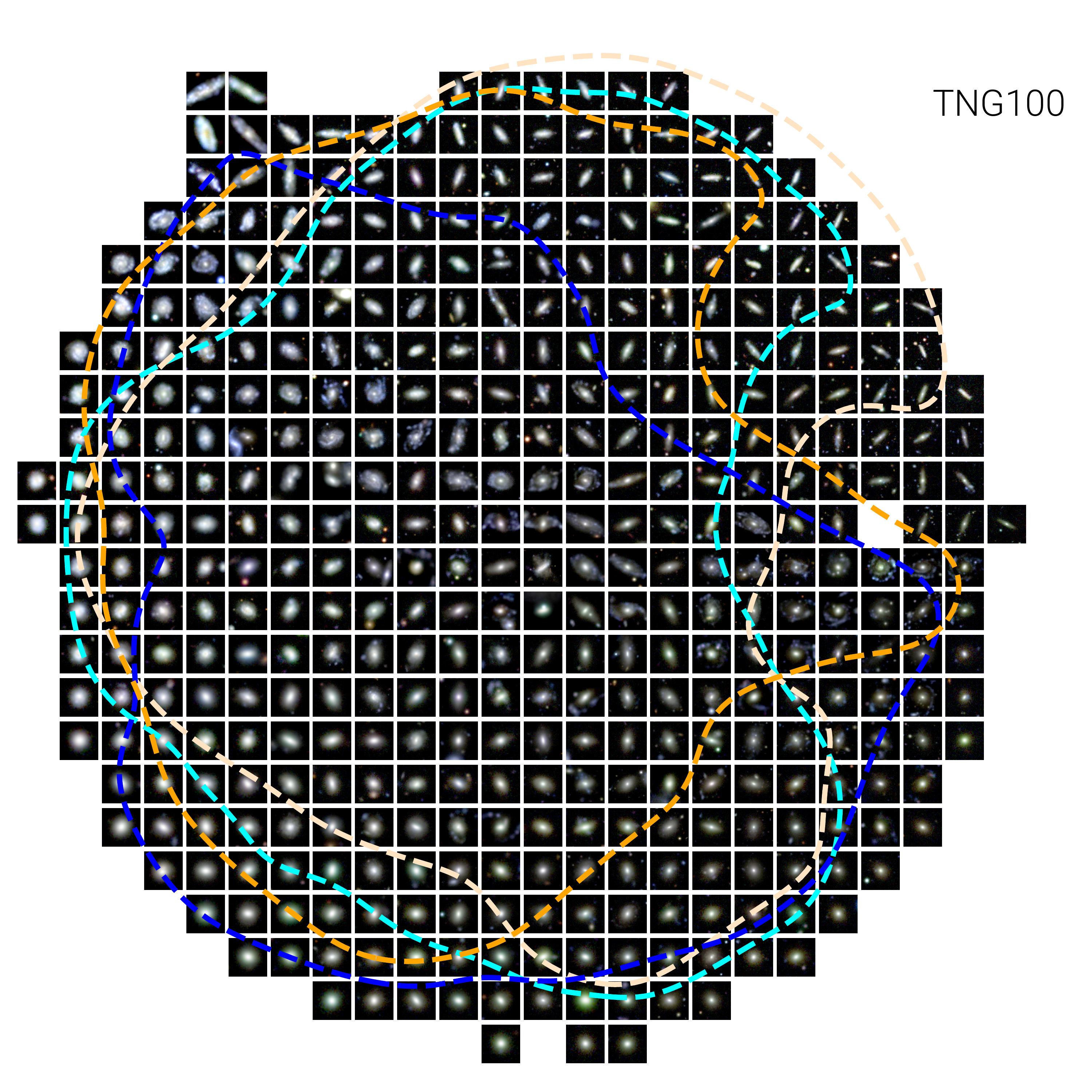}
	\caption{{\bf What images underlie the UMAP representations?} We align TNG100 galaxy images to their positions in the UMAP representations from Figure~\ref{fig:kde_umap_compare}. For each node on a rectangular grid in 2D UMAP space, we choose the closest data point (i.e. image representation) and plot its corresponding image. As a guideline, we report the contours as in Figure~\ref{fig:kde_umap_compare} covering $80$ per cent of the datasets: TNG100 in blue, TNG50 in orange, and the corresponding HSC matched samples in light blue and light orange. With this visual inspection, we see that the galaxy structures are indeed well grouped together.}
	\label{fig:umap_images_TNG100}
\end{figure*}

\begin{figure*}
	\centering
	\includegraphics[width=16cm]{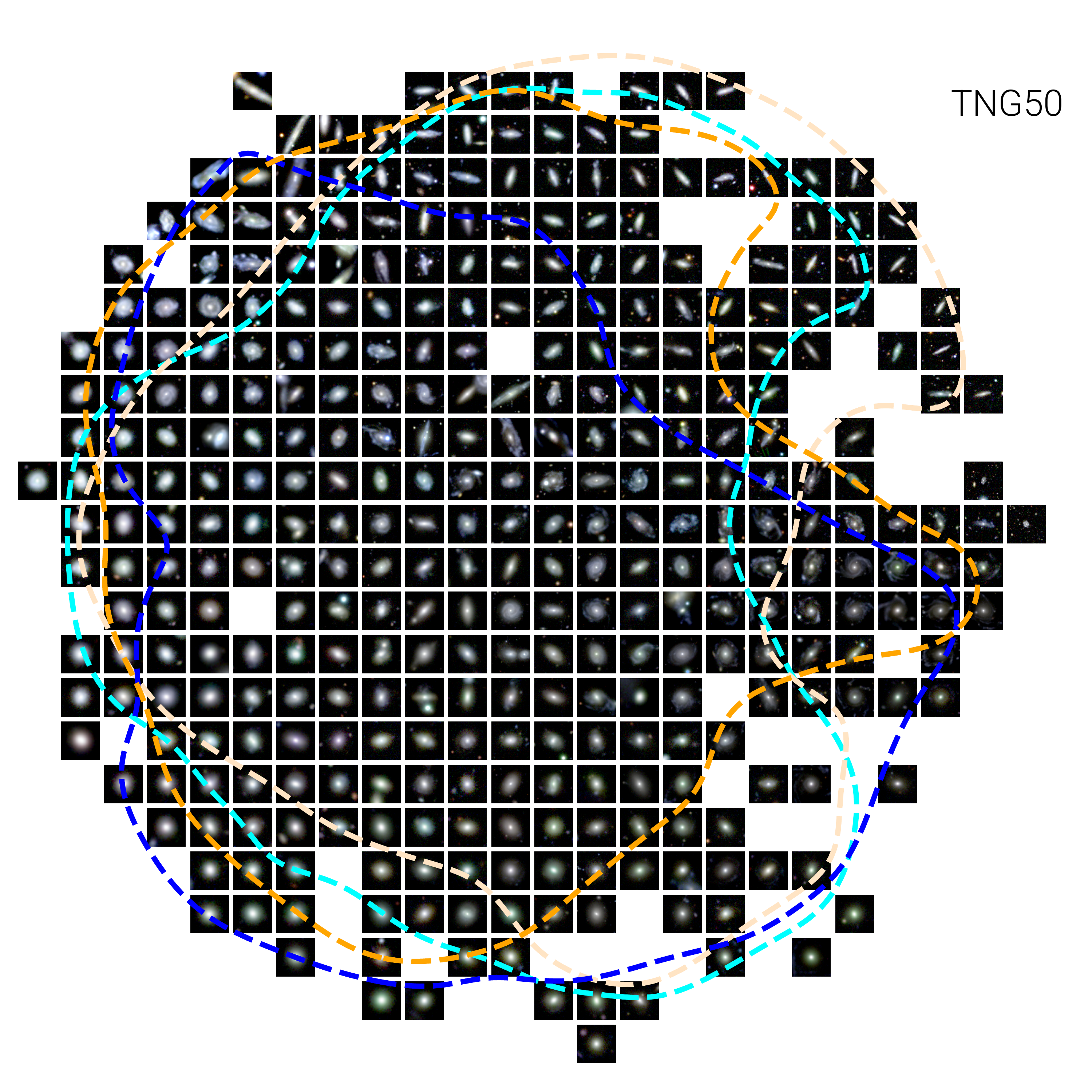}
	\caption{Same as Figure \ref{fig:umap_images_TNG100} but for TNG50 galaxies. Note that there are more empty regions due to the smaller sample size.}
	\label{fig:umap_images_TNG50}
\end{figure*}

\begin{figure*}
	\centering
	\includegraphics[width=16cm]{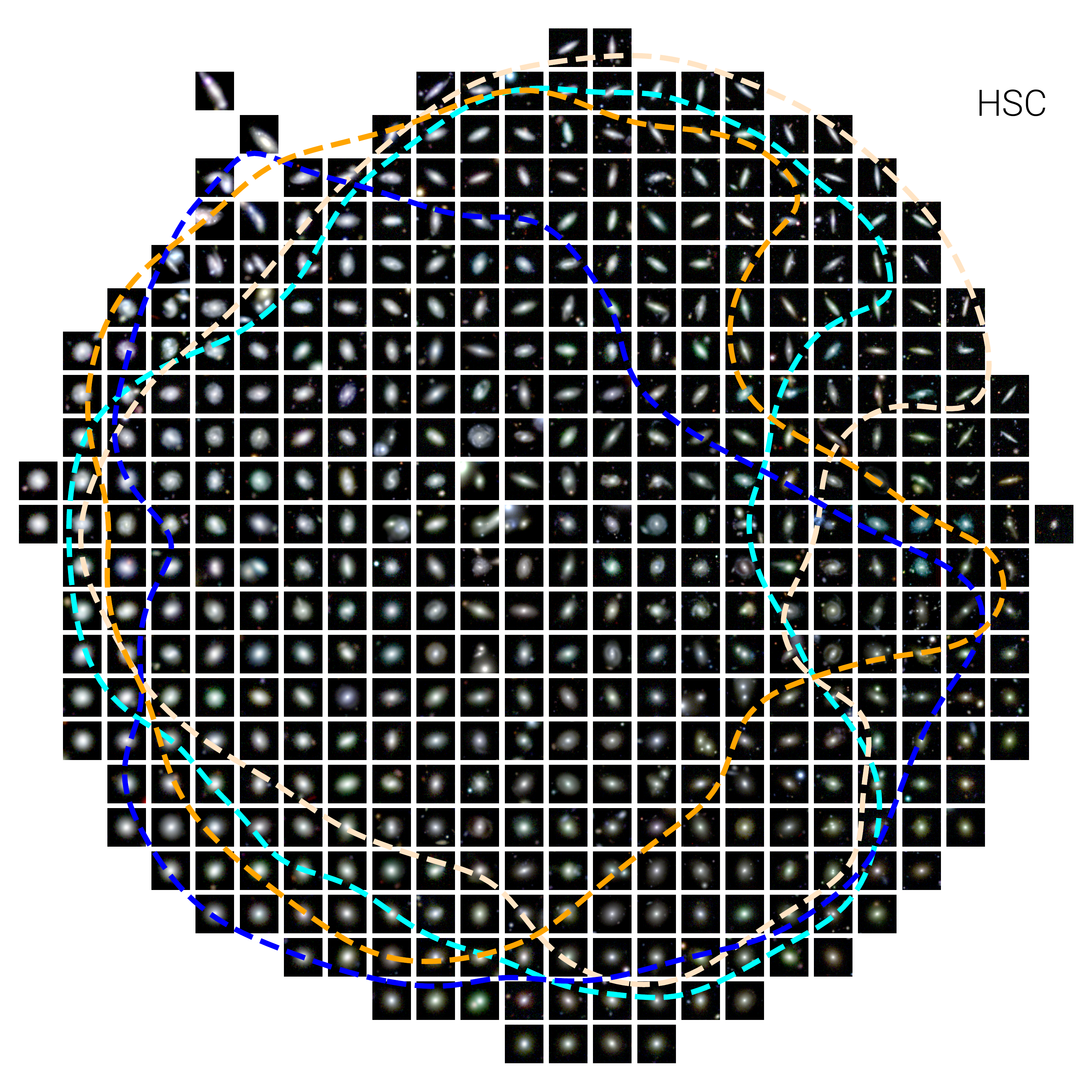}
	\caption{Same as Figure \ref{fig:umap_images_TNG100} but for HSC galaxies, specifically the HSC sample matched to TNG100. For brevity, we omit showing the analog figure for the HSC sample matched to TNG50, which would appear very similar.}
	\label{fig:umap_images_HSC}
\end{figure*}

In Figures \ref{fig:umap_images_TNG100} (TNG100), \ref{fig:umap_images_TNG50} (TNG50) and \ref{fig:umap_images_HSC} (HSC, matched to TNG100) we show the images that are related to the representations of Figure~\ref{fig:kde_umap_compare}: namely, on a grid in 2D UMAP space, we choose the closest data point of the representations and plot the affiliated galaxy image. 

By comparing visually the trends for the three sets, we can see that the similarities are astonishing. For example, in all figures: (1) In the top-right, we see edge-on spiral galaxies; (2) in the top-left, we see spirals from the top, i.e. face on; (3) at the bottom, we see elliptical galaxies. 
Comparing this to the results of the quantities in Figure~\ref{fig:tng_umaps} that where directly extracted from the simulated data, we can conclude that the overall concept of diskyness is well understood by the model.

However, there are also a clear differences between TNG50/100 and HSC: blue stellar clumps around and in disks of TNG galaxies are not prominently observed in HSC. 
Features like this have been previously noted and commented upon in stellar light mocks of TNG100 galaxies \citep{Gomez_2019} and other simulations like Illustris and Eagle as well \citep{Torrey_2015, Trayford_2017}. It might be caused by young stellar populations that are insufficiently resolved by the particle scheme, i.e. a bright and fast evolving stellar population is treated by only a single particle.
Furthermore the strength of these blue regions might be also affected by dust assumptions for the light-map creation with SKIRT. These simulated galaxies with blue stellar clumps are very prominent in the top-left to bottom-right diagonal, which corresponds to the 'central and diagonal segment' we have noticed in Figure~\ref{fig:kde_umap_compare}. We find that it is somewhat model dependent if these blue clumps have a strong influence on the shape of the UMAPs. However, as the UMAP is just a 2D projection of the 256-dimensional representation space, the blue clumps might still be well outside the HSC sample. For the ML model adopted throughout, the blue clumps are expressed by a certain clustering of these galaxies in the aforementioned segment. 

{\change In the top-left, we see very extended edge-on spirals, some of which are partially even larger than the FOV. In Figure~\ref{fig:kde_umap_compare}, we can see that the fitted Petrosian radii, used to determine the FOV of the images, are comparatively small in this corner. Contrary to this, the physical (simulation-based) mass radii in Figure~\ref{fig:tng_umaps} and the apparent Sersic sizes in Figure~\ref{fig:tng_umaps_2}, suggest that the spirals are indeed large. This area is, therefore, populated by massive spirals whose Petrosian radii fit, and therefore FOV calculation, failed. Interestingly, this area is not heavily populated by HSC galaxies. We hypothesize that they are missing, as those large spirals have been removed from the HSC sample due to a crossmatch failure between HSC and SDSS sources \citep{Shimakawa_2023}.}

\subsubsection{Quantifying the similarity via the normalized 8th-nearest-neighbour distance and the OOD Score}
\label{sec:nnd}
While the visual inspection of the UMAPs is a rather intuitive way to compare two distributions, it only takes a compressed 2D map of the full 256 dimensions into account. This might let galaxies appear closer/farther to each other than they are in the higher-dimensional space. We see this problem, for example, by the fact that the simulated clumpy galaxies that clearly differ from the HSC observations are in the center of the UMAP. {\change Furthermore, relative volumes are not conserved by the non-linear mapping. The areas enclosed by contours in Figure~\ref{fig:kde_umap_compare} are therefore not one to one transferable to the contours in the high dimensional representation space.} In this section, we therefore proceed with a more quantitative way to compare simulated and observed galaxies by defining and calculating a scalar that tells us how far a TNG galaxy is from the HSC domain (or the other way round), i.e. an out-of-domain score.

Firstly, we introduce the normalized 8th-neighbour distance, which is obtained by the following steps:
\begin{enumerate}
    \item Normalize the dimensions; i.e. scale each dimension of the representation to a mean of 0 and a variance of 1 to ensure that the variance in all dimensions have the same amplitude. We perform the normalization for all datasets (HSC and TNG50/100) together.
    \item Calculate the distance to the 8th nearest neighbour for each galaxy image in its own set ($d_{\rm{self}}$) and to the 8th nearest neighbour in the other set ($d_{\rm{other}}$); {\change here we use the Euclidean norm.} Here we use these distances as a measure for the point density. To avoid too noisy statistics we smooth the distances out by using a $k^{\mathrm{th}}$ nearest neighbour instead of choosing the nearest.
    \item Define the normalized 8th-neighbour distance as the ratio between the two distances $d_{\rm{other}} / d_{\rm{self}}$. The normalization by $d_{\rm{self}}$ is necessary to ensure that the distance is not mainly dominated by varying point densities. Note that each distance is calculated separately for each galaxy; it is therefore a quantity assigned to each galaxy separately and not for the distribution at whole.
\end{enumerate} 
For a given sample positioned somewhere in representation space, the normalized 8th-neighbour distance thereby describes the local ratio of point densities between another dataset and the sample's own dataset. For example, for a TNG galaxy that resides far outside of the HSC domain but still has many TNG neighbours, $d_{\rm{self}}$ will be small and $d_{\rm{other}}$ will be large, resulting in a normalized 8th-neighbour distance that is greater than unity. In contrast, for a TNG galaxy that resides in a space that is heavily populated by HSC galaxies and few other TNG galaxies, $d_{\rm{other}}$ will be smaller than $d_{\rm{self}}$ and the 8th-neighbour distance will be much smaller than unity.
Very large distances therefore indicate discrepancies between the \emph{individual images} and the populations to which they are compared to. Very small distances on the other hand indicate that the \emph{individual images} are well within the domain of the population they are compared to. {\change Note that this normalized 8th-neighbour distance is only meaningful if it is calculated between two sets of equal size. In the following we therefore always select a random subset from the larger set which matches the number of samples in the smaller set.}
\begin{figure*}
	\centering
	\includegraphics[width=0.45\linewidth]{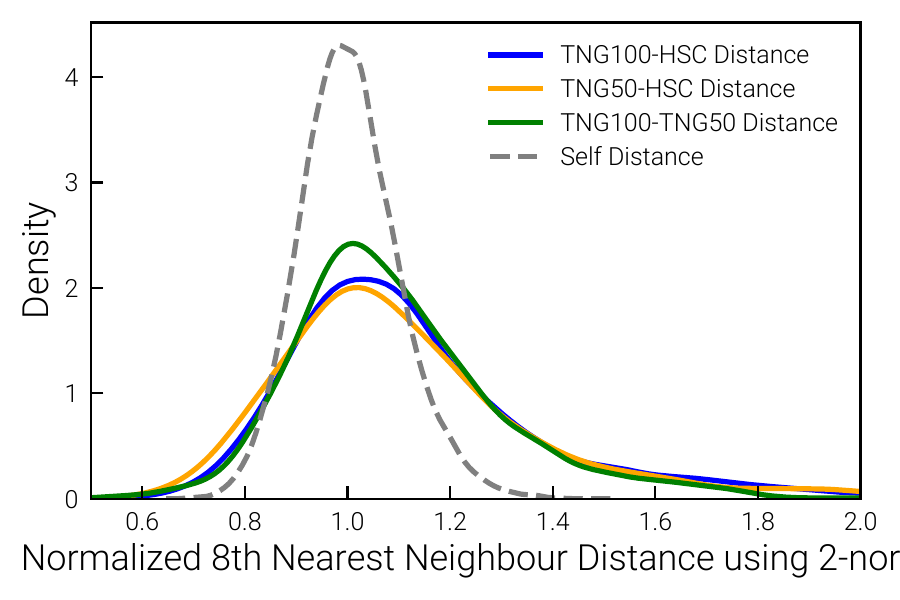}
	\includegraphics[width=0.45\linewidth]{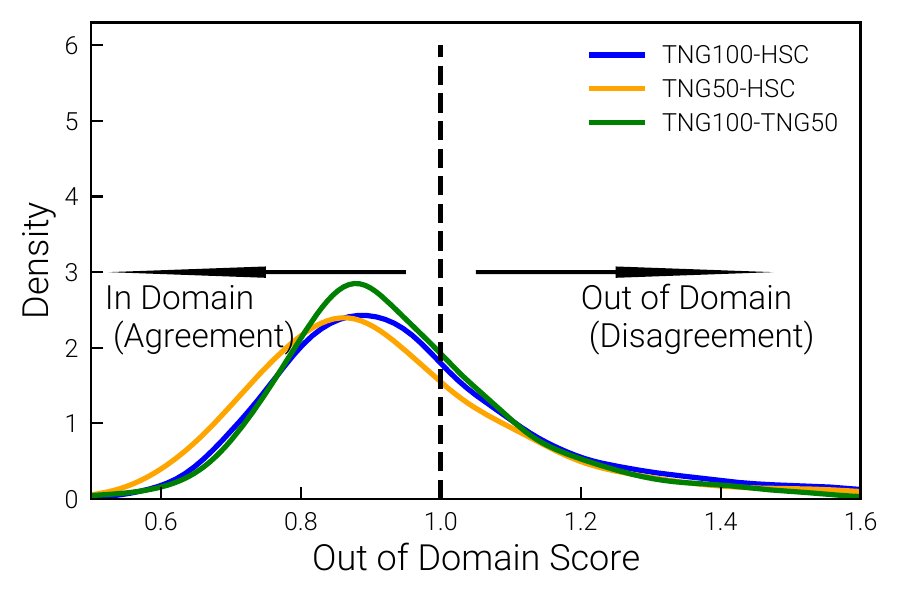}
	\caption{{\bf Our measure to quantify the discrepancy between the individual galaxies and a target dataset: the Out of Domain Score (OOD).} Here we show the distributions for the measures that we have introduced and that aim at quantifying the overlap between the domains of the observed (HSC) and simulated (TNG50 and TNG100) datasets. Left: Distributions of the `raw' normalized 8th nearest neighbour distance. We measure this quantity for various scenarios: between pairs of the three data sets and also between two random splits of each dataset to investigate the intrinsic scatter that persists even if two distributions are drawn from the same population. For convenience we only show the HSC self-distance as the self-distances are qualitatively similar for all three sets. Note that there is a difference from which 'side' this quantity is measured: E.g. TNG100-HSC is the distribution of the distances for the TNG100 galaxies relative to the HSC galaxies. HSC-TNG100 is the distribution of the distances for the HSC galaxies relative to the TNG100 galaxies. However, for convenience we only show the distances in one direction as the distances are also qualitatively similar for both directions. Right: The measure that we use in the rest of the paper: the Out of Domain Score (OOD). It is overall equal to the left panel but normalized by the 95th percentile of the self distances. In doing so we define galaxies with scores larger than $1$ as out-of-domain while scores smaller than 1 as in-domain. With this definition we measure that $33$ per cent of the TNG100 and $30$ per cent of the TNG50 galaxies are out-of-domain relative to the HSC images. Overall we measure that the comparison in 256 dimensions yield similar results as the ones in the 2D UMAP representations in Figure \ref{fig:kde_umap_compare}. There is a large overlap between all three datasets. However, it is also noteworthy that there are fewer extreme outliers between TNG50 and TNG100 than between the simulations and HSC. This is a hint for a general offset between TNG and reality.}
	\label{fig:similarity_distribution}
\end{figure*}

For two equal-sized sets with identical distributions this ratio is anticipated to be equal to $1$. However we have to expect a certain intrinsic scatter which is caused by the fact that $d_{\rm{self}}$ itself is affected by a certain selection noise from the nearest neighbour search. To interpret the similarities of galaxy images, we hence need to quantify the amplitude of this scatter. Therefore, for each dataset, we also calculate the normalized 8th-neighbour distance between two equal-sized splits of that dataset to estimate the `baseline' scatter. In other words, this is the scatter about unity that one would get when calculating the scores between two equal-sized samples drawn from the same population. In total, we measure the distances for the following scenarios:
\begin{enumerate}
    \item Between the TNG100 set and the HSC set
    \item Between the HSC set and the TNG100 set
    \item Between the TNG50 set and the HSC set
    \item Between the HSC set and the TNG50 set
    \item Between the TNG100 set and the TNG50 set
    \item Between the TNG50 set and  the TNG100 set
    \item Between two random same-size splits of the TNG50/100 sets
    \item Between two random same-size splits of the HSC set
\end{enumerate}
In all cases, by HSC set we mean the one matched to the corresponding simulation sample, as per Section~\ref{sec:matching}. The last three scenarios are aimed at estimating the baseline scatter for each dataset. Note, that there is a priory a difference in which direction this quantity is calculated; e.g. for 'Between the TNG100 set and the HSC set' we calculate $d_{\rm{self}}$ using the TNG100 galaxies and $d_{\rm{other}}$ using the HSC galaxies. While we derived the quantities in both directions, the differences between the two directions are qualitatively negligible and we therefore focus in the following only on the scores that are calculated for the simulations relative to HSC (cases i and ii above). 

We plot the distributions of the obtained measures in the left panel of Figure \ref{fig:similarity_distribution}. The modal value is indeed close to $1$ for all scenarios. Looking at the variance, we see that the distributions of the distances between HSC and TNG galaxy images (blue and yellow curves) extend to values larger than 1, as the underlying samples are not identical, as we already saw qualitatively with the UMAPs. We have checked and this picture remains unchanged if we had used a distance to the $n$th nearest neighbour with $n=4, 16, 32$ instead of 8 (Figure~\ref{fig:nn_variation}).  

Now, comparing the distributions between HSC and TNG galaxy images with the distribution between the three self-splits of TNG/HSC galaxies (all roughly equivalent and indicated with the one dashed gray curve), we can now quantify the intrinsic scatter. We incorporate this in the definition of the so called out-of-domain score (OOD), defined by dividing each distance by the $95$th percentile of the self-distances. The resulting distributions are shown in the right panel of Figure \ref{fig:similarity_distribution}. 

We define galaxies with an OOD Score smaller than $1$ as ``within the domain'' of the target distribution. Galaxies with an OOD Score larger than $1$ are ``outside of the domain''. Naturally, this definition is arbitrary but should serve for now as a start point to classify the galaxies into the two regimes. However, we should ensure that the OOD is not too dependent on the model and the choice of the number of neighbours taken into account. Otherwise the main takeaways of the following sections would not reflect properties of the galaxy samples but mainly features of our fiducial model. We investigate and discuss this topic in Appendix \ref{sec:ood_dependence}. All in all, there is a large overlap between all three datasets. With our definitions, the large majority of the simulated galaxies are well within the domain of HSC observed galaxies: we measure that $33$ per cent of the TNG100 and $30$ per cent of the TNG50 images are out-of-domain relative to the HSC images, or conversely that about 70 per cent of the simulated galaxies are realistic, with little difference between the two simulations. However, it is also noteworthy that there are fewer extreme outliers between TNG50 and TNG100 than between the simulations and HSC. This is a hint for an offset between TNG and reality.

\begin{figure*}
	\centering
	\includegraphics[width=0.45\linewidth]{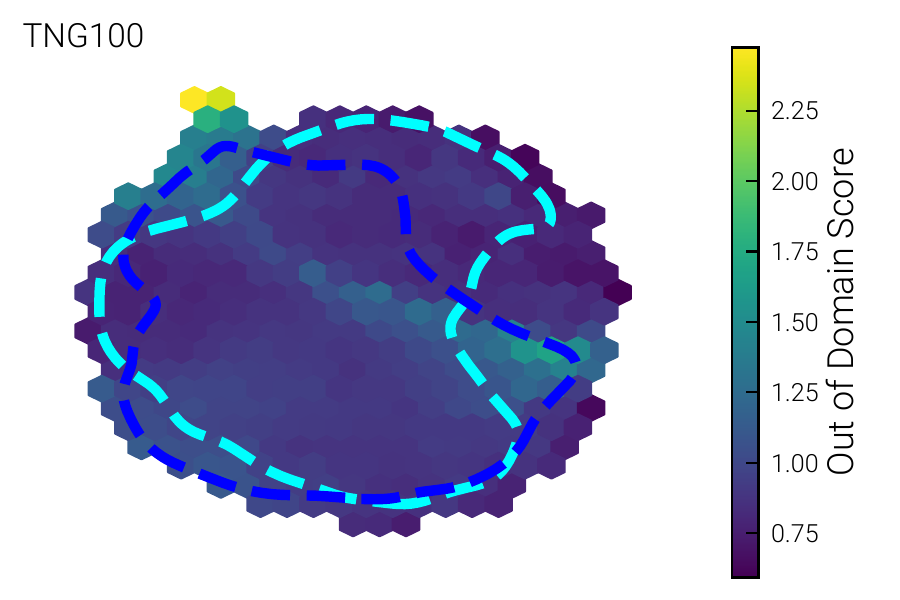}
	\includegraphics[width=0.45\linewidth]{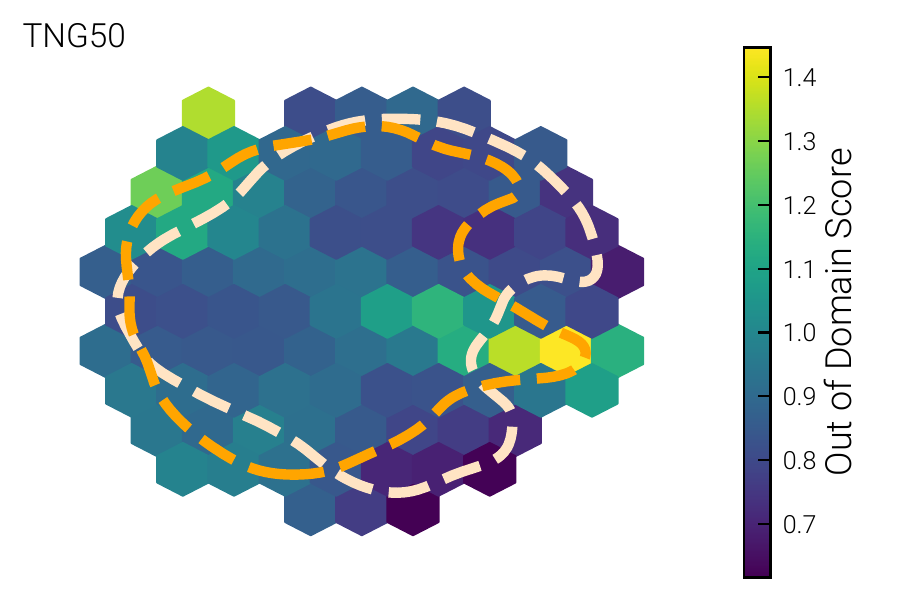}
    \includegraphics[width=0.45\linewidth]{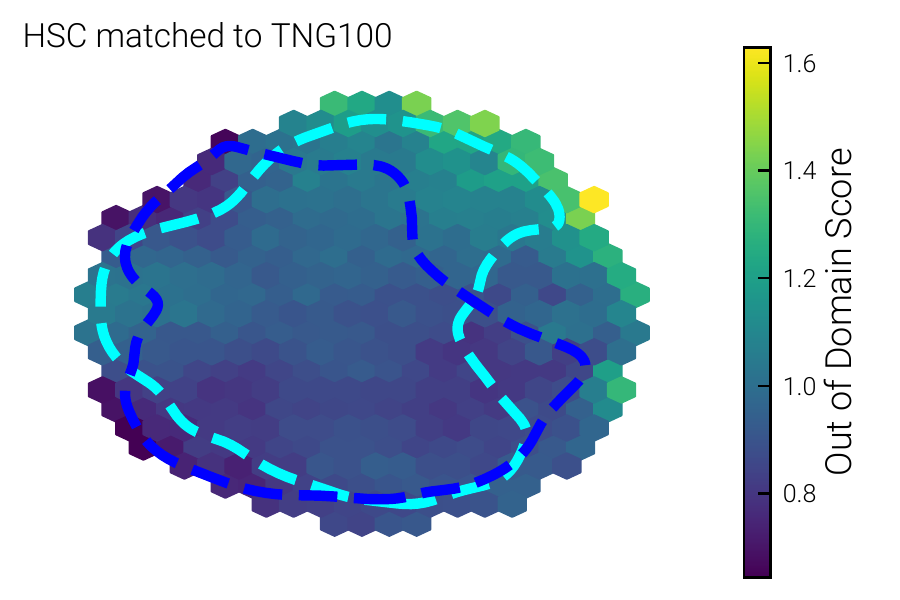}
    \includegraphics[width=0.45\linewidth]{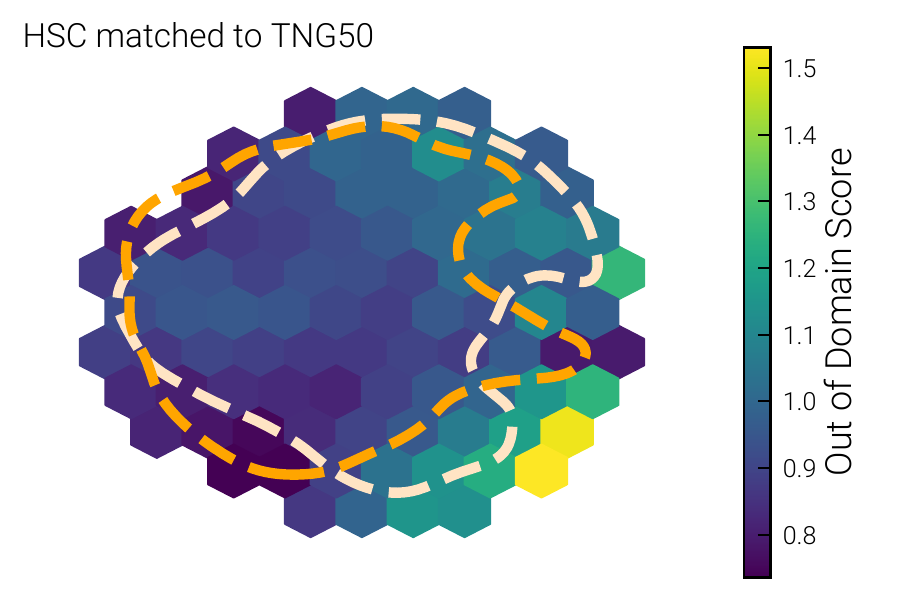}
	\caption{{\bf The Out of Domain Score as measurement for the offset between the TNG and HSC datasets.} Top-Left: UMAP of the representations of TNG100 galaxies. Bottom-Left: UMAP of the representations of HSC galaxies matched to TNG100. Top-Right: UMAP of the representations of TNG50 galaxies. Bottom-Right: UMAP of the representations of HSC galaxies matched to TNG50. In contours we show the $80$ per-cent coverings of TNG50 (orange), TNG100 (blue), HSC matched to TNG50 (light orange) and HSC matched to TNG100 (light blue) galaxies. The UMAPs are coloured according to the bin-wise median of the Out of Domain Score assigned to the galaxies relative to the matched HSC galaxies (for the TNG sets) and relative to the TNG50/100 galaxies (for the HSC sets). As a reminder, a OOD Score larger than 1 denotes discrepancy.}
	\label{fig:umap_similarity}
\end{figure*}

To connect the OOD Score also to the UMAPs of the previous Sections, we plot in Figure~\ref{fig:umap_similarity} the UMAPs coloured according to the twice-normalized 8th-neighbour distance, i.e. the OOD Score, for the TNG and the HSC sets. The measure can be easily used to identify the subsets that are disjoint/matching. Here it is noteworthy that for TNG there is a diagonal strip of galaxies that are not comparable to HSC; looking at the images of Figures~\ref{fig:umap_images_TNG100} and \ref{fig:umap_images_TNG50}, we see that they might overlap with the clumpy TNG galaxies identified earlier.

\subsubsection{Connecting the OOD Score to galaxy properties}
\label{sec:similarity_score}
With the OOD measure defined and calculated in Section~\ref{sec:nnd}, we have now a tool to identify which kind of galaxies are similar and which ones differ between simulations and observations based on meaningful image-based representations. But so, which TNG galaxies are more or less realistic compared to HSC galaxies?

\begin{figure*}
	\centering
	\includegraphics[width=0.33\linewidth]{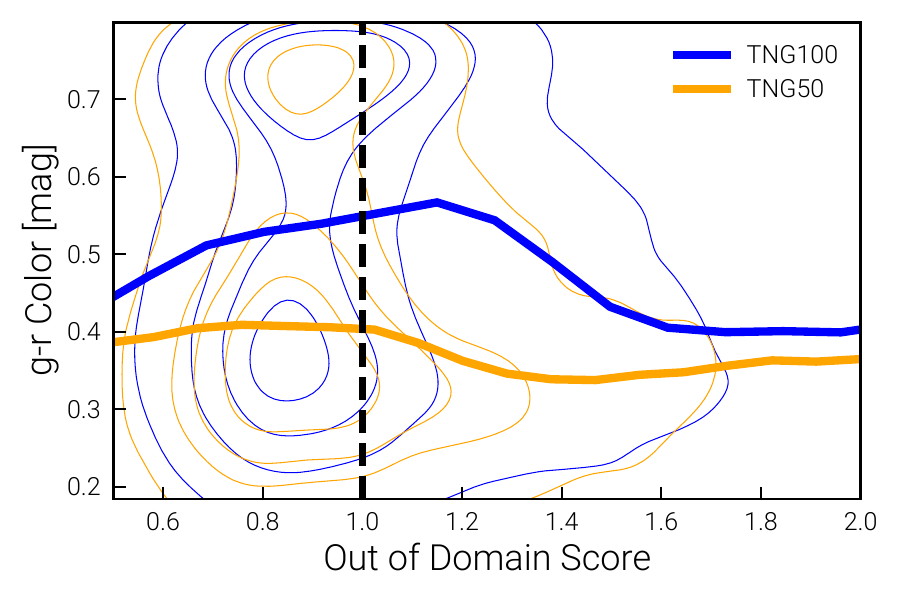}
	\includegraphics[width=0.33\linewidth]{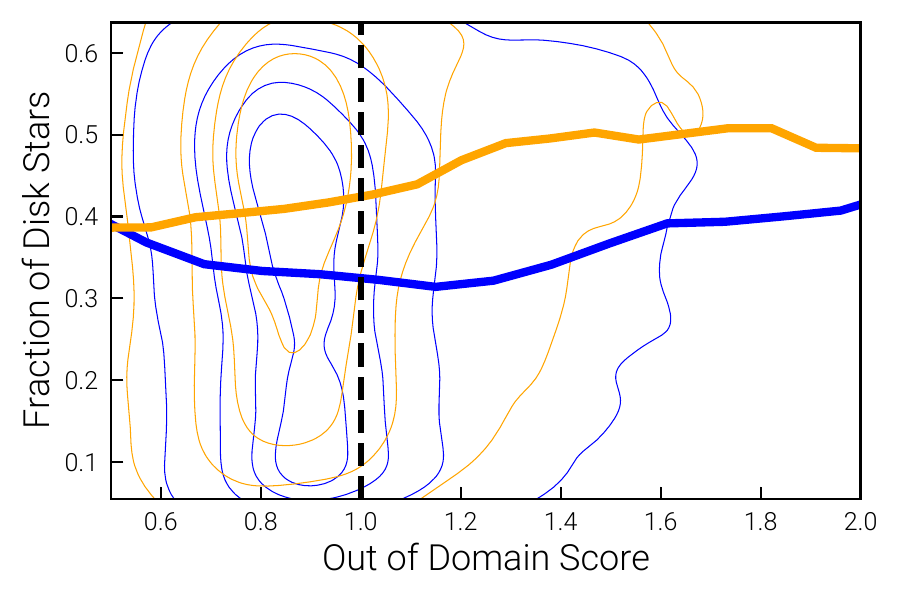}
	\includegraphics[width=0.33\linewidth]{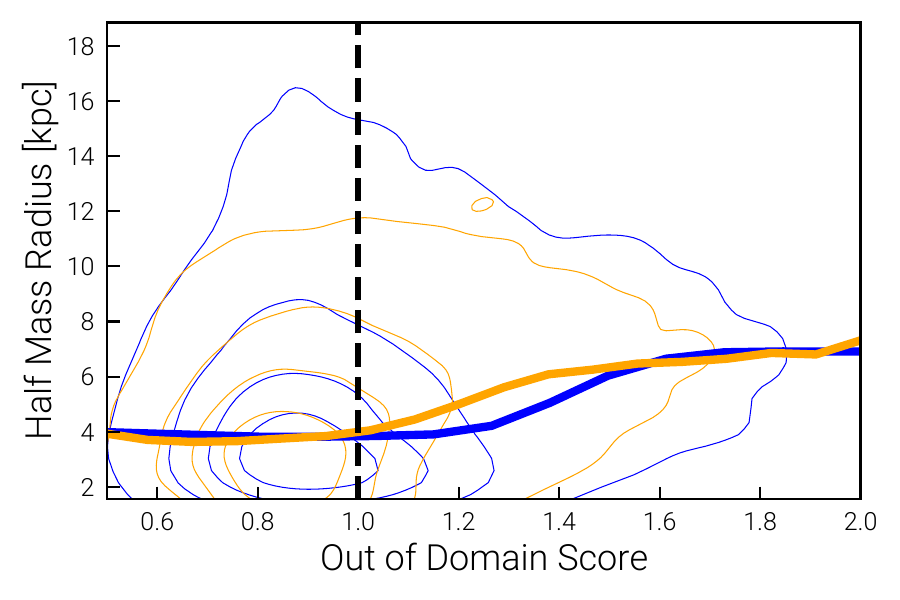}
	\includegraphics[width=0.33\linewidth]{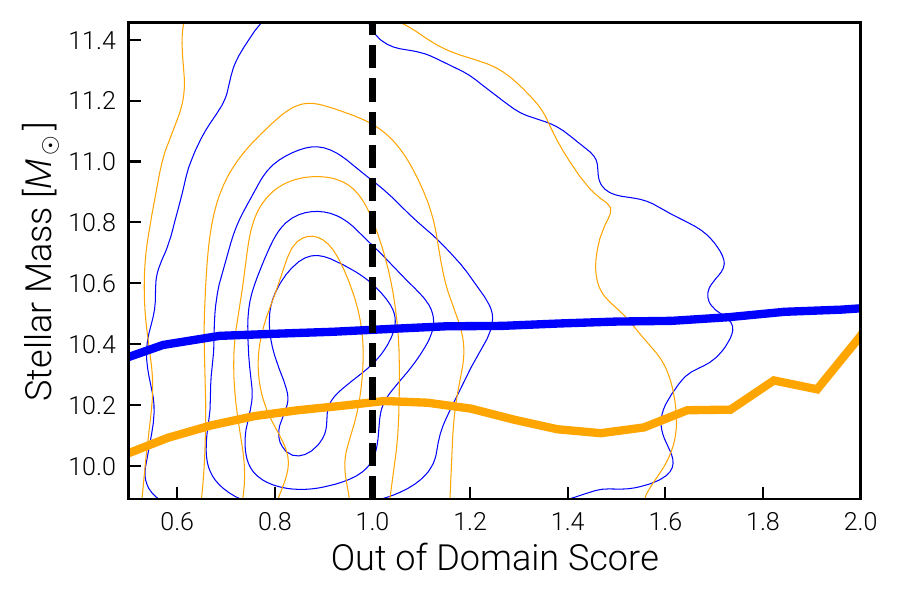}
  	\includegraphics[width=0.33\linewidth]{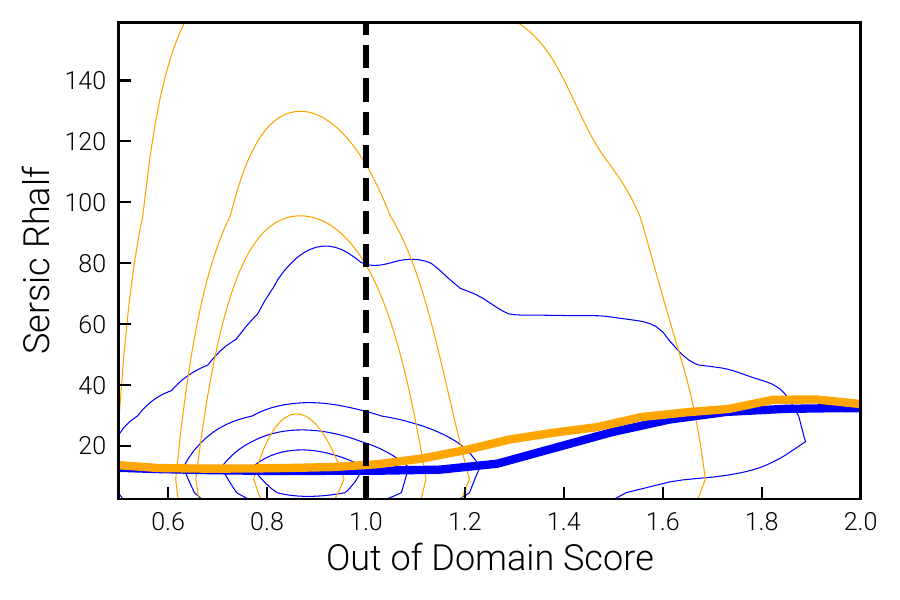}
 	\includegraphics[width=0.33\linewidth]{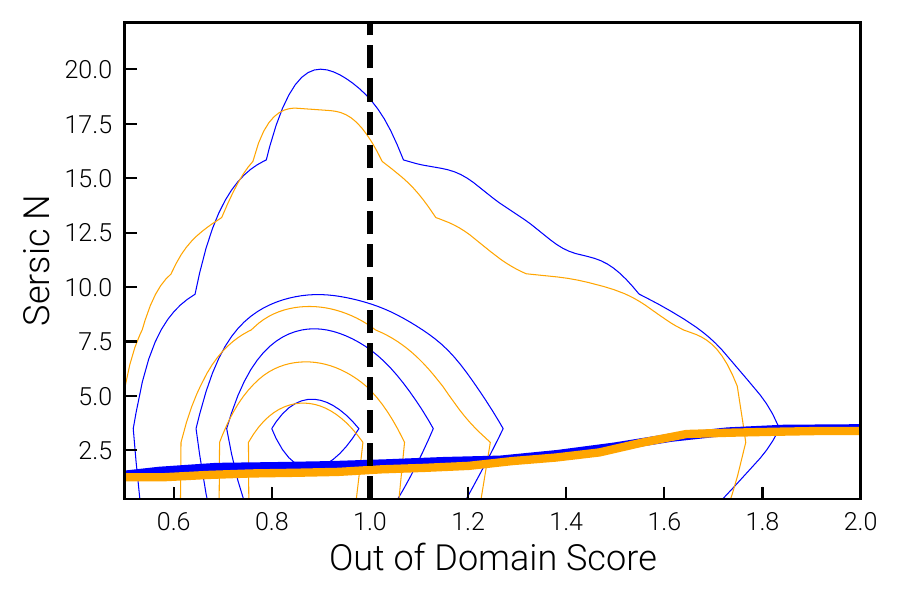}
  	\includegraphics[width=0.33\linewidth]{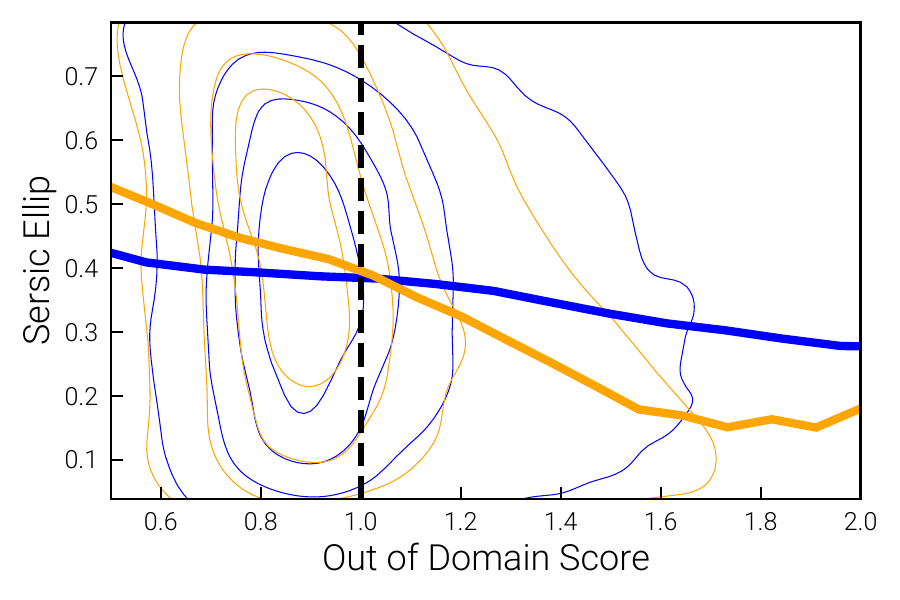}
  	\includegraphics[width=0.33\linewidth]{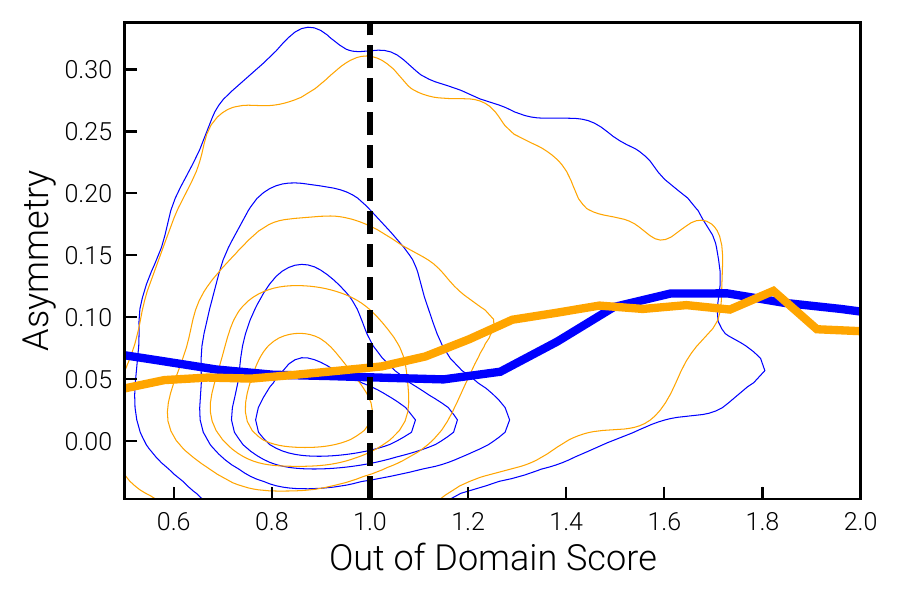}
  	\includegraphics[width=0.33\linewidth]{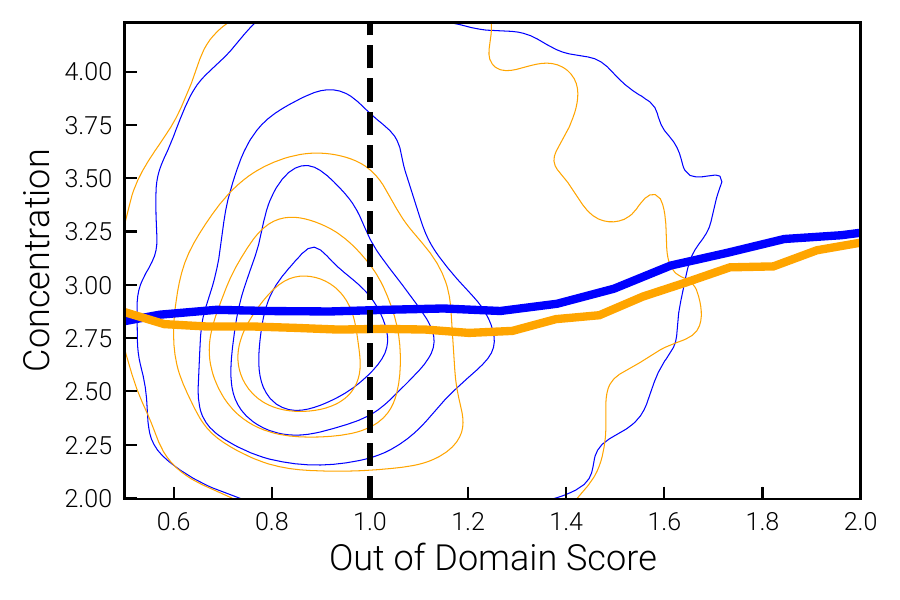}
  	\includegraphics[width=0.33\linewidth]{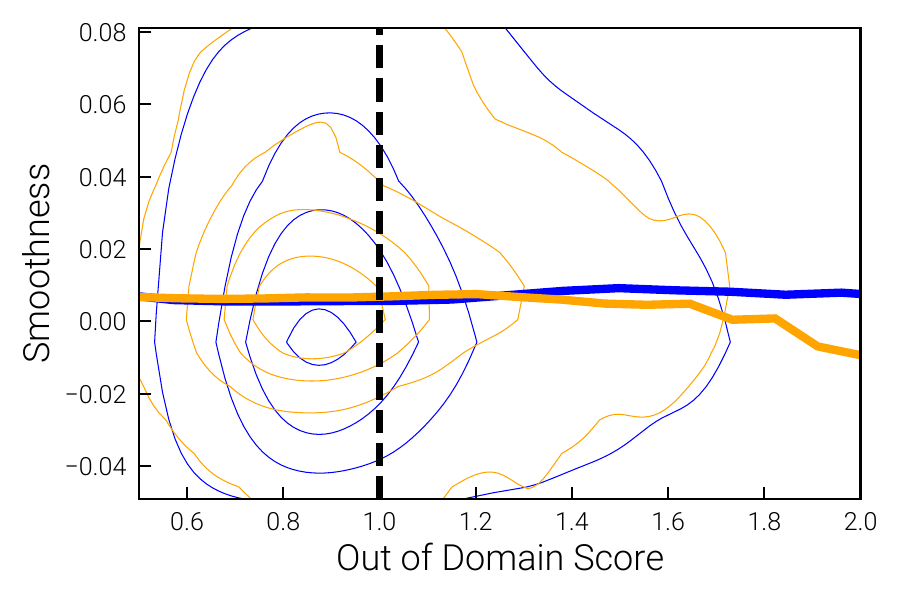}
  	\includegraphics[width=0.33\linewidth]{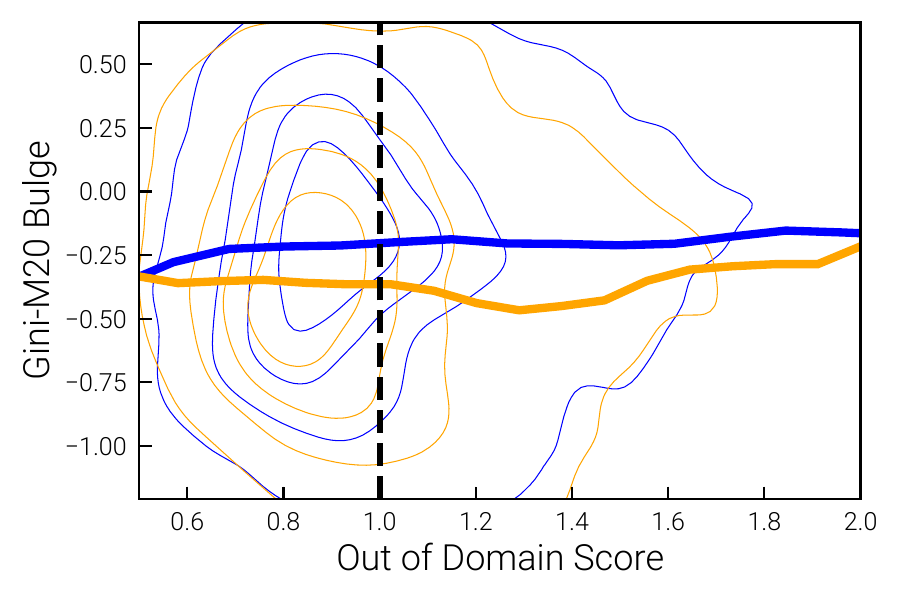}
  	\includegraphics[width=0.33\linewidth]{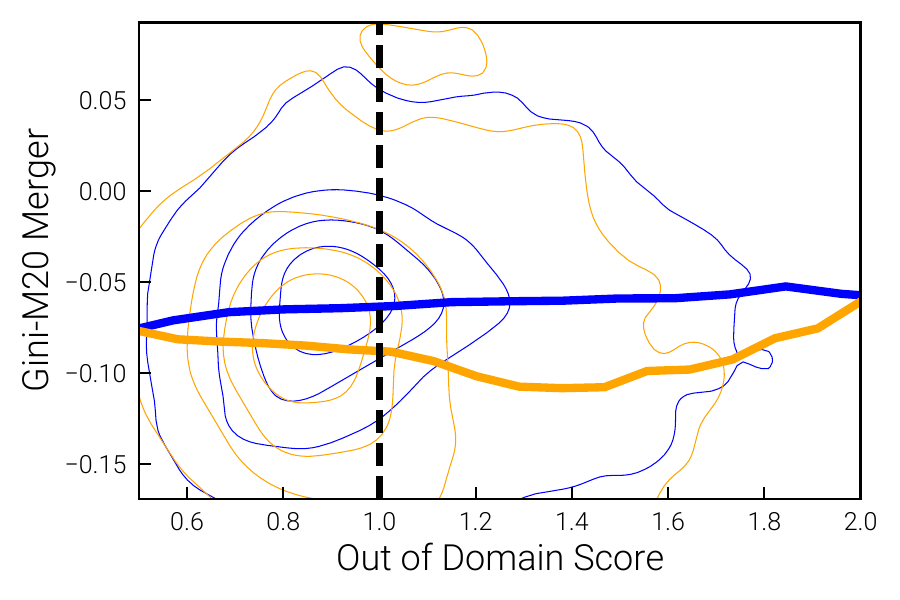}
	\caption{{\bf What galaxy observables drive the OOD Scores of the TNG images relative to the HSC images?} We plot 'in-principle' observable galaxy properties against the OOD Score; these are: total stellar mass, fraction of disk stars, integrated galaxy colour index (g-r), half mass radius, Sérsic half light radius, Sérsic index, Sérsic ellipticity, asymmetry of the light distribution, concentration of the light, smoothness of the light distribution, Gini-M20 bulge parameter and Gini-M20 merger parameter (these are the same as in Section~\ref{sec:umap}). We use these distributions to investigate which types of TNG galaxies fall into the domain of HSC galaxies and which types do not. We perform this study on TNG100 galaxies (blue) and TNG50 galaxies (orange), relative to their respective matched HSC samples. The dashed lines show the distributions while the solid lines show the median in 10 fixed bins of OOD Score. We define all galaxies with OOD Scores smaller than $1$ as within the domain, while all galaxies larger than $1$ outside of the domain. Note that the TNG100 and TNG50 samples shown here represent distinct galaxy image populations: e.g. the stellar mass of the TNG50 sample is significantly lower than the one of TNG100.}
	\label{fig:similarity_properties}
\end{figure*}

In Figure~\ref{fig:similarity_properties} the OOD score is plotted against each of the galaxy properties introduced in Section \ref{sec:umap} and Figures~\ref{fig:tng_umaps},\ref{fig:tng_umaps_2}, and \ref{fig:tng_umaps_3}. As those quantities are derived from the TNG simulations, these plots include TNG50 (orange) and TNG100 (blue) representations only. The following trends are visible: (1) The OOD scores correlate with: larger radii (half mass and sersic half light), steeper sersic profiles, smaller sersic ellipticities and larger asymmetries. (2) There is also a correlation with the fraction of disk stars, however especially prominent for TNG50. We see a general trend for both simulations that brighter/more massive galaxies fall outside of the HSC domain. There may be multiple reasons for this result, including ones that are not related to the simulation output itself: (1) Brighter and more massive galaxies might have also a more unique halo including stellar halo features, mergers and/or large satellite galaxies. (2) Overall the region of higher mass/lower surface brightness is less well-sampled and could therefore cause that the model is less well suited for this regime.
The same reasoning is also valid to some extent to the Sersic radius (which is a relative measure for the galaxy size) and the stellar half-mass radius (which is an absolute measure for the galaxy size).

The visually-observed clumpiness discussed above is captured by the trend in asymmetry but not by the smoothness. We assume that this is caused by the fact that the blue clumps have less of an effect on the r-band derived \textsc{Statmorph} features with the smoothness focusing only on the central part of the galaxy in question. Furthermore as we have removed unreliable \textsc{Statmorph} fits, we also might have removed images showing a very unique morphology.

\section{First applications, insights, discussion and outlook}
\label{sec:discussion}
We conclude this work with two initial applications of our trained model and a discussion about our findings. Again, all analyses discussed so far have been performed on the test set only. First, we use the model to identify counterparts between TNG and HSC galaxies. This will serve as a visual validation of the work we presented in the previous sections. Second, we infer the redshift, Petrosian radius and i-band-magnitude for the HSC galaxies, by using only the information from TNG, to get a first idea of the performance of this method for simulation-based inference.

\subsection{Identifying counterparts between TNG and HSC galaxies}
\label{sec:counterparts}
\begin{figure*}
	\centering
        \includegraphics[trim={0 3cm 0cm 2.5cm},clip,width=0.45\linewidth]{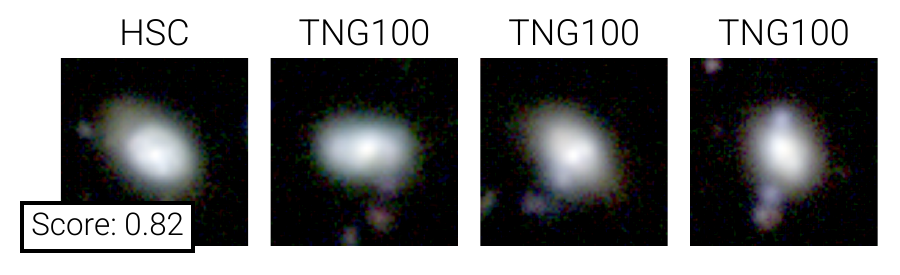}
        \includegraphics[trim={0 3cm 0 2.5cm},clip,width=0.45\linewidth]{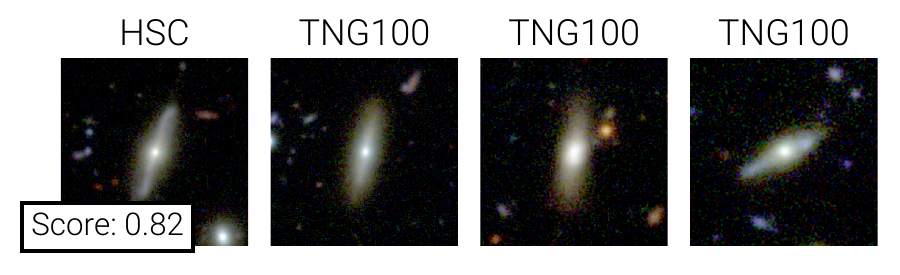}
        \includegraphics[trim={0 3cm 0 2.5cm},clip,width=0.45\linewidth]{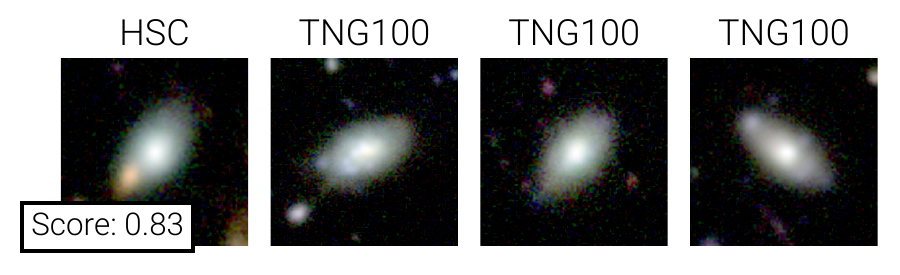}
        \includegraphics[trim={0 3cm 0 2.5cm},clip,width=0.45\linewidth]{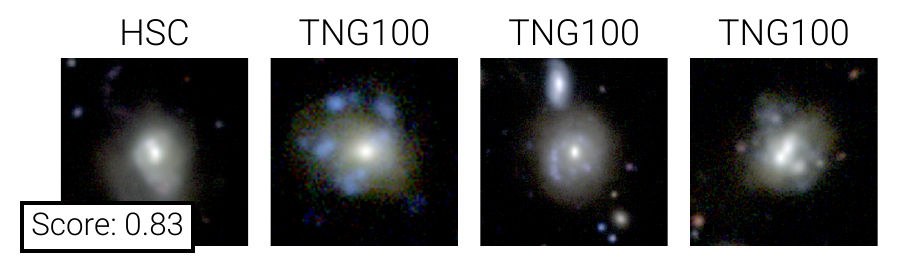}
        \includegraphics[trim={0 3cm 0 2.5cm},clip,width=0.45\linewidth]{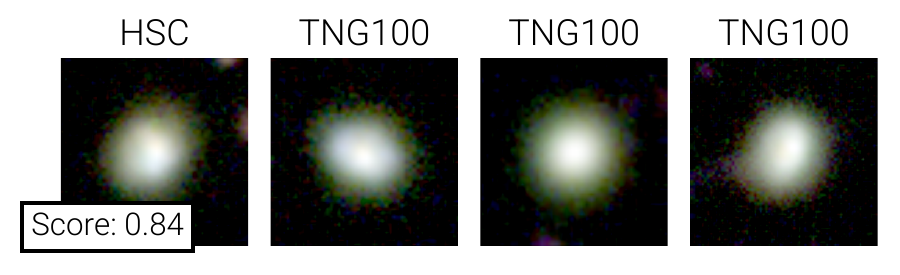}
        \includegraphics[trim={0 3cm 0 2.5cm},clip,width=0.45\linewidth]{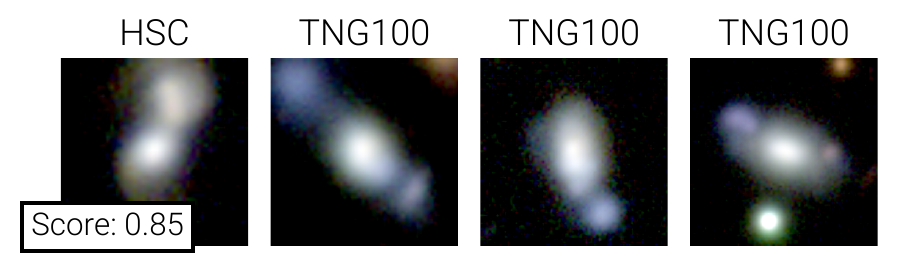}
        \includegraphics[trim={0 3cm 0 2.5cm},clip,width=0.45\linewidth]{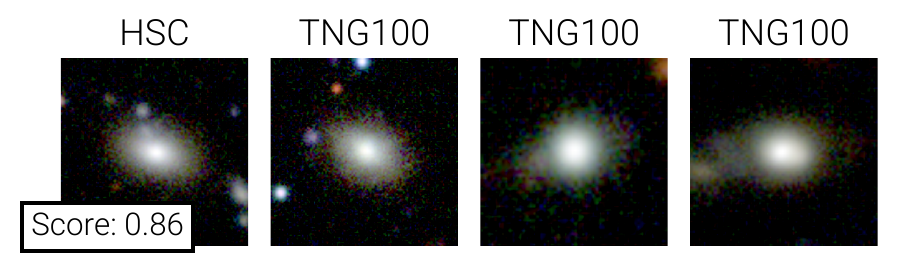}
        \includegraphics[trim={0 3cm 0 2.5cm},clip,width=0.45\linewidth]{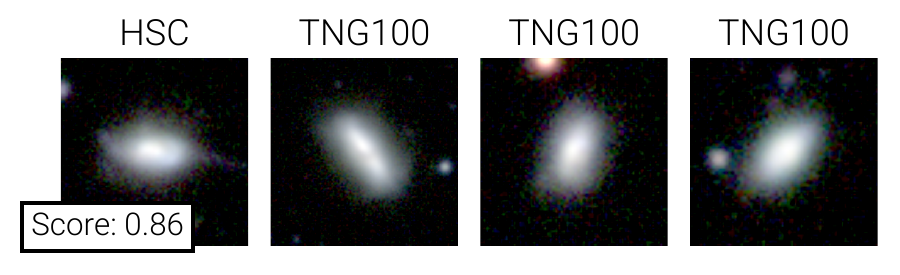}
        \includegraphics[trim={0 3cm 0 2.5cm},clip,width=0.45\linewidth]{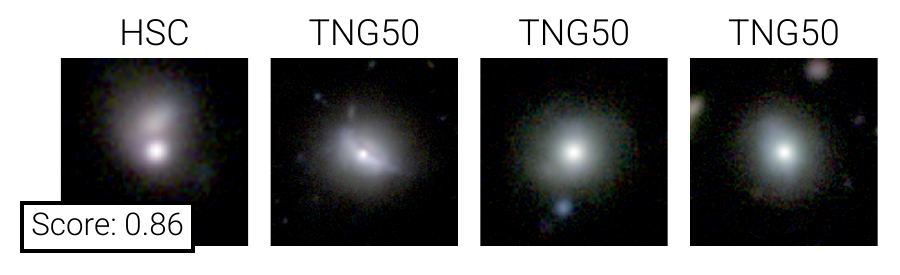}
        \includegraphics[trim={0 3cm 0 2.5cm},clip,width=0.45\linewidth]{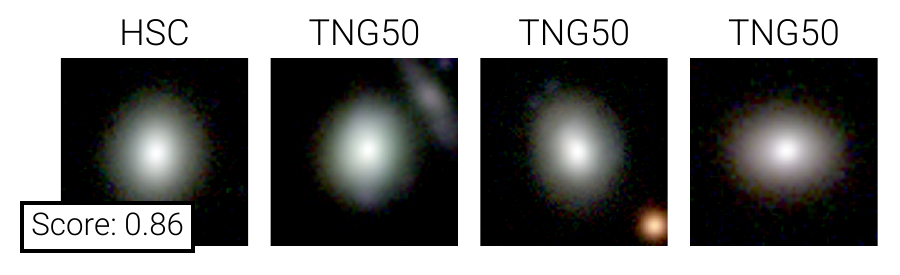}
        \includegraphics[trim={0 3cm 0 2.5cm},clip,width=0.45\linewidth]{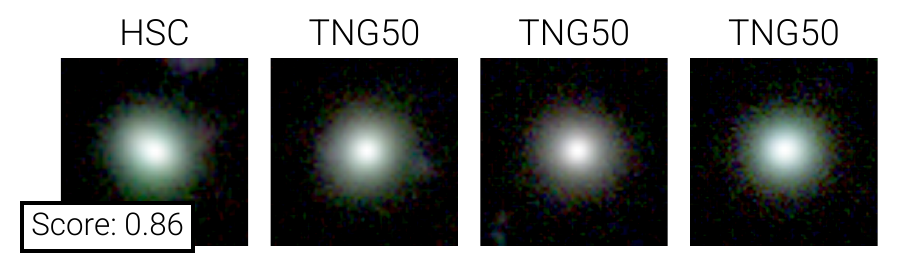}
        \includegraphics[trim={0 3cm 0 2.5cm},clip,width=0.45\linewidth]{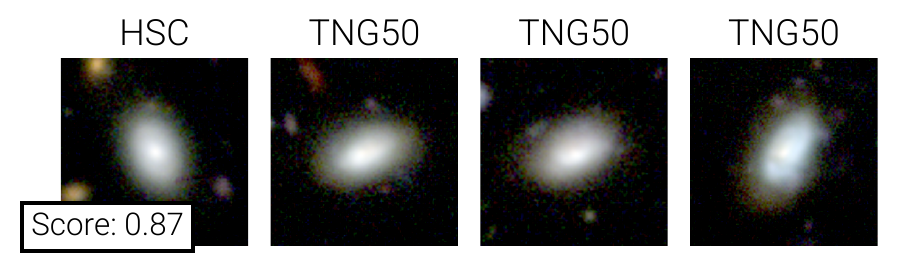}
        \includegraphics[trim={0 3cm 0 2.5cm},clip,width=0.45\linewidth]{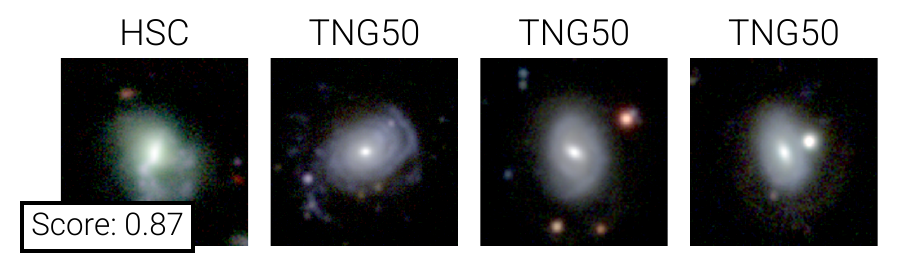}
        \includegraphics[trim={0 3cm 0 2.5cm},clip,width=0.45\linewidth]{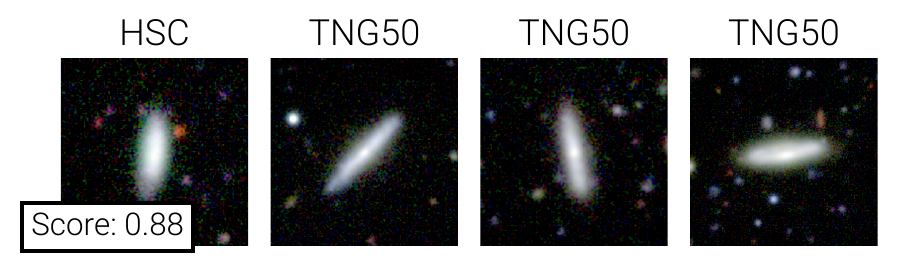}
        \includegraphics[trim={0 3cm 0 2.5cm},clip,width=0.45\linewidth]{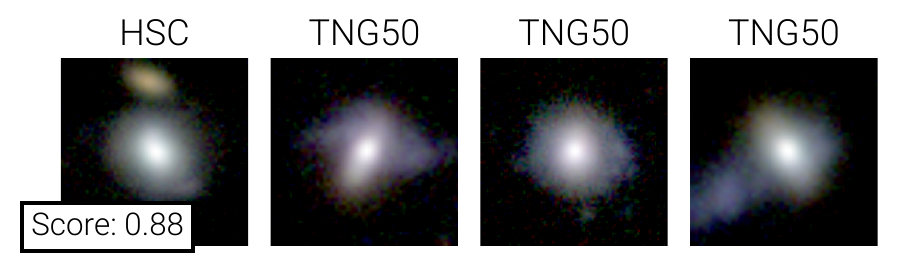}
        \includegraphics[trim={0 3cm 0 2.5cm},clip,width=0.45\linewidth]{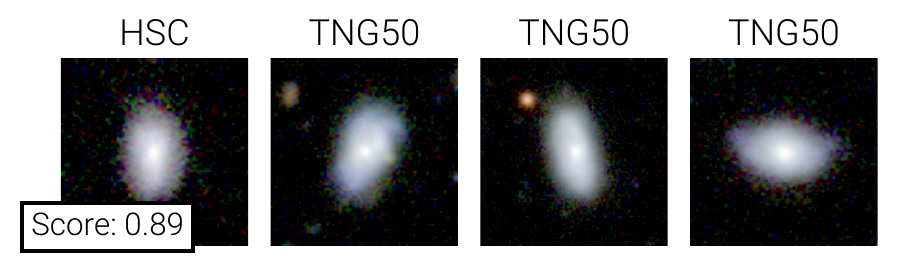}
	\caption{{\bf Identified TNG50/100 counterparts of HSC galaxy images.} We show TNG50/100 counterparts of randomly selected HSC galaxies which have been identified as the closest neighbour in the representation space (using the sum-norm). Additionally we show the OOD Score for each HSC galaxy. We see overall a good visual match among the galaxies.}
	\label{fig:counterparts}
\end{figure*}

\begin{figure*}
	\centering
        \includegraphics[trim={0 3cm 0cm 2.5cm},clip,width=0.45\linewidth]{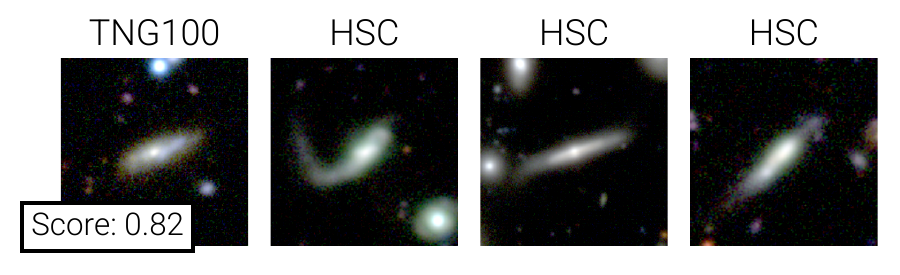}
        \includegraphics[trim={0 3cm 0 2.5cm},clip,width=0.45\linewidth]{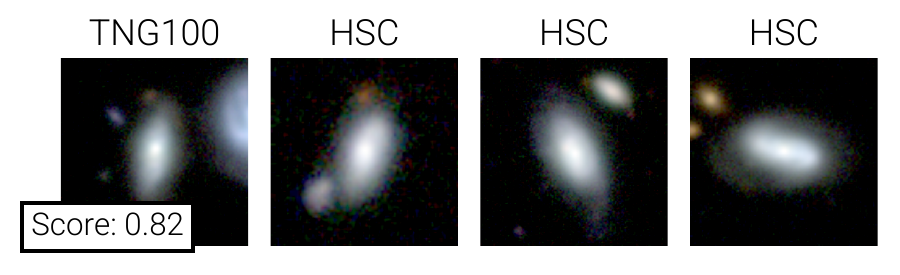}
        \includegraphics[trim={0 3cm 0 2.5cm},clip,width=0.45\linewidth]{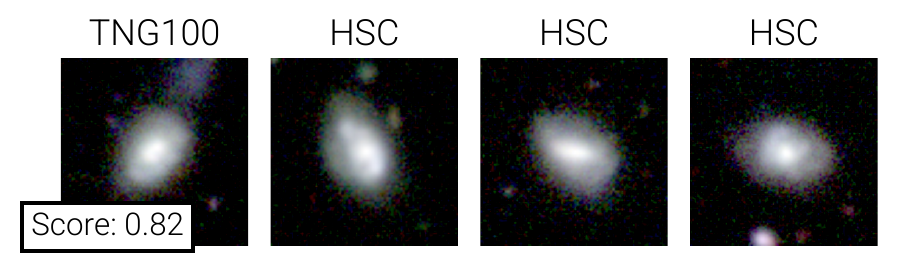}
        \includegraphics[trim={0 3cm 0 2.5cm},clip,width=0.45\linewidth]{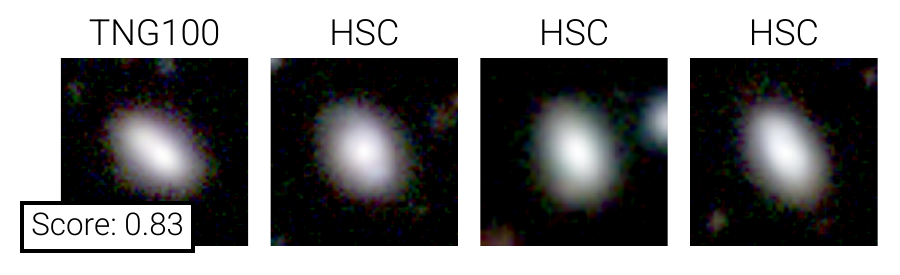}
        \includegraphics[trim={0 3cm 0 2.5cm},clip,width=0.45\linewidth]{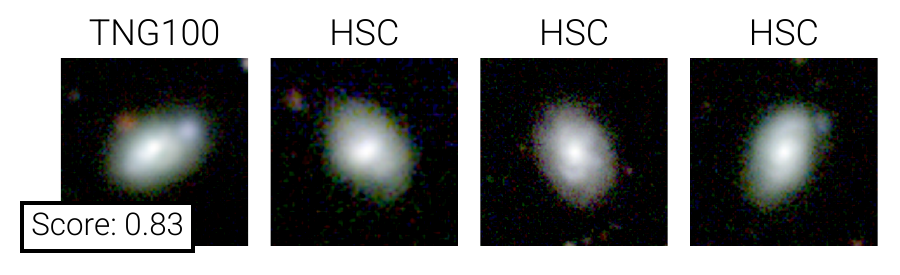}
        \includegraphics[trim={0 3cm 0 2.5cm},clip,width=0.45\linewidth]{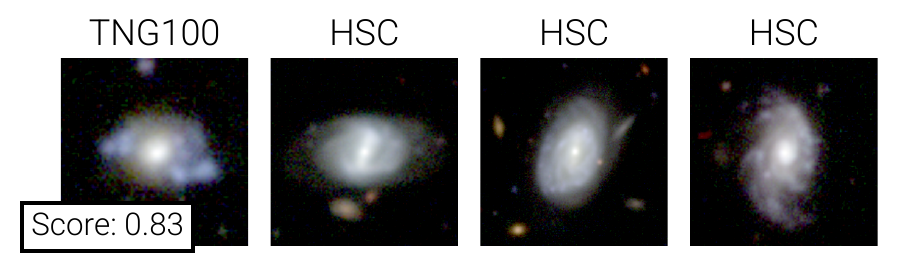}
        \includegraphics[trim={0 3cm 0 2.5cm},clip,width=0.45\linewidth]{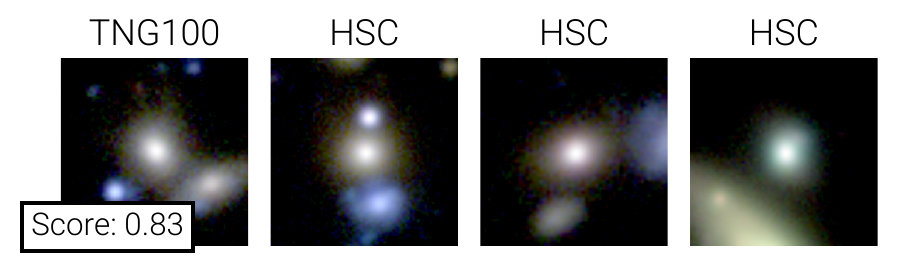}
        \includegraphics[trim={0 3cm 0 2.5cm},clip,width=0.45\linewidth]{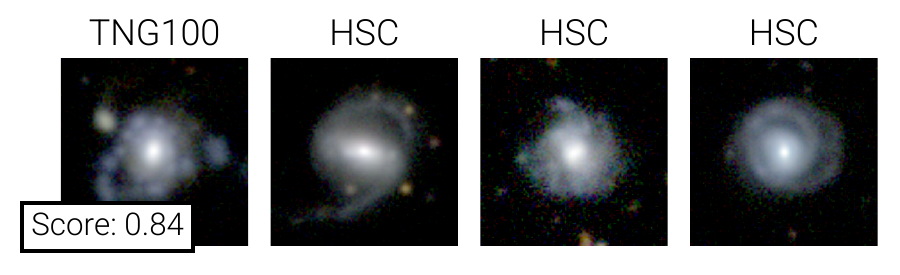}
        \includegraphics[trim={0 3cm 0 2.5cm},clip,width=0.45\linewidth]{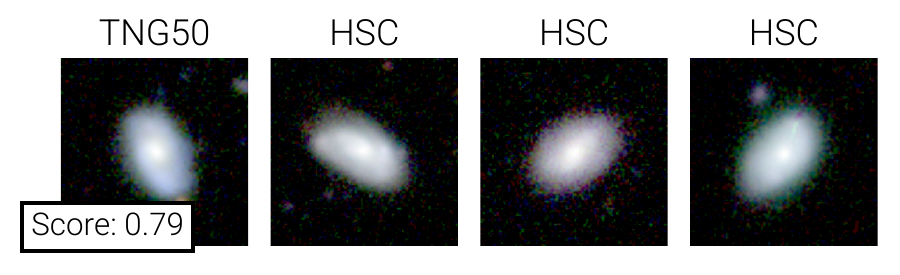}
        \includegraphics[trim={0 3cm 0 2.5cm},clip,width=0.45\linewidth]{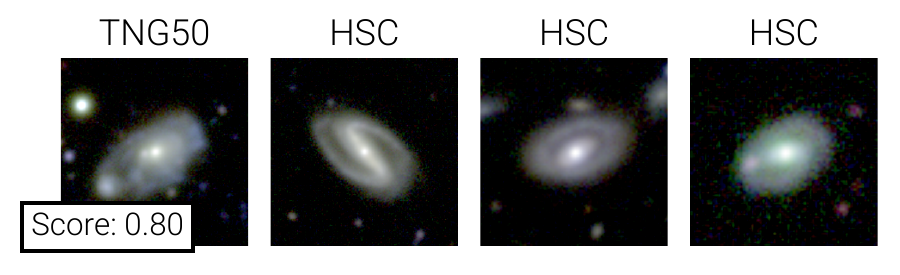}
        \includegraphics[trim={0 3cm 0 2.5cm},clip,width=0.45\linewidth]{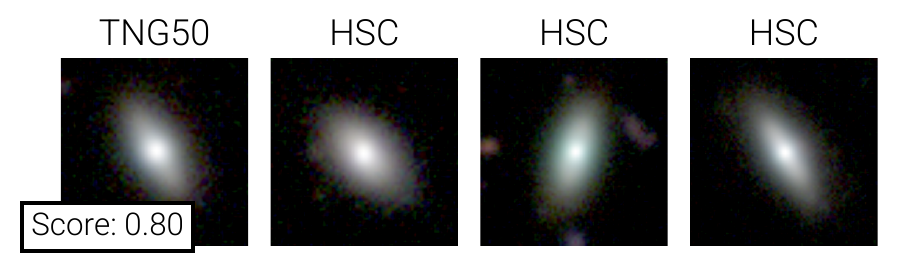}
        \includegraphics[trim={0 3cm 0 2.5cm},clip,width=0.45\linewidth]{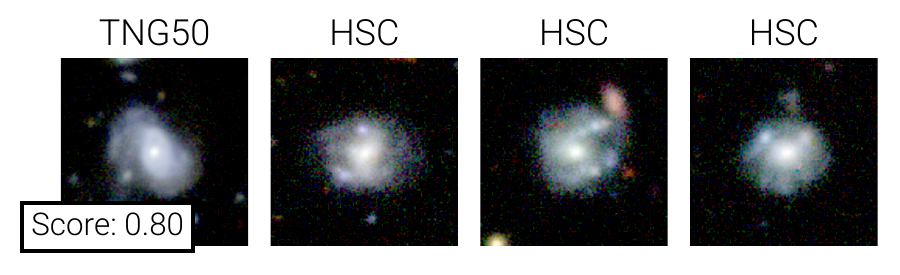}
        \includegraphics[trim={0 3cm 0 2.5cm},clip,width=0.45\linewidth]{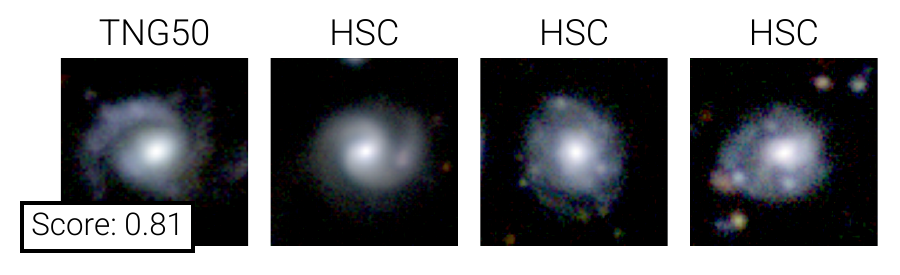}
        \includegraphics[trim={0 3cm 0 2.5cm},clip,width=0.45\linewidth]{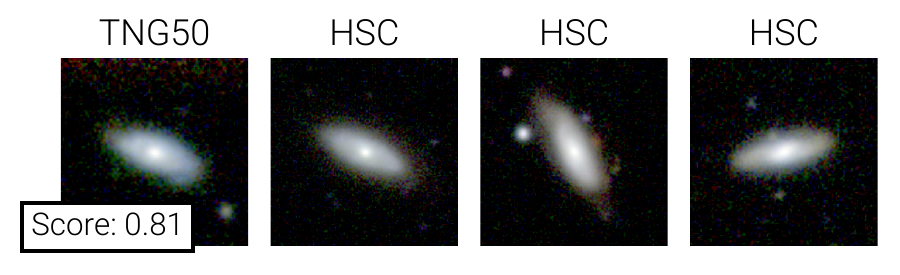}
        \includegraphics[trim={0 3cm 0 2.5cm},clip,width=0.45\linewidth]{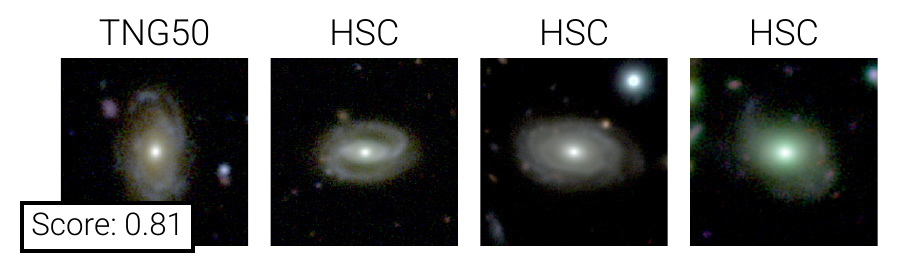}
        \includegraphics[trim={0 3cm 0 2.5cm},clip,width=0.45\linewidth]{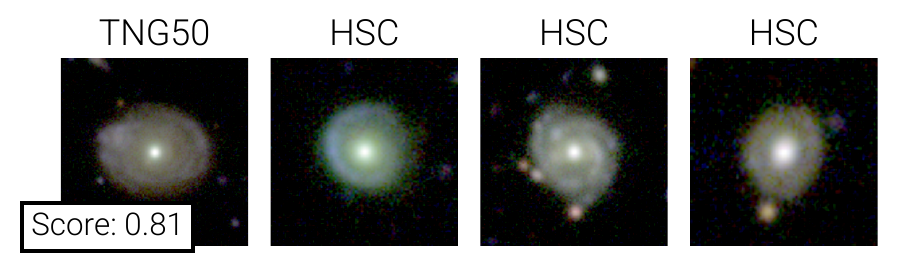}
	\caption{{\bf Identified HSC counterparts of TNG50/100 galaxy images.} Same as Figure \ref{fig:counterparts} but with HSC counterparts for randomly selected TNG50/TNG100 galaxies.}
	\label{fig:counterparts_2}
\end{figure*}

\begin{figure*}
	\centering
         \includegraphics[trim={0 3cm 0 2.5cm},clip,width=0.45\linewidth]{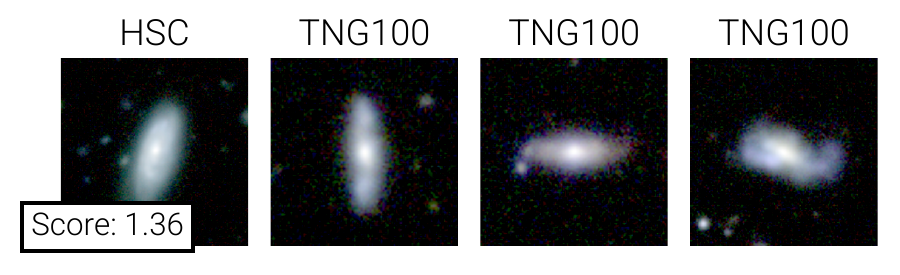}
        \includegraphics[trim={0 3cm 0 2.5cm},clip,width=0.45\linewidth]{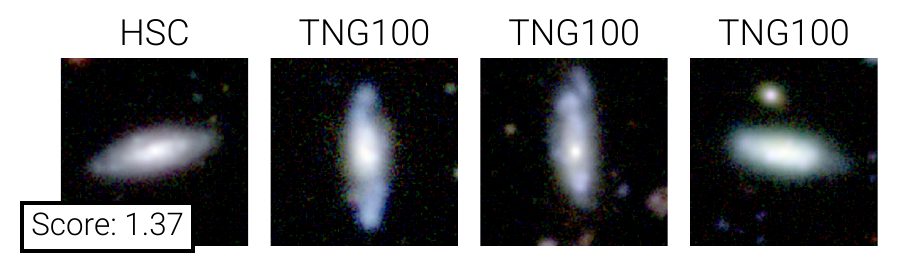}
        \includegraphics[trim={0 3cm 0 2.5cm},clip,width=0.45\linewidth]{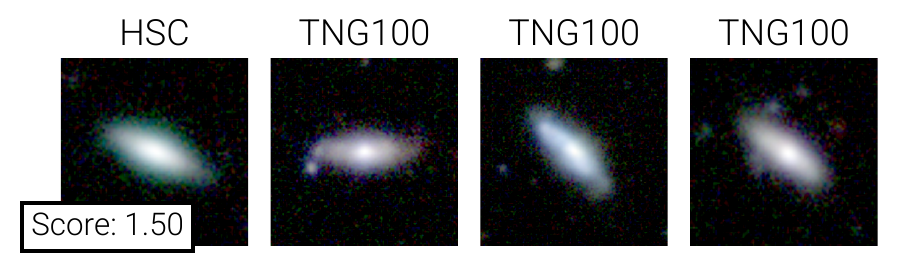}
        \includegraphics[trim={0 3cm 0 2.5cm},clip,width=0.45\linewidth]{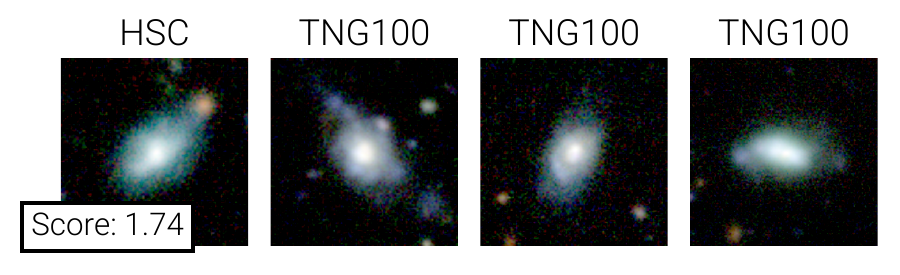}
        \includegraphics[trim={0 3cm 0 2.5cm},clip,width=0.45\linewidth]{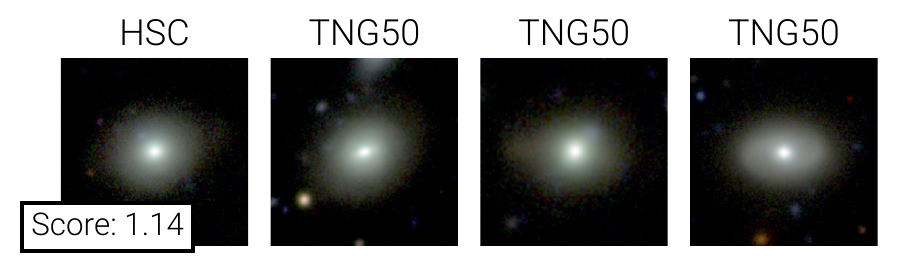}
        \includegraphics[trim={0 3cm 0 2.5cm},clip,width=0.45\linewidth]{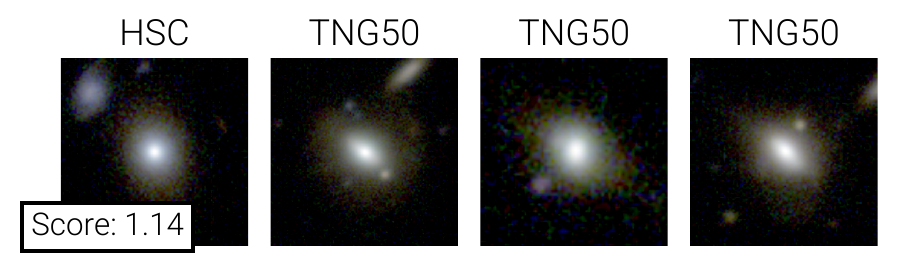}
        \includegraphics[trim={0 3cm 0 2.5cm},clip,width=0.45\linewidth]{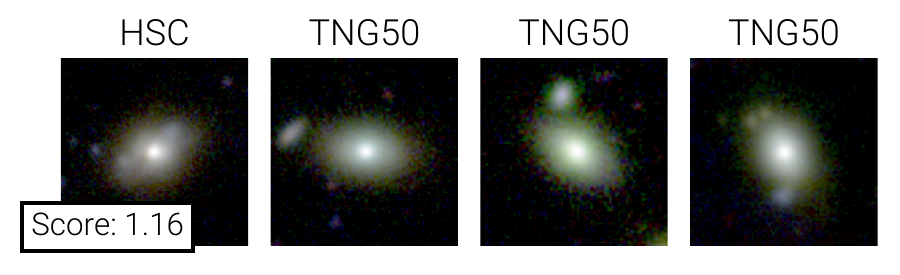}
        \includegraphics[trim={0 3cm 0 2.5cm},clip,width=0.45\linewidth]{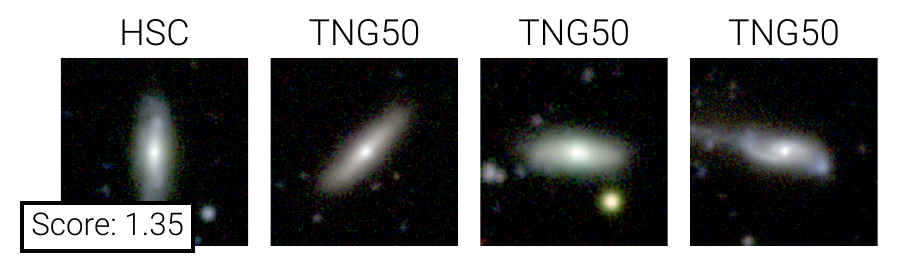}
	\caption{Same as Figure \ref{fig:counterparts} but with randomly selected HSC galaxies with high OOD Scores. We do not see any significant differences, which instead would have been expected given the high OOD Scores. So, based on this visual comparison, it appears that, in fact, the TNG sets seem to extrapolate well beyond their own domains.}
	\label{fig:counterparts_3}
\end{figure*}

\begin{figure*}
	\centering
        \includegraphics[trim={0 3cm 0 2.5cm},clip,width=0.45\linewidth]{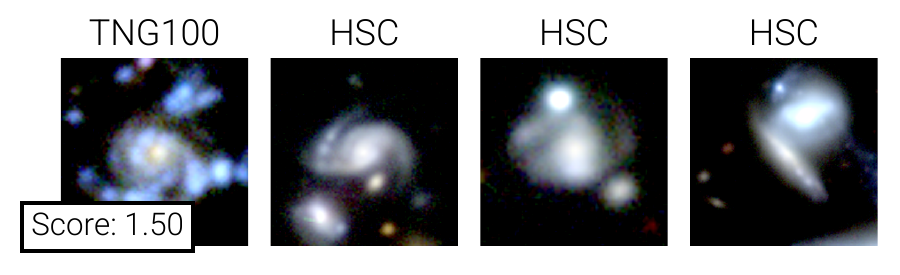}
        \includegraphics[trim={0 3cm 0 2.5cm},clip,width=0.45\linewidth]{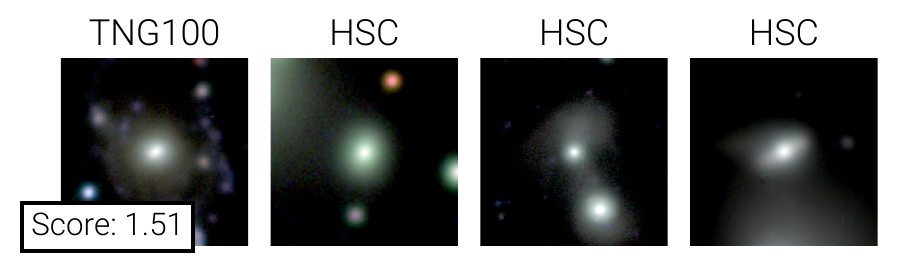}
        \includegraphics[trim={0 3cm 0 2.5cm},clip,width=0.45\linewidth]{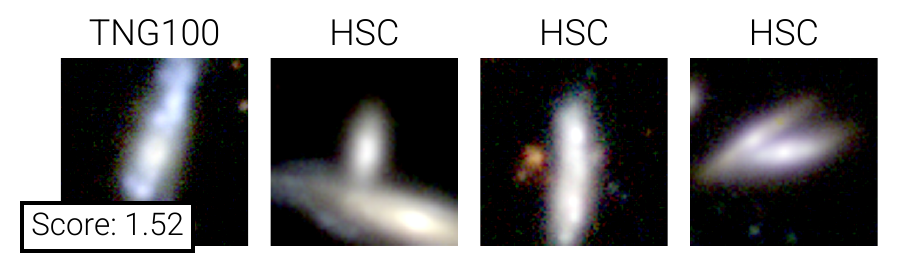}
        \includegraphics[trim={0 3cm 0 2.5cm},clip,width=0.45\linewidth]{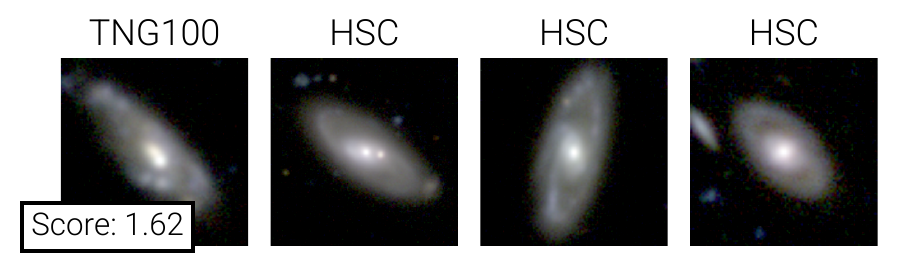}
        \includegraphics[trim={0 3cm 0 2.5cm},clip,width=0.45\linewidth]{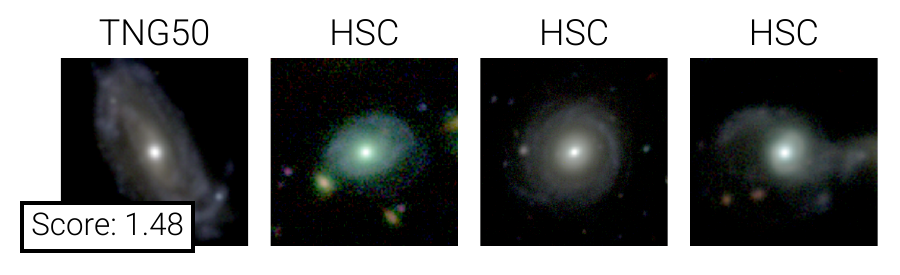}
        \includegraphics[trim={0 3cm 0 2.5cm},clip,width=0.45\linewidth]{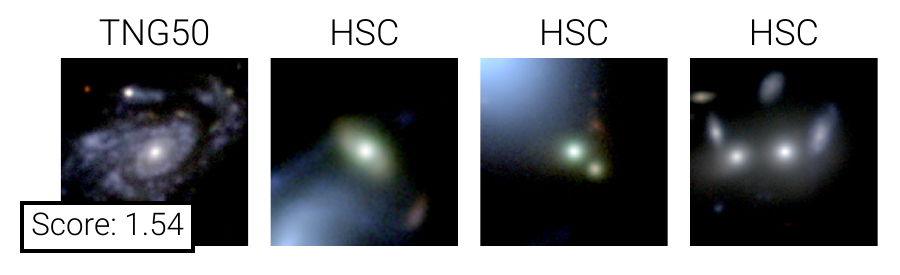}
        \includegraphics[trim={0 3cm 0 2.5cm},clip,width=0.45\linewidth]{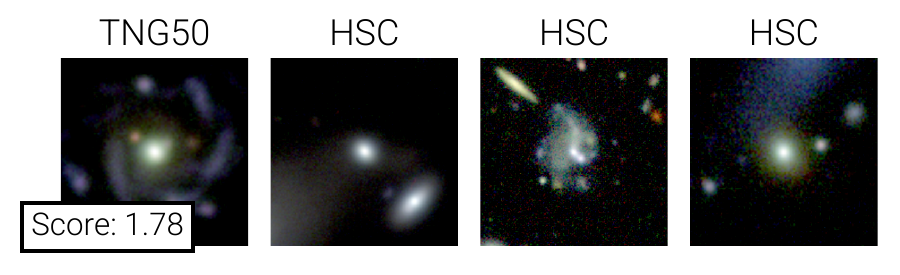}
        \includegraphics[trim={0 3cm 0 2.5cm},clip,width=0.45\linewidth]{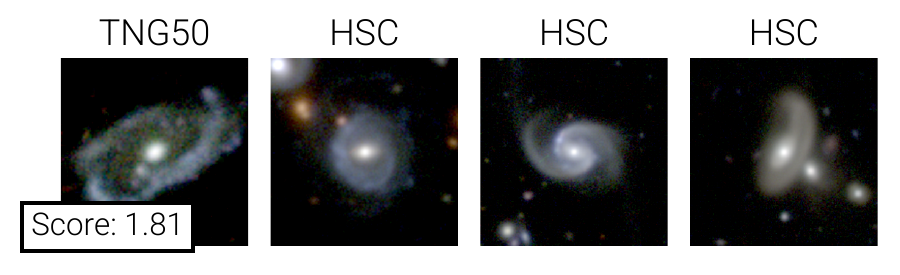}
	\caption{Same as Figure \ref{fig:counterparts_2} but with very high OOD scores. Here we clearly see that the identification of out-of-domain galaxies works. In particular, we see galaxies that are in a very crowded environment and/or for which the Petrosian fitting did not work i.e. captured not only the light profile of the central galaxy but a whole cluster leading therefore to a very large FOV. We also find cases of very large neighbouring galaxies and galaxies with a very clumpy appearance. These galaxies could be excluded successfully for scientific applications by using the OOD Score.}
	\label{fig:counterparts_4}
\end{figure*}

It is possible to use the representation spaces constructed in this paper to identify counterparts between sets of galaxy images. In particular, to identify the most similar TNG/HSC galaxy, we choose the nearest neighbour (using again the sum-norm) in representation space for the HSC/TNG galaxy in question. To visualize the variance in potential counterparts we not only show the closest neighbours but the three closest neighbours. To also investigate the importance of the OOD Score we show multiple examples of counterparts for various ranges of OOD Scores. The following is a compilation of visuals:
\begin{itemize}
    \item In Figure~\ref{fig:counterparts} we show HSC galaxies with OOD Scores $< 1$ and their TNG50/100 counterparts; i.e. each HSC galaxy is well within the domain of the TNG galaxies (i.e. within the domain, by our definition).
    \item In Figure~\ref{fig:counterparts_2} we show TNG50/100 galaxies with OOD Scores $< 1$ and their HSC counterparts; i.e. each TNG50/100 galaxy is well within the domain of the HSC galaxies (i.e. within the domain, by our definition).
    \item In Figure~\ref{fig:counterparts_3} we show HSC galaxies with OOD Scores $> 1$; i.e. each HSC galaxy is in a region that is significantly underpopulated by TNG galaxies (i.e. out of domain, by our definition).
    \item Finally, in Figure~\ref{fig:counterparts_4} we show examples as in Figure~\ref{fig:counterparts_3} but with OOD Scores $> 1.5$, to focus on counterparts of HSC galaxies that come from a region really outside the domain of the simulated ones. 
\end{itemize}
As the three bands relating to the images' RGB have been stretched independently in the preprocessing, the colours of the images might appear different from the original data: differences in the apparent colour should be therefore not a direct concern.

To first order, we see a good agreement among the galaxy images identified as counterparts, in terms of their type and shape. In fact, based on the visual inspection of Figure~\ref{fig:counterparts_3}, choosing an OOD Score value of 1 to discriminate between in- and out-of-domain may be rather conservative: as shown in Figure~\ref{fig:counterparts_3}, we can find a number of galaxy counterparts with OOD Scores $> 1$ (and usually smaller than about 1.5) that actually appear rather similar. By placing the out-of-domain threshold at e.g. 1.3 instead of 1.0, we find that only 8 (instead of 33) per cent of the TNG100 and 6 (instead of 30) per cent of the TNG50 images are out-of-domain relative to the HSC images, or conversely that as many as $\approx 90$ per cent of the simulated galaxies could be considered realistic.

This said, for TNG50 and TNG100, we are clearly able to identify problematic cases by using the OOD Score, particularly for cases with  OOD Score $> 1.5$ as those of Figure~\ref{fig:counterparts_4}: the method and metric developed in this paper can therefore be applied as a first cleaning step for simulation-based inference.

\subsection{A preview: are we able to infer the redshift, Petrosian radii and i-band-magnitude from the HSC images?}
\label{sec:infer}
In the previous sections we have mainly focused on how to \emph{compare} the survey-realistic mocks of TNG simulated galaxies with HSC galaxies. However, the ultimate goal of our project ERGO-ML is to infer physical parameters from observed galaxies by using the information provided by simulated mock galaxies. 
In future work we want to expand on this idea by training an ML model on the representation space to transfer the properties from the simulated galaxies to the observed ones. Here, as a preview, we want to present a simpler application: we perform a nearest neighbour interpolation among the TNG galaxies. 

For each HSC galaxy (in the test set) we choose the nearest TNG galaxy (in the test set) and use the property of that TNG galaxy as a `prediction'. We consider the same galaxy properties we used for the sample matching in Section \ref{sec:matching}: redshift, Petrosian radius and i-band-magnitude. We already ensured that their definition is as similar as possible between the two domains of simulated and observed images. Furthermore we ensured with the galaxy-sample matching, that their prior distributions are the same for both datasets. And although these galaxy parameters are still very basic, their prediction from the images is not trivial as, for example, we normalize the images in brightness and FOV.

\begin{figure*}
	\centering
        \includegraphics[width=0.33\linewidth]{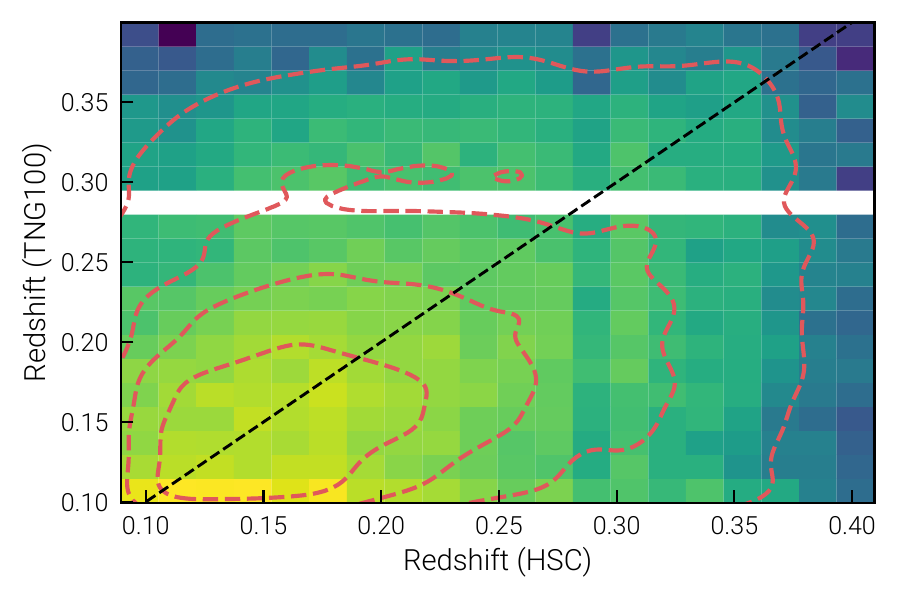}
        \includegraphics[width=0.33\linewidth]{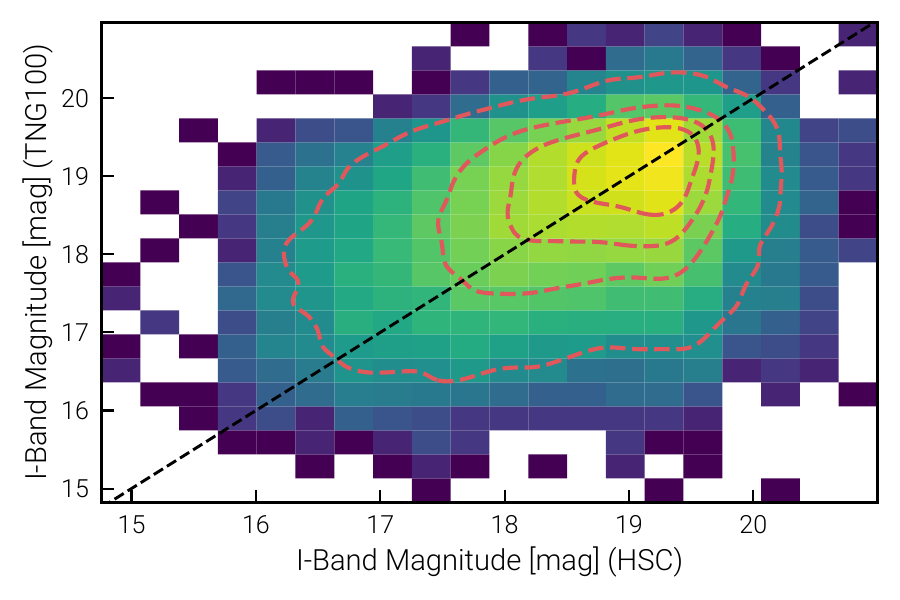}
        \includegraphics[width=0.33\linewidth]{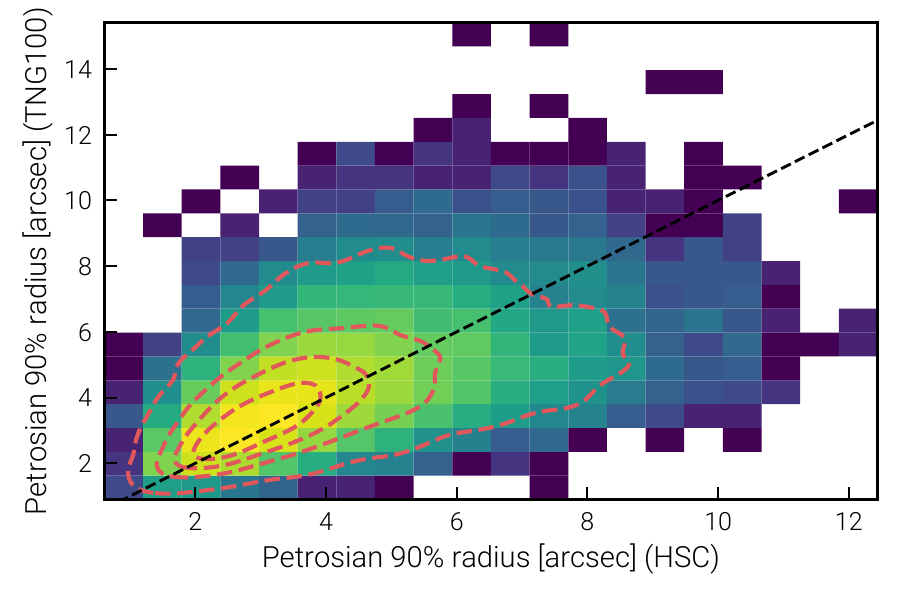}
        \includegraphics[width=0.33\linewidth]{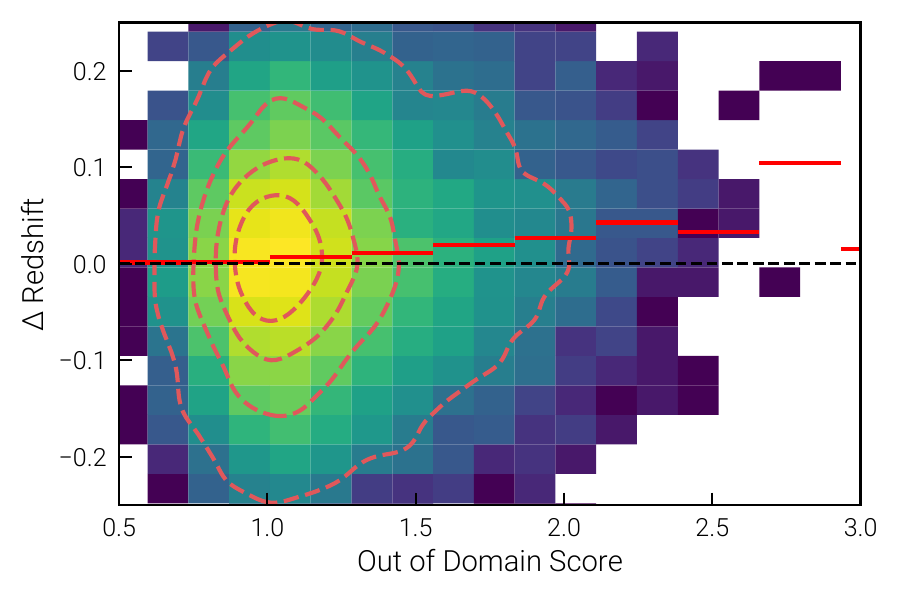}
        \includegraphics[width=0.33\linewidth]{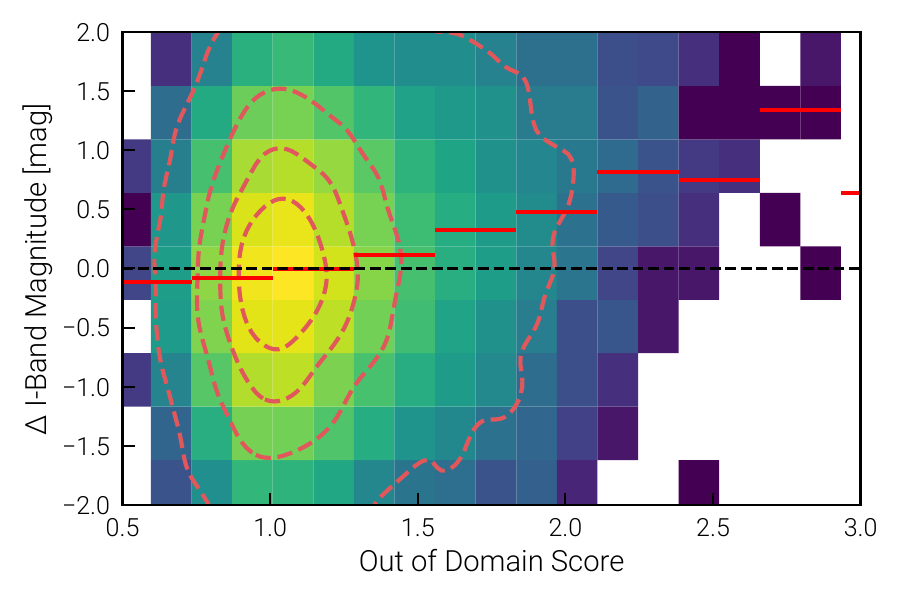}
        \includegraphics[width=0.33\linewidth]{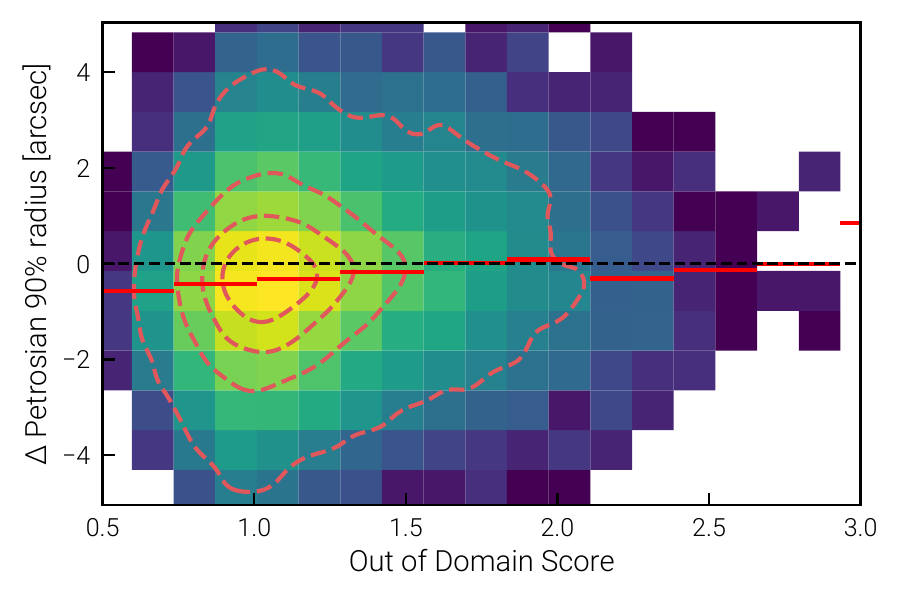}

	\caption{{\bf How well can we reproduce galaxy properties of HSC by interpolating from TNG100 galaxies in representation space?} For this exercise we assign to each HSC galaxy the properties of the closest TNG100 neighbour in the 256-dimensional representation space and we focus on redshift, $i$-band magnitude and Petrosian radius. In the top panel we show the 2D histogram of the results for each property: on the x axis is the HSC ground truth while the y axis shows the quantities derived from the interpolation from TNG100. In the lower panel, we show the dependence of the prediction error (HSC Ground Truth - TNG100-based prediction) on the Out of Domain Score. Additional to the histogram we show in each plot the 'ideal' result as black dotted line and outline the distribution of the histogram also with red dotted lines. We see that with our model we can nicely reproduce the $i$-band magnitude and the Petrosian radius while the quality of the prediction is also dependent on the OOD Score. Note that this result is fully unsupervised -- at no point was any active training performed to reproduce these quantities. {\change The large scatter in the prediction error is especially caused by the coarse interpolation to the nearest neighbor in a high dimensional -- and sparsely populated - space: the results of this figure represent only a lower estimation for the overall capabilities of an inference model which is based on these representations.}}
	\label{fig:infer_properties}
\end{figure*}

We show the results of this exercise in Figure~\ref{fig:infer_properties}. In the top we show the interpolated quantities from TNG100 against the ground truth quantities from the HSC survey. We see that overall the prediction follows the ground truth for all three properties of interest, albeit with a rather large scatter. However, the reader is to be reminded that we did not train on any labels and this prediction is fully self-supervised, i.e. this prediction is just based on the positioning of galaxies in representation space according to their appearance. By incorporating these representations into a supervised machine learning method, which is able to perform higher order interpolations, we can expect much better results. {\change We also emphasize that the photometric redshifts used for the HSC galaxy sample have a comparatively large error of $\sigma[\Delta z_{phot}/(1 + z_{phot})] \approx 0.05$ \citep{2018PASJ...70S...9T}.}

In the bottom we show the absolute prediction error against the similarity score of each HSC galaxy, i.e. the OOD Score. In doing so we check if the scores we introduced and derived in Section~\ref{sec:nnd} are meaningful in terms of prediction capability. HSC galaxies that are far out of the domain populated by the TNG galaxies, i.e. have no similar TNG counterparts, can not be expected to get a meaningful prediction for their properties (although an extrapolation might be possible of course).
We see that this is indeed the case, and especially for the i-band magnitude, which is systematically over-predicted for larger OOD Scores. However, for the redshift and the radii the method extrapolates to the HSC domain without significant systematic biases. We also checked, although we do not show, that the inference works also between the two simulations, TNG50 and TNG100: the prediction error is much smaller than between TNG100 and HSC, which is expected as the galaxies have on average a smaller OOD Score and as we actually compare the exactly-known properties from the simulations. In conclusion, the extraction of galaxy properties based on the learned representations is promising: we will further improve when the unsupervised learning is augmented with a supervised training towards quantities of interest. 

\subsection{Discussion}
\subsubsection{Are IllustrisTNG galaxies sufficiently realistic for simulation-based inference?}
In this paper we have compared three distinct galaxy image datasets: one observed and two simulated. As this work is meant as a step towards a simulation-based inference framework, the relevant question now is: how realistic are the galaxy-image mocks from TNG?

On one hand, this comparison is important insofar as it will establish the degree to which the TNG simulations can produce realistic galaxy morphologies and populations. In other words, are the model assumptions, numerical recipes, and numerical resolution sufficient for this task? But it is also important for us to ensure that the identified TNG counterparts of HSC galaxies are indeed similar enough so that we can safely assume that also their (not directly observable) physical properties are the `same'. If this is the case, then we can map the knowledge from the simulation on to the observations.

The problem of this task, determining whether two galaxies are similar or not, is threefold: (1) Two images which are (nearly) exactly the same still show two different galaxies with not necessarily the same properties. I.e. there can be multiple formation pathways that can lead to similar galaxy morphologies. We showed this already for an inference based on scalars \citep{Eisert_2023}. However, also with 2D images this intrinsic uncertainty is present. (2) On the other hand, two viewing-angles of the same galaxy can look completely different. As our model is not aware of this, spirals e.g. split themselves into a face-on and edge-on region (see Figure \ref{fig:umap_images_HSC}). (3) The number of TNG galaxies is limited: We might not find for a given HSC image a very similar TNG galaxy. The model must therefore learn to interpolate among the various galaxy images and their morphological features. The quality of the model is therefore another source of uncertainty.

Previous works have shown that TNG galaxies reproduce, to first approximation, many fundamental properties and statistics of observed galaxies \citep[e.g.][]{Genel_2018, Torrey_2019, Gomez_2019, Donnari_2019, Huertas_2019}. In the previous Section~\ref{sec:nnd}, we have also shown that the largest majority of the simulated galaxies from TNG are well within the domain of observed HSC galaxies, without any remarkable difference between TNG100 and TNG50 \citep[differently than what found by ][with an ML-based comparison of galaxy images from SDSS]{Zanisi_2021}. However, details of the stellar structures of TNG galaxies have also been shown to deviate in certain aspects and regimes from observations. In particular:
\begin{itemize}
    \item Tension in the morphology-colour-size relation was found between observed Pan-STARRS and simulated TNG100 galaxies \citep{Gomez_2019}.
    \item It was shown with previous machine learning approaches that the fine-grained details in the structures of certain classes of galaxies, specifically small, more spheroidal, and quenched galaxies, are different between observed SDSS and simulated TNG galaxies \citep{Zanisi_2021}.
    \item Higher concentrations and asymmetries were found in low-mass TNG50 galaxies compared to Kilo-Degree Survey (KiDS) observations \citep{Ortega_2023}.
    \item Finally, in this work we have seen that certain TNG galaxies exhibit blue clumps in the images that are not visible for HSC (Figures \ref{fig:umap_images_TNG100} and \ref{fig:umap_images_TNG50}).
\end{itemize}
For the purposes of a simulation-based inference framework, what needs to be done is therefore to develop a method that only maps from similar and realistic-enough galaxies and that allows us to identify and exclude non-similar galaxies. Or even better, what is needed is a metric that allows to place a confidence score on the predictions based on some identifiable galaxy properties.

In this work, by matching the datasets according to redshifts, radii and luminosities, we have already ensured that the three datasets have the same priors in terms of very basic parameters. We have then trained the contrastive learning method NNCLR on all datasets (simulated and observed) at the same time: in this way, the model is agnostic about the fact that it is looking at images from different domains and can find similarities and differences among the images.
We have then compared the images based on the location of their respective representations derived by this model. To this aim, we have introduced an Out of Domain (OOD) Score that describes the local difference in point density in the high dimensional-representation space. In doing so we are able to identify regions in the representation space where galaxies of a dataset are over- or under-represented. As a success we have demonstrated that simulated galaxies with TNG and observed galaxies with HSC overall land roughly in the same space. The simulation methods and the mocking procedure are therefore able to produce large samples of sufficiently realistic galaxies. We can for example also see that our model is e.g. able to understand the concept of a disk galaxy across the domains (see the counterpart examples of Figure~\ref{fig:counterparts} and following). We can further use the OOD Score to gauge the level of realism, as we expand upon next.

\subsubsection{The relation between OOD Score and the degree of realism}
We now want to further elaborate about the meaning of the OOD Score introduced in this work in terms of realism of the TNG set: is a low OOD Score equivalent to a high degree of realism? And vice versa?

A low OOD Score means that a TNG galaxy is mapped together with HSC galaxies to the same location in the representation space. Based on our checks, we know that there is an overall correlation between representation space and observable properties (e.g. Figures~\ref{fig:tng_umaps}, \ref{fig:tng_umaps_2}, \ref{fig:tng_umaps_3}). Furthermore our search for counterparts (Section~\ref{sec:counterparts}) and our basic interference exercise (Section~\ref{sec:infer}) showed consistent results. 

However, one of the uncertainties in the proposed approach are the features that the model looks at: are those the image properties and galaxy features that we are actually interested in? For example, there is a small minority of cases whereby a low OOD Score is actually assigned to galaxies that do not resemble their neighbours in representation space. We have found galaxies with a very low OOD Score that are not in fact similar, upon visual inspection, to their identified counterparts. This occurs for example when the Petrosian fit fails and the FOV is too large or too small (Figure~ \ref{fig:counterparts_5}). The ensuing difference in galaxy size with respect to the FOV is then identified as a major feature by the model and HSC and TNG galaxies are grouped according to it, mostly ignoring other observable features. This grouping then leads to a small OOD Score as the failing FOV determination is the common and defining issue.

Keeping these uncertainties in mind, we can provide an answer to the opening question (``is a low OOD Score for simulated galaxies equivalent to a high degree of realism?'') with a conservative yes. 
But it is always advisable to validate the OOD Score against basic galaxy properties and parameters that are available from the simulations, to double check the learned features and to eliminate outliers.

On the other hand, is a high OOD Score for simulated galaxies equivalent to a low degree of realism? A high OOD Score means that a TNG galaxy is in an area with few or no HSC galaxies. It is therefore identified as peculiar by the model. For TNG, we have seen that this is the case for e.g. galaxies with larger radii (half mass and sersic half light), steeper sersic profiles, smaller sersic ellipticities and larger asymmetries in comparison to the rest of the simulated galaxy population. Even though we should remember that, because of unreliable fits with \textsc{Statmorph}, we might lack some peculiar galaxies, we have confirmed that the high OOD Scores typically correspond to peculiar and non-realistic cases (Figure \ref{fig:counterparts_4}). Interestingly, as noted above in the discussion of Figure~\ref{fig:counterparts_3}, we have also found that HSC galaxies that have been assigned an OOD Score larger than $1$ can still be seemingly similar to their TNG counterparts. This could be caused by the fact that some disky TNG galaxies with clumpy features are repelled from the smoother HSC disks, leaving in their region more HSC than TNG galaxies and causing therefore an OOD Score of $1$ or more: there may still be realistic TNG galaxies in that region but ultimately we do not grasp the full diversity of galaxy formation as not all TNG disks land in the same region.

Keeping these uncertainties in mind, we can answer to the second question (``is a high OOD Score for simulated galaxies equivalent to a low degree of realism?'') also with a conservative yes. But it is advisable to consider various threshold values in OOD Score, i.e. various operational definitions of OOD Score, to discriminate between within and out of domain.

\subsubsection{Implications for simulation-based inference}
Finally, we want to comment about the implications of our results on the broader goal of implementing simulation-based inference from galaxy images. The starting point of this paper was the idea to replace the scalar set of inputs of the inference model presented in \citealt{Eisert_2023} with images. We can now replace the scalar inputs with the results of the ResNet pre-trained and validated in this paper, as we know that the resulting representation space faithfully encodes observational features of the imaged galaxies. Most notably, this is the case not only for TNG galaxies but also HSC galaxies. Therefore, with this paper, we have not only skipped the definition and calculation of summary statistics, but we also have a tool to gap the bridge between observations and simulations.

In future applications, to deal, on an first order, with the discrepancies between TNG and HSC galaxies, we can e.g. exclude all TNG/HSC galaxies from training that have an OOD Score above $1$.
Furthermore, we suggest to validate this method also by including mock images of other cosmological simulations like e.g. EAGLE and SIMBA. This would provide additional sets of a ground truth we can test against, similar to our inference test between TNG50 and TNG100 in Section~\ref{sec:infer}, and would greatly lower the chance to overfit towards unique details of the TNG mocks. Moreover, this would allow us to automatically marginalize over multiple galaxy formation and evolution models, as the posterior distributions inferred for any given parameter or property would also include the uncertainty caused by different numerical realizations.

Another possible application and avenue allowed by the tools developed in this work could be a check of the inferred values against scalars that are derived `traditionally' i.e. by analytical modelling. This can for example be done for properties like the photometric redshift (like done in Section~\ref{sec:infer} or the galaxy stellar mass. The ability to derive those statistics from stellar light images with a sufficiently high precision would give the model a higher credibility and a broad applicability. 

\section{Summary and Conclusions}
\label{sec:summary}

Modern cosmological hydrodynamical simulations for galaxy formation and evolution are now able to return tens of thousands of synthetic, reasonably realistic galaxies across cosmic time. However, quantitatively assessing the level of realism of simulated universes in comparison to the real one is difficult and, to a certain extent, open ended.
The main goal of this work was to develop a framework to compare simulated (IllustrisTNG) galaxies with observed (HSC: Hyper Suprime-Cam) galaxies at the image/map level without the need to define and calculate summary statistics. In particular, we have used a large sample of hundreds of thousands synthetic HSC-SSP images of galaxies from the TNG50 and TNG100 simulations \citep{Bottrell_2023} and compared them with real HSC-SSP galaxies over $0.1 \leq z \leq 0.4$ in the galaxy stellar mass range of $10^{9-12}\MSUN$.

To conduct the comparison, we have used imaging data from both HSC and TNG50/100 sets to train a Wide-ResNet model, by using the NNCLR contrastive learning method. Our model maps the images onto a common 256 dimensional representation space, whereby representations of galaxy images that are identified to be similar/different are close/far to each other. Using these representations, we have examined the degree of similarity among the datasets. In Section~\ref{sec:umap}, we showed that the representations correlate strongly with physical galaxy properties derived from the simulations. We showed that they also correlate with image-derived morphological properties like Sersic parameters. The network is therefore able to encode the morphological structures of galaxies in the representations and is by that also able to distinguish galaxies based on their morphological properties. In Section \ref{sec:similarity} we have investigated how the representations of mocks from TNG and HSC galaxies are related to each other. We have shown that there is overall a large intersection between the TNG and HSC datasets in the representation space, which also implies that the mocking procedure is overall working and that the simulation itself produces sufficiently realistic galaxies.

To quantify the level of agreement between simulated and observed galaxy images, we have introduced in Section~\ref{sec:similarity_score} a score, the so-called Out Of Domain Score, that aims at quantifying the similarity of a given galaxy relative to the galaxies in another set. We have connected the predicted values of the OOD Score of galaxies to a selection of --in principle observable-- galaxy properties.

In Section~\ref{sec:discussion}, we have shown first applications of our framework and methodology: for example, it can be used to find similar TNG galaxies of HSC galaxies; or it could be used in the opposite direction, to find HSC galaxies with certain properties, for example disk galaxies. As a first-order approach, we have shown that it is possible to infer redshift, i-band-mag and Petrosian radii from HSC galaxy images by taking the prediction from their closest TNG neighbour in representation space.

Our main findings are:
\begin{enumerate}
    \item Contrastive learning is an efficient way to learn morphological features of galaxy images (Figure~\ref{fig:tng_umaps} and following) without the need for prior labeling.
    \item HSC galaxies cluster close to the TNG galaxies in representation space: our model is not able to distinguish between simulated and observed galaxies for a large subset ($\gtrsim 70$ per cent) of the overall adopted galaxy population (Figures~\ref{fig:kde_umap_compare} and \ref{fig:similarity_distribution}). This indicates that the galaxies produced in the IllustrisTNG simulations are overall realistic also at the stellar-light image level.
    \item At the same time, we can identify types and subsets of TNG galaxies that have not been observed by HSC (Figures~\ref{fig:similarity_properties} and \ref{fig:counterparts_4}), and that hence are not realistic. For example, especially problematic are blue clumps in and around TNG50 and TNG100 galaxies that are less common in HSC observations. This outlier, or 'out-of-distribution', detection allows us to assess the realism of the simulated galaxies and their stellar light images.
    \item We are able to identify galaxy counterparts between HSC observations and the TNG simulations (Figure~\ref{fig:counterparts} and following).
    \item It is possible and practically feasible to transfer galaxy properties from the simulated galaxies on the observed HSC galaxy images (Figure~\ref{fig:infer_properties}).
\end{enumerate}

In the next work of our ERGO-ML series, we will use the representation space obtained in this work as input to a conditional Invertible Neural Network (cINN) to infer properties of observed galaxies by using the underlying connections predicted by the cosmological galaxy simulations. We will aim to predict posteriors for observable quantities (e.g. magnitudes, redshift, galaxy sizes/profiles), semi-observable quantities (e.g. stellar mass, fraction of disk stars) and unobservable quantities (e.g. ex-situ fraction, merger mass and times) directly from the images for observed HSC galaxies. The Out-of-Domain Score that we have introduced and studied in this paper will allow us to e.g. excluding from the training subsets of simulated galaxies that are deemed to be not sufficiently realistic. 

\section*{Acknowledgements}
LE and this work are supported by the Deutsche Forschungsgemeinschaft (DFG, German Research Foundation) under Germany’s Excellence Strategy EXC 2181/1-390900948, Exploratory project EP 3.4 (the Heidelberg STRUCTURES Excellence Cluster). CB gratefully acknowledges support from the Forrest Research Foundation and the Natural Sciences and Engineering Research Council of Canada (NSERC) via their post-doctoral fellowship program [PDF-546234-2020]. DN acknowledges funding from the Deutsche Forschungsgemeinschaft (DFG) through an Emmy Noether Research Group (grant number NE 2441/1-1). The primary TNG simulations were carried out with compute time granted by the Gauss Centre for Supercomputing (GCS) under Large-Scale Projects GCS-ILLU and GCS-DWAR on the GCS share of the supercomputer Hazel Hen at the High Performance Computing Center Stuttgart (HLRS). Part of this research was carried out using the High Performance Computing resources at the Max Planck Computing and Data Facility (MPCDF) in Garching, operated by the Max Planck Society. We thank the Max Planck computing support teams for their valued assistance. 

The Hyper Suprime-Cam (HSC) collaboration includes the astronomical communities of Japan and Taiwan, and Princeton University. The HSC instrumentation and software were developed by the National Astronomical Observatory of Japan (NAOJ), the Kavli Institute for the Physics and Mathematics of the Universe (Kavli IPMU), the University of Tokyo, the High Energy Accelerator Research Organization (KEK), the Academia Sinica Institute for Astronomy and Astrophysics in Taiwan (ASIAA), and Princeton University. Funding was contributed by the FIRST program from the Japanese Cabinet Office, the Ministry of Education, Culture, Sports, Science and Technology (MEXT), the Japan Society for the Promotion of Science (JSPS), Japan Science and Technology Agency (JST), the Toray Science Foundation, NAOJ, Kavli IPMU, KEK, ASIAA, and Princeton University. This paper makes use of software developed for Vera C. Rubin Observatory. We thank the Rubin Observatory for making their code available as free software at \url{http://pipelines.lsst.io/}. This paper is based on data collected at the Subaru Telescope and retrieved from the HSC data archive system, which is operated by the Subaru Telescope and Astronomy Data Center (ADC) at NAOJ. Data analysis was in part carried out with the cooperation of Center for Computational Astrophysics (CfCA), NAOJ. We are honoured and grateful for the opportunity of observing the Universe from Maunakea, which has the cultural, historical and natural significance in Hawaii. 

\section*{Data Availability}

Data directly related to this publication and its figures will be made available on request from the corresponding author. In fact, the IllustrisTNG simulations are already publicly available and accessible in their entirety at \href{www.tng-project.org/data}{www.tng-project.org/data} \citep{nelson2019illustristng}. The associated code and ML models developed here are also available upon request. The data pertaining the survey-realistic mocks of TNG50 and TNG100 galaxies have been made available on the same IllustrisTNG website \citep{Bottrell_2023}.



\bibliographystyle{mnras}
\bibliography{references} 



\appendix

\section{Hyperparametersearch with Optuna}
\label{sec:optuna}
To find the best working set of hyperparameters for the NNCLR machine learning method, we perform a hyperparameter search using \textsc{Optuna} \citep{Optuna}. Optuna is a bayesian hyperparametersearch framework that allows to identify the best set of parameters by iteratively probing and narrowing the posterior of the parameterspace to find the set of parameters that minimizes the final validation NNCLR i.e. contrastive-loss for the model.

We optimize the model architecture by training each model with random parameters (Batch Size, Dropout-rate of the ResNet, ResNet Depth and Width, Linear Layer Depth and Width, dimension of the representation space) given by the Optuna optimizer.  
We also optimize the model for the learning rate, learning rate patience and decay, patience of the early stopping, the augmentation parameters and the $L_2$ Decay. The parameters are summarized in Table~\ref{tab:optuna} together with the search range and the fiducial choices. Parameters that are not mentioned here are fixed.
Because of the large number of images, we perform the validation after every $10000$ seen images. Furthermore we stop the training if the validation loss has not improved for twice the Learning Rate patience, i.e. when the learning rate has decreased twice without an effect.

\begin{table*}
\centering
\begin{tabularx}{\linewidth}{{>{\hsize=0.4\hsize}X
                              >{\hsize=0.2\hsize}X 
                              >{\hsize=0.2\hsize}X}}
Parameter & Searchspace  & Best Value/Fiducial Choice\\ 
\hline
Batch Size & [4, 128] & 32\\ 
Depth of the ResNet & [10,16] & 16\\ 
Width of the ResNet & [1, 3] & 2\\ 
Dropout of the ResNet & [0.0, 0.5] & 0.3\\ 
Representation Dimension & [64, 512] & 256\\ 
Representation MLP Depth & [1, 3] & 2\\ 
Projection MLP Dimension & [64, 512] & 256\\ 
Projection MLP Depth & [1, 3] & 2\\ 
Initial Learning Rate & [0.0001, 0.005] & 0.001\\ 
Learning Rate Decay & [0.2, 0.9] & 0.8\\ 
Learning Rate Patience & [2, 10] & 3\\ 
L2 Decay & [0.000001, 0.001] & 0.000001\\ 
NCE Temperature & [0.01, 0.1] & 0.04\\ 
NNCLR Queue Size (in number of Batches) & [64, 4096] & 512\\ 
NNCLR Translation Transformation & [0.05, 0.5] & 0.25\\ 
NNCLR Scaling Transformation & [1.0, 2.5] & 1.1\\
NNCLR Gaussian Noise Transformation & [0.01, 0.1] & 0.08\\ 
NNCLR Gaussian Blur Transformation & [0.001, 10] & 0.1
\end{tabularx}
\caption{{\bf Overview over the model and training parameters that have been optimized by using the hyperoptimization tool Optuna.}}
\label{tab:optuna}
\end{table*}

This said, as the training of a single model can take more than 24 hours on a single GPU, it is computationally very expensive to probe the whole space at once. We therefore iterate with our optimization run:
\begin{enumerate}
    \item As the first step we use a fixed training time of 2h (By previous testing, we ensured that the model start to converge within this time-range). In doing so we get an overall rough idea about how the performance of the model is in general related to the choices without spending too much time for single runs.
    \item At a second step we narrow the search space based on the previous step and train until convergence is reached i.e. the validation loss is not improving anymore for a certain interval (this interval is part of the optimized parameters) or is pruned because it is under-performing. Under performance is triggered by an above median validation loss after 2h.
    \item Additionally we run a selection of runs with larger architectures as they might be deprived by the time-based training/pruning. 
\end{enumerate}
We perform each model training on a single GPU with 40GB Memory, which limits the possible model size, batch size and NNCLR queue size. We therefore can't ensure that larger models might not perform even better.

In the following we will briefly comment on our findings. First, the Batchsize has to be equal or larger than approx 32 to give good results; however with larger batchsizes we impose restrictions on the NNCLR Queue Size and model architecture because of limited GPU Memory. The length of the NNCLR Queue should be 256 (although the queue could be larger we see no significant improvement anymore from this point on. I.e. with $256 \times 32 = 8192$ images in the queue, our dataset is sufficiently sampled). We find that we should favour larger and wider ResNet model although again a bigger model is again limited by the GPU Memory. The dropout rate of the ResNet layers should be around 0.3 although the loss is not very sensitive to this parameter. For the linear layers we see for the depth not a very big sensitivity, we choose to use $2$ layers for each.
The dimension of the projection network clearly favours $256$ dimensions. For the dimension of the representation space we see best validation losses for $128$ or $256$. {\change We suspect that a smaller representation space prunes the information content too much while a too large representation space probably benefits over-fitting because of the additional degrees of freedom.} However, because of the results of previous study \citep{Company_2023}, which favour larger representation spaces, we settle for a dimension of $256$. 
Regarding the transformations we see that we better use small zoom ins of $1.1$, i.e. $10$ per cent. This might be due to the already quite narrow FOV. The translation parameter should be between $0.2 - 0.3$ (as fraction of the image edge length). The sigma for the Gaussian blur is rather small with $0.1$ pixels while the added noise is considerably larger with an upper standard deviation of $0.8$ (considering pixel values normalized to $1$).
With an initial learning rate of $0.001$ and a decay by a factor of $0.8$ every $3$ epochs without improvement we train the model with a slow but steady decrease in learning rate.

\section{Dependence of the Out of Domain Score on the model and parameters}
\label{sec:ood_dependence}
Each model training is a stochastic process and the resulting machine learning model is dependent on the architectural and parametric choices made. Here we investigate how much the results of this paper, especially the Out of Domain Score, depend on these choices and stochastic influences. 

\begin{figure*}
	\centering
        \includegraphics[width=0.45\linewidth]{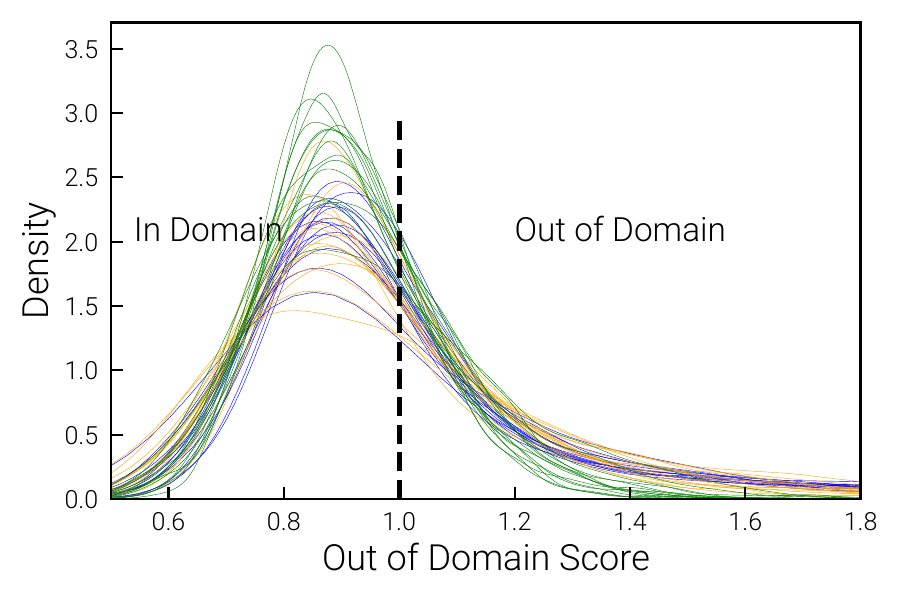}
        \includegraphics[width=0.45\linewidth]{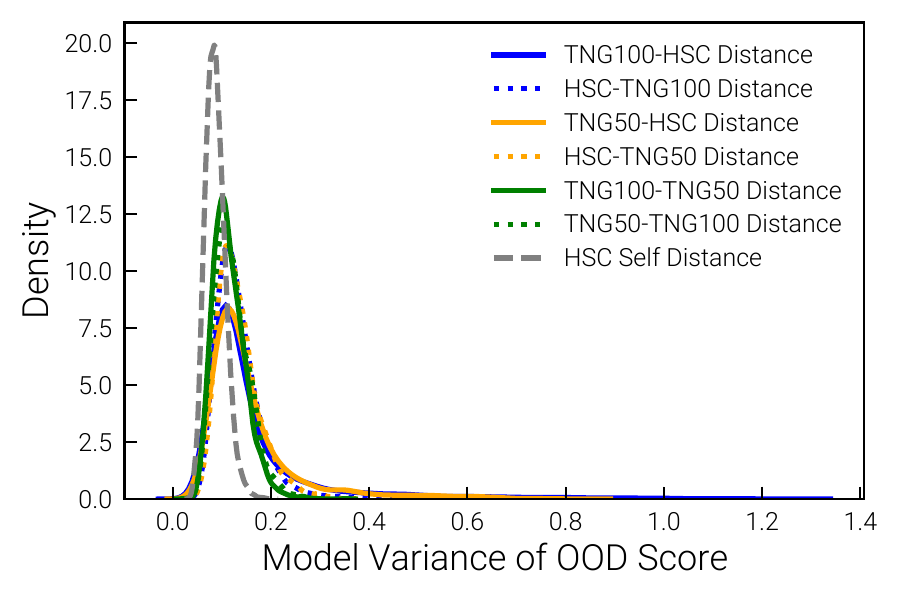}
	\caption{We show the influence of the model to model variation on the OOD Score defined and calculated in Section \ref{sec:nnd}. To do so we recalculate the OOD Scores for the 15 best Optuna runs and compare them at the galaxy-population level (left plot) and for individual galaxies (right plot). On the left we reproduce Figure \ref{fig:similarity_distribution} but showing the curves for all 15 runs. We see that the run to run scatter is too small to actually influence the plot qualitatively. On the right we show the model to model variance in OOD Score for each individual galaxy. The most galaxies have a variance below 0.2, which is small relative to the overall range in OOD values. } 
	\label{fig:run_variation}
\end{figure*}

To do so, we use the 15 best runs (in terms of validation loss) made for the hyper parameter optimization in Appendix \ref{sec:ood_dependence} and calculate representations with their resulting models like the fiducial run presented in this work. We show the variation in OOD Score among the runs in Figure~ \ref{fig:run_variation}. Although there is scatter among the 15 runs, the overall trends are clearly visible; e.g. there are more high OOD Score galaxies in TNG100 relative to HSC than in TNG50 relative to HSC, while the agreement between the two TNG runs is in general identified to be higher. Looking at the OOD Score variations for each galaxy individually, we see that the OOD calculation for most galaxies does not have a variance above $\pm 0.2$ across the 15 runs. I.e. galaxies with a large OOD Score are almost always consistently identified as OOD by the 15 models. //

\begin{figure*}
	\centering
        \includegraphics[width=0.45\linewidth]{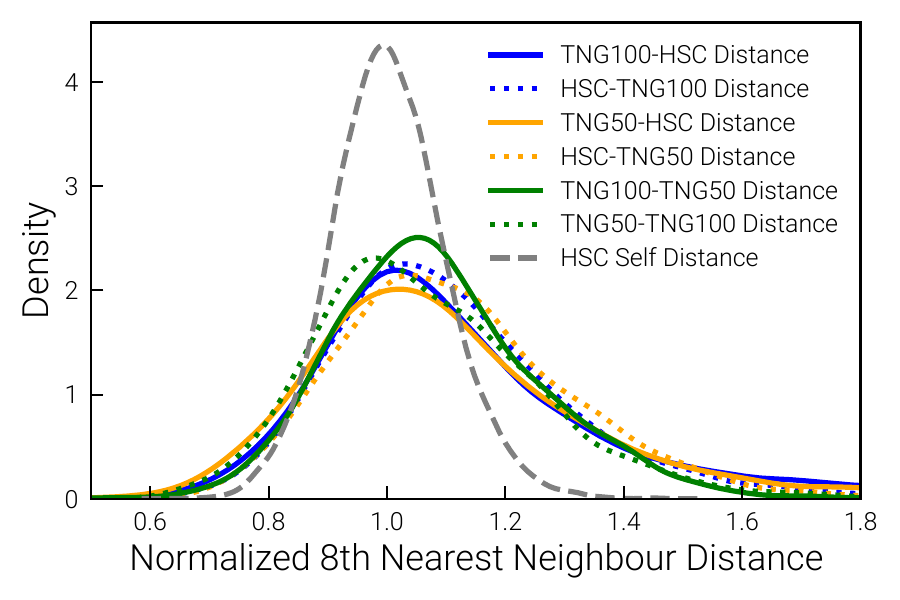}
        \includegraphics[width=0.45\linewidth]{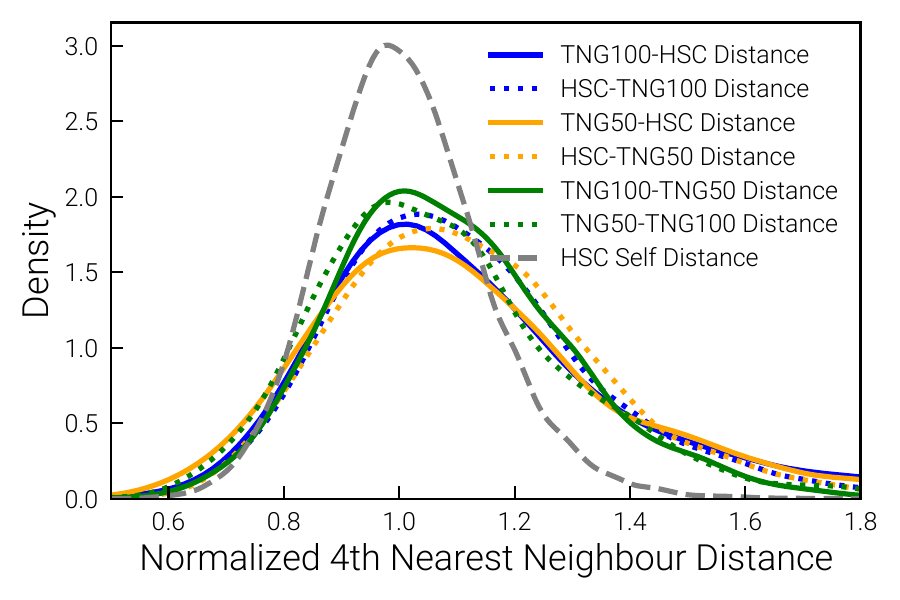}
        \includegraphics[width=0.45\linewidth]{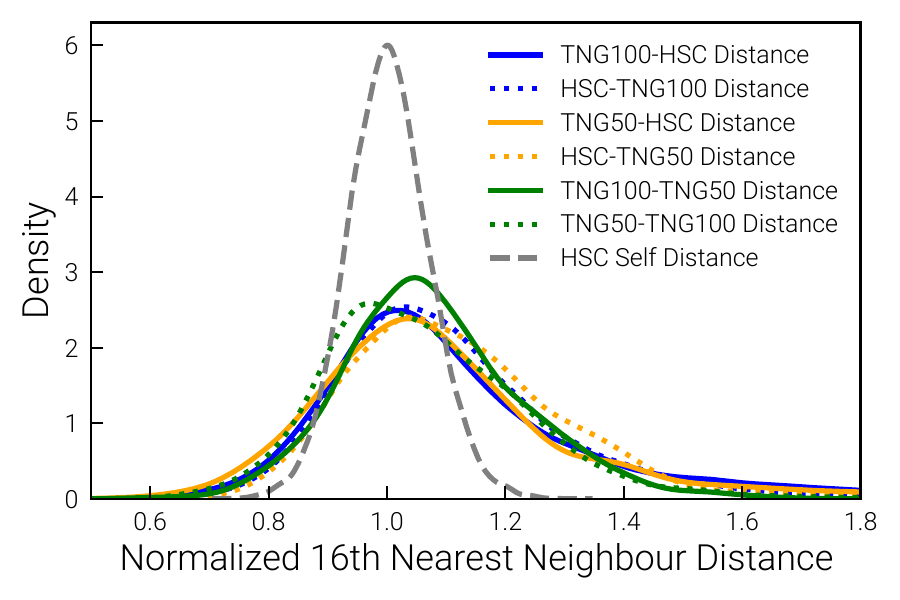}
        \includegraphics[width=0.45\linewidth]{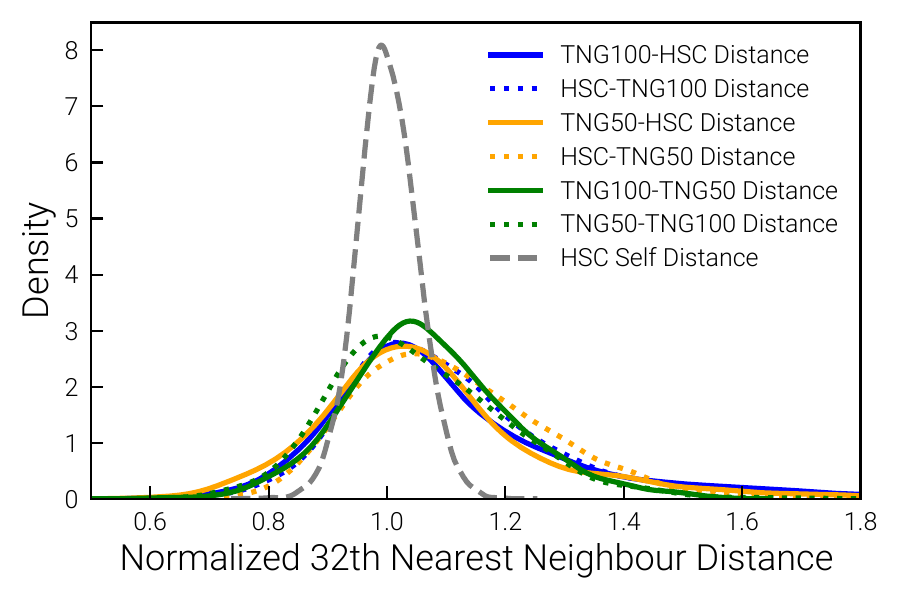}
	\caption{We show the influence of the choice in nearest neighbour on the calculation of the Normalized Nearest Neighbour Distance introduced in Section \ref{sec:nnd}. We show the same plot for (from top-left to bottom-right) 4 choices in nearest neighbour: 8 (our fiducial choice), 4, 16 and 32. While the variation has an influence on the curve, the differences are too marginal to influence the claims made in this work.}
	\label{fig:nn_variation}
\end{figure*}

For the calculation of the OOD Score we also perform a nearest neighbour search. Our fiducial choice is to use the 8th nearest neighbour; as validation, we also recalculate the distances using the 4th, 16th and 32th nearest neighbour in Figure~\ref{fig:nn_variation}. While such a parameter has an influence, i.e. especially causes a smoothing of the curves, we do not see a significant change among the choices.

\section{Example for False Friends because of a failing FOV determination}
\label{sec:failed_FOV}
\begin{figure*}
	\centering
        \includegraphics[trim={0 3cm 0cm 2.5cm},clip,width=\linewidth]{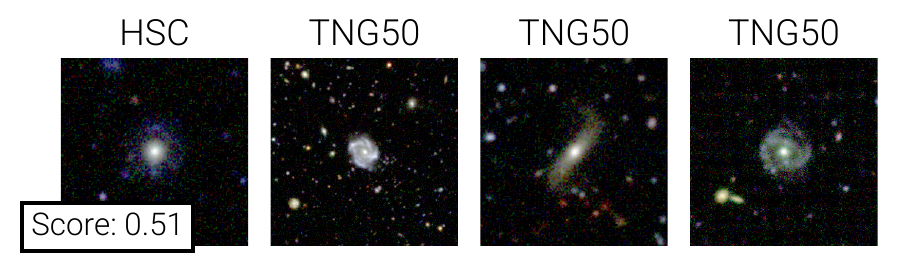}
	\caption{Examples for ``False Friends'' because of a failing FOV determination. The depicted HSC galaxy on the left is given a very low score as it got closely grouped together with other galaxies (from HSC and TNG) whose FOV is significantly larger than the average. It is possible to identify such failing cases either by visual inspection or by comparing the known galaxy properties for the TNG mock images.}
	\label{fig:counterparts_5}
\end{figure*} 
We find galaxies with very small values of the OOD Score, which we have introduced in the main text of this paper as an indicator for high similarity between any given galaxy and a set of galaxies for comparison. However, after closer inspection, we found that some of those imaged galaxies are not similar at all: rather, the FOV of these galaxy images are much larger than the average of the images. An example for such a case can be seen in Figure~\ref{fig:counterparts_5}. The model captured this large FOV as a significant similarity/point of tension with the other galaxies. This is a good example of how the definition of similarity by a machine learning model can differ from a human definition. Furthermore it underlines the importance to use validation checks and to not completely rely on the contrastive learning in finding a meaningful representation.


\bsp	
\label{lastpage}
\end{document}